\documentclass[sort&compress]{elsarticle}

\usepackage{amsfonts}
\usepackage{graphicx}
\usepackage{latexsym}
\usepackage{amsmath,amssymb}
\usepackage{mathrsfs}
\usepackage{hhline}
\usepackage{dcolumn}
\usepackage{bm}
\usepackage[usenames,dvipsnames]{xcolor}
\usepackage{supertabular}

\usepackage{esvect} 

\usepackage[normalem]{ulem}

\usepackage{url}

\begin{document}
\begin{frontmatter}
\title{Random walks and diffusion on networks}

\author{Naoki Masuda}
\ead{naoki.masuda@bristol.ac.uk}
\address{Department of Engineering Mathematics, University of Bristol, Bristol, UK}

\author{Mason A. Porter}
\address{Department of Mathematics, University of California Los Angeles, Los Angeles, USA}
\address{Mathematical Institute, University of Oxford, Oxford, UK}
\address{CABDyN Complexity Centre, University of Oxford, Oxford, UK}

\author{Renaud Lambiotte}
\address{Department of Mathematics/Naxys, University of Namur, Namur, Belgium}
\address{Mathematical Institute, University of Oxford, Oxford, UK}


\begin{abstract}
Random walks are ubiquitous in the sciences, and they are interesting from both theoretical and practical perspectives. They are one of the most fundamental types of stochastic processes; can be used to model numerous phenomena, including diffusion, interactions, and opinions among humans and animals; and can be used to extract information about important entities or dense groups of entities in a network. Random walks have been studied for many decades on both regular lattices and (especially in the last couple of decades) on networks with a variety of structures. In the present article, we survey the theory and applications of random walks on networks, restricting ourselves to simple cases of single and non-adaptive random walkers. We distinguish three main types of random walks: discrete-time random walks, node-centric continuous-time random walks, and edge-centric continuous-time random walks. We first briefly survey random walks on a line, and then we consider random walks on various types of networks. We extensively discuss applications of random walks, including ranking of nodes (e.g., PageRank), community detection, respondent-driven sampling, and opinion models such as voter models.

\end{abstract}

\maketitle
\tableofcontents


\end{frontmatter}

\section{Introduction\label{sec:introduction}}

Random walks (RWs) are popular models of stochastic processes with a very rich history \cite{Aldous2002book,Feller-book1, Feller1971book2,Hughes1995book,Kutner2017EurPhysJB}. \footnote{See \url{https://www.youtube.com/watch?v=stgYW6M5o4k} for an introduction to random walks for a public audience from the U.S. Public Broadcasting Service (PBS).}
 The term ``random walk'' was coined by Karl Pearson \cite{Pearson1905Nature}, and the study of RWs dates back to the ``Gambler's Ruin'' problem analyzed by Pascal, Fermat, Huygens, Bernoulli, and others \cite{Ore1960AmMathMonthly}. Additionally, Albert Einstein formulated stochastic motion (in the form of ``Brownian motion'') of particles in continuous time due to their collisions with atoms and molecules \cite{Einstein1905AnnPhys-brownian}. Theoretical developments have involved mathematics (especially probability theory), computer science, statistical physics, operations research, and more. RW models have also been applied in various domains, ranging from locomotion and foraging of animals \cite{Viswanathan1999Nature,Codling2008JRSocInterface,Humphries2010Nature,Okubo2001book}, the dynamics of neuronal firing \cite{Tuckwell1988book2,Gabbiani2010book} and decision-making in the brain \cite{Usher2001PsycholRev,Gold2007AnnuRevNeurosci} to population genetics \cite{Ewens2010book}, polymer chains \cite{fisher1966,isic1992}, descriptions of financial markets \cite{CampbellLoMackinlay1996book,Mantegna1999book}, evolution of research interests (through RWs on problem space) \cite{jia2017}, ranking systems \cite{Gleich2015SiamRev}, dimension reduction and feature extraction from high-dimensional data (e.g., in the form of ``diffusion maps'') \cite{Coifman2005PNAS-1,Coifman2006ApplComputHarmonAnal},
and even sports statistics \cite{clauset2015,Godreche2017arxiv}. RW theory can also help predict arrival times of diseases spreading on networks \cite{Iannelli2017PhysRevE}.
There exist several monographs and review papers on RWs. Many of them treat RWs on classical network topologies, such as regular lattices (e.g., ${\mathbb Z}^d$) and Cayley trees (i.e., trees in which each node has the same number of neighboring nodes, which we henceforth call the node ``degree'')
\cite{Spitzer1976book,Weiss1994book,Hughes1995book,Redner2001book,Burioni2005JPhysA,Krapivsky2010book,Klafter2011book,Ibe2013book}. Other monographs and surveys focus on RWs on fractal structures, revealing diffusion properties that are ``anomalous'' compared to RWs on regular lattices or Euclidean spaces (i.e., $\mathbb{R}^d$) \cite{Rammal1984JStatPhys,Havlin1987AdvPhys,Bouchaud1990PhysRep,Benavraham2000book,Burioni2005JPhysA,Benichou2014PhysRep}. Other literature treats RWs on finite networks, which are equivalent to a finite Markov chain (in the discrete-time case) \cite{Doyle1984book,Lovasz1993Boyal,Aldous2002book,Burioni2005JPhysA} and are at the core of several stochastic algorithms.

In parallel, ``network science'' has emerged in recent years as a central approach to the study of complex systems \cite{ejam-special-2016,Newman2010book,Barabasi2016book,Boccaletti2006PhysRep}. Networks are a natural representation of systems composed of interacting elements and allow one to examine the impact of structure on the dynamics and function of a system (as well as the impact of dynamics and function on network structure). Examples include friendship networks, international relationships, gene-regulatory networks, food webs, airport networks, the internet, and myriad more. In each case, one can represent the system's connectivity structure as a set of nodes (representing the entities in the system) and edges (representing interactions among those entities). The study of networks is highly interdisciplinary, and it integrates theoretical and computational tools from subjects such as applied mathematics, statistical physics, computer science, engineering, sociology, economics, biology, and other domains. Many networks exhibit complex yet regular patterns that are explainable (sometimes arguably) by simple mechanisms. Network science has also had a strong impact on the understanding of dynamical processes because of the critical role of structure on spreading processes, synchronization, and others \cite{strogatz01,Barrat2008book,Porter2016book}.
%
%
As with RWs, numerous books and review papers have been written on networks, including textbooks \cite{Dorogovtsev2003book,Newman2010book,CohenHavlin2010book,Estrada2012book,Barabasi2016book}, general review articles
\cite{Newman2003SiamRev,Boccaletti2006PhysRep}, and more specialized reviews on topics such as dynamical processes on networks \cite{Arenas2008PhysRep,Barrat2008book,Porter2016book},
connections to statistical physics \cite{Albert2002RevModPhys,Dorogovtsev2008RevModPhys}, temporal networks \cite{HolmeSaramaki2012PhysRep,Holme2015EurPhysJB,MasudaLambiotte2016book}, multilayer networks \cite{Boccaletti2014PhysRep,Kivela2014JCompNetw,naturephysicsspreading}, and community structure \cite{porter2009,Fortunato2010PhysRep,santo2016}.


The main purpose of the present review is to bring together two broad subjects --- RWs and networks --- by discussing their many interconnections and their ensuing applications. RWs are often used as a model for diffusion, and there has been intense research on the impact of network architecture on the dynamics of RWs. Moreover, nontrivial network structure paves the way for different definitions of RWs, and different definitions can be ``natural'' from some perspective, while leading to different diffusive processes on the same network. Finally, RWs are at the core of several algorithms to uncover structural properties in networks. We will discuss these points further in the next three paragraphs.

First, RWs are often used as a model for diffusion, and there has been intense research on the impact of network architecture on the dynamics of RWs. The finiteness of a network --- along with properties such as degree heterogeneity, community structure, and others --- can make diffusion on networks both quantitatively and even qualitatively different from diffusion on regular or infinite lattices. RWs on networks are an example of a Markov chain in which the set of nodes is the state space and the transition probabilities depend on the existence and weights of the edges between nodes. In this review, we will include a summary of results on the dependence of dynamical properties --- including stationary distribution and mean first-passage time --- on structural properties of an underlying network.

Second, the irregularity of underlying network structure opens the door for different definitions of RWs. Each is ``natural'' from some perspective, but they lead to different diffusive processes even when considering the same network. For example, it is useful to distinguish between discrete-time and continuous-time RWs. On networks in which degree (i.e., the number of neighbors) is heterogeneous (i.e., it depends on the node), one needs to subdivide continuous-time RWs further into two major types, depending on whether the random events that induce walker movement are generated on nodes or edges and corresponding to different types of propagators (normalized versus unnormalized Laplacian matrices). Different literatures use different variants of RWs, often implicitly. We distinguish different types of RWs and clarify the relationship between them, and we discuss formulations and results that are informed by empirical networks (such as networks with heavy-tailed degree distributions, multilayer networks, and temporal networks).

Finally, RWs lie at the core of many algorithms to uncover various types of structural properties of networks. Consider the notion of identifying ``central'' nodes, edges, or other substructures in networks \cite{Newman2010book}. A powerful set of diagnostics (e.g., PageRank \cite{Brin1998conf,Gleich2015SiamRev} and eigenvector centrality \cite{bonacich1972}) are derived based on recursive arguments of the type ``a node is important if it is connected to many important nodes'', and such derivations often rely on the trajectories of random walkers. Similarly, flow-based algorithms, based on trajectories of dynamical processes (e.g., RWs) being trapped within certain sets of node for a long time, are helpful for discovering mesoscale patterns in networks \cite{santo2016,Jeub2015PhysRevE}. These techniques and algorithms open a wealth of applications that go well beyond classical applications of RWs. Their design benefits both explicitly and implicitly from developing an understanding of how RW dynamics are influenced by network structure and how different types of RWs behave on the same network.

There has been a vast amount of research on RWs on networks, and it is scattered across disparate corners of the scientific literature. It is impossible to cover everything, and we choose specific subsets of it to make our review cohesive, although we will occasionally include pointers to other parts of the landscape.
%
%
First, we focus on the most standard types of RWs, in which a random walker moves to a neighbor with a probability proportional to edge weight, and their very close relatives.
We only very rarely mention some of the numerous other types of RWs, which include correlated RWs \cite{Gillis1955ProcCambPhilSoc}, self-avoiding RWs \cite{Domb1983JStatPhys,Madras1993book,Hughes1995book}, 
zero-range processes \cite{EvansHanney2005JPhysA}, multiplicative random processes \cite{Schenzle1979PhysRevA,Havlin1988PhysRevLett}, adaptive RWs (including reinforced RWs \cite{Pemantle2007ProbSurv}), branching RWs \cite{Schinazi1999book}, L\'{e}vy flights \cite{Klafter2011book,Ibe2013book}, elephant RWs \cite{SchutzTrimper2004PhysRevE}, quantum walks \cite{Kempe2003ComtempPhys,Mulken2011PhysRep}, intermittent RWs \cite{Benichou2011RevModPhys}, persistent RWs \cite{Tejedor2012PhysRevLett}, starving RWs \cite{Benichou2014PhysRevLett,Benichou2016JPhysA,Bhat2017arxiv}, mortal RWs \cite{mortal}, and so on.
These processes are of course fascinating, and many of the different flavors of RWs are often developed with specific motivation from an application (e.g., a Pac-Man-like ``hungry RW'' \cite{hungry2016} has been used as a model for chemotaxis in a porous medium), are often inspired by applications, such as animal movement \cite{Codling2008JRSocInterface,Okubo2001book} or financial markets \cite{Mantegna1999book}, and one can find discussions of different flavors of RWs in Refs.~\cite{Hughes1995book,Klafter2011book,Ibe2013book}.
Second, we will not cover many results for RWs on particular generative models of networks, except that we do give extensive attention to first-passage times for fractal and pseudo-fractal network models (see Section~\ref{sub:MFPT}).
Third, we will not discuss various important, rigorous results from mathematics and theoretical computer science. For such results, see \cite{Doyle1984book,Lovasz1993Boyal,Weiss1994book,Hughes1995book,Aldous2002book}. We focus instead on results that we believe give physical insight on RW processes and their applications.

As a final warning, we focus exclusively on diffusive processes in which the total number of walkers (or, equivalently, the total probability of observing a walker) is a conserved quantity \footnote{We thus consider ``conservative'' processes, though non-conservative processes are also interesting \cite{lerman,Yan2016PeerJComputSci}.}. The only exception is in Section~\ref{sub:voter model}, where we use ``coalescing RWs'' as an analytical tool. As we will see, this conservation rule translates into certain properties of the operator that drives the RW process. When transposed, the operator leads naturally to linear models for consensus dynamics (see Sections~\ref{sub:voter model} and \ref{sub:DeGroot model}). Among notable non-conservative processes, which we do not cover in this review, are classical epidemic processes \cite{Anderson1991book,Barrat2008book,Pastorsatorras2015RevModPhys,Porter2016book}, in which the number of entities (e.g., viruses or infected individuals) varies over time. In the linear regime, corresponding to a small number of infected nodes, the propagator of infection events in simple epidemic processes such as susceptible--infected (SI) and susceptible--infected--recovered (SIR) models are the adjacency matrix \cite{Wang2003SRDS,Klemm2012SciRep}. In contrast, a propagator of an RW is a type of Laplacian matrix, as we will discuss in detail in Section~\ref{sec:nets}. If all nodes have the same degree, these Laplacian and adjacency matrices are related linearly, and their dynamics are essentially the same \cite{Godsil2001book,MasudaLambiotte2016book}.
%
%
However, they are generically different for heterogeneous networks, such as when degree depends on node identity. Therefore, the difference between conservative dynamics (described by a Laplacian matrix) and non-conservative dynamics (described by the adjacency matrix) tends to be more striking for heterogeneous than for homogeneous networks. Other spreading models that are also beyond the scope of this work include threshold models of social contagions \cite{valente-book,Porter2016book} (e.g., for modeling adoption of behaviors) and reaction--diffusion dynamics \cite{ReactionDiffusion}.

The rest of our review proceeds as follows. In Section \ref{line}, we discuss RWs on the line. In Section \ref{sec:nets}, we give a lengthy presentation of RWs on networks. We then discuss RWs on multilayer networks in Section \ref{sec:multilayer} and RWs on temporal networks in Section \ref{sub:temporal networks}. We discuss applications in Section \ref{sec:applications}, and we conclude in Section \ref{sec:outlook}.


\section{Random walks on the line} \label{line}

In this section, we review some basic properties of RW processes on one-dimensional space (i.e., the infinite line). This section serves as a primer to later sections, in which we examine RWs on general networks. In this and later sections, we carefully distinguish between discrete-time and continuous-time models.


\subsection{Discrete time\label{dtrw}}

Consider a discrete-time RW (DTRW) process on the infinite line, which we identify with $\mathbb{R}^1 \equiv \mathbb{R}$. There is a single walker. At each discrete time step, it moves from some point
to some other point, including the case of moving from a point to itself. The length and direction of the move are both random variables. We assume that the probability that a walker located at $x$ moves to the interval $[x+r, x+r+\Delta r]$ in one step is equal to $f(r)\Delta r$. The normalization is $\int_{-\infty}^{\infty} f(r) {\rm d}r=1$, and we assume that moves at different times are independent.

Let's derive the probability density $p(x; n)$ that a random walker is located at a point $x \in \mathbb{R}$ after $n$ steps. (For emphasis, we sometimes use the term ``discrete time'' or ``event time'' for $n$.)
%
%
The master equation is given by
\begin{equation}
	p(x; n) =  \int_{-\infty}^{\infty} f(x-x^{\prime}) p(x^{\prime}; n-1) {\rm d}x^{\prime}\,.
\label{eq:RW master equation 1dim}
\end{equation}
It is convenient to solve Eq.~\eqref{eq:RW master equation 1dim} for general $x$ and $n$ in the Fourier domain. We define the Fourier transform by
\begin{equation}
	\hat{p}(k; n) \equiv  \int_{-\infty}^{\infty} p(x; n) e^{-i k x} {\rm d}x
\label{eq:def FT}
\end{equation}
and the inverse Fourier transform by
\begin{equation}
	p(x; n) \equiv \frac{1}{2 \pi} \int_{-\infty}^{\infty} \hat{p}(k; n) e^{i k x} {\rm d}k\,.
\label{eq:inverse FT}
\end{equation}
Note that $\hat{p}(-k; n)$ is the ``characteristic function'' of a random variable $x$ with probability density $p(x; n)$. The Fourier transform $\hat{f}(k)$ of $f(x)$ is sometimes called the ``structure function'' of the RW. The Taylor expansion of $\hat{p}(k; n)$ around $k=0$ yields
\begin{align}
	\hat{p}(k; n) =&  \langle e^{-i k x}\rangle\notag\\
	=& 1- i k \langle x\rangle - \frac{1}{2} k^2 \langle x^2\rangle + O(k^3)\,,
\label{smallk}
\end{align}
where $\langle \cdot\rangle$ is the expectation unless we state otherwise.
One can thereby obtain moments of $p(x; n)$ from the derivatives of $\hat{p}(k; n)$ at $k=0$.

The Fourier transform maps a convolution, such as Eq.~\eqref{eq:RW master equation 1dim}, to a product; and Eq.~\eqref{eq:RW master equation 1dim} thus yields
\begin{equation}
	\hat{p}(k; n) = \hat{f}(k) \hat{p}(k; n-1)\,.
\end{equation}
If a random walker is located initially at $x=0$, we obtain $p(x;0)=\delta(x)$, where $\delta(x)$ is the Dirac delta function, which has Fourier transform $\hat{p}(k;0)=1$. We thereby obtain
\begin{equation}\label{four}
	\hat{p}(k; n) = \left[\hat{f}(k) \right]^n\,.
\end{equation}
Using the inverse Fourier transform in Eq.~\eqref{eq:inverse FT}, we obtain a formal solution for $p(x; n)$ in the time domain:
\begin{equation}
	p(x; n) = \frac{1}{2 \pi} \int_{-\infty}^{\infty} \left[\hat{f}(k) \right]^n e^{i k x} {\rm d}k\,.
\label{eq:inverse FT RW}
\end{equation}

The qualitative behavior of the solution in Eq.~\eqref{eq:inverse FT RW} depends on the details of the structure function $\hat{f}(k)$. However, the asymptotic behavior of the RW as $n \rightarrow \infty$ depends only on some of the properties of $\hat{f}(k)$. When the first two moments of $\hat{f}(k)$ are finite, the solution converges to the Gaussian profile
\begin{equation}
	p(x; n) = \frac{1}{(4 \pi D n)^{1/2}} e^{- \frac{(x-v n)^2}{4Dn}}\,,
\label{eq:Gaussian profile}
\end{equation}
where $v\equiv\langle r\rangle$ and $D\equiv \langle (r-\langle r\rangle)^2\rangle / 2$.
Equation~\eqref{eq:Gaussian profile} implies that the variance of $x$ grows linearly with time. This result is the ``central limit theorem'' for the sum of the sizes of the moves, which are independent random variables. This asymptotic regime is well-defined because the underlying space (i.e., the line) is infinitely large.
One can derive these results in a similar manner when the underlying space is discrete (e.g., a one-dimensional lattice) \cite{Feller-book1,Weiss1994book,Hughes1995book,Redner2001book}. In situations in which the second moment of the structure function diverges, the process exhibits superdiffusion and the probability profile converges to so-called ``L\'evy distributions'' \cite{Klafter2011book,Ibe2013book}.


\subsection{Continuous time\label{sub:CTRW}}

In this section, we consider continuous-time RWs (CTRWs), which incorporate the timing of moves \cite{Montroll1965JMathPhys,Weiss1994book,Hughes1995book,Klafter2011book,Ibe2013book,Kutner2017EurPhysJB}. We assume that a walker waits betweegn two moves for a duration $\tau$ that independently obeys the probability density function $\psi(\tau)$. In other words, the move events are generated by a renewal process \cite{Feller1971book2}. If $\tau=1$ with probability $1$, the CTRW reduces to the DTRW described in Section \ref{dtrw}. In a standard CTRW, one assumes that the time of a move event and the selection of a destination in a given move are independent. Therefore, a combination of $\psi(\tau)$ and $f(r)$, where $r$ is the displacement in a single move, completely determines the dynamical properties of a random walker.

\begin{figure}[tb]
\begin{center}
\includegraphics[scale=0.37]{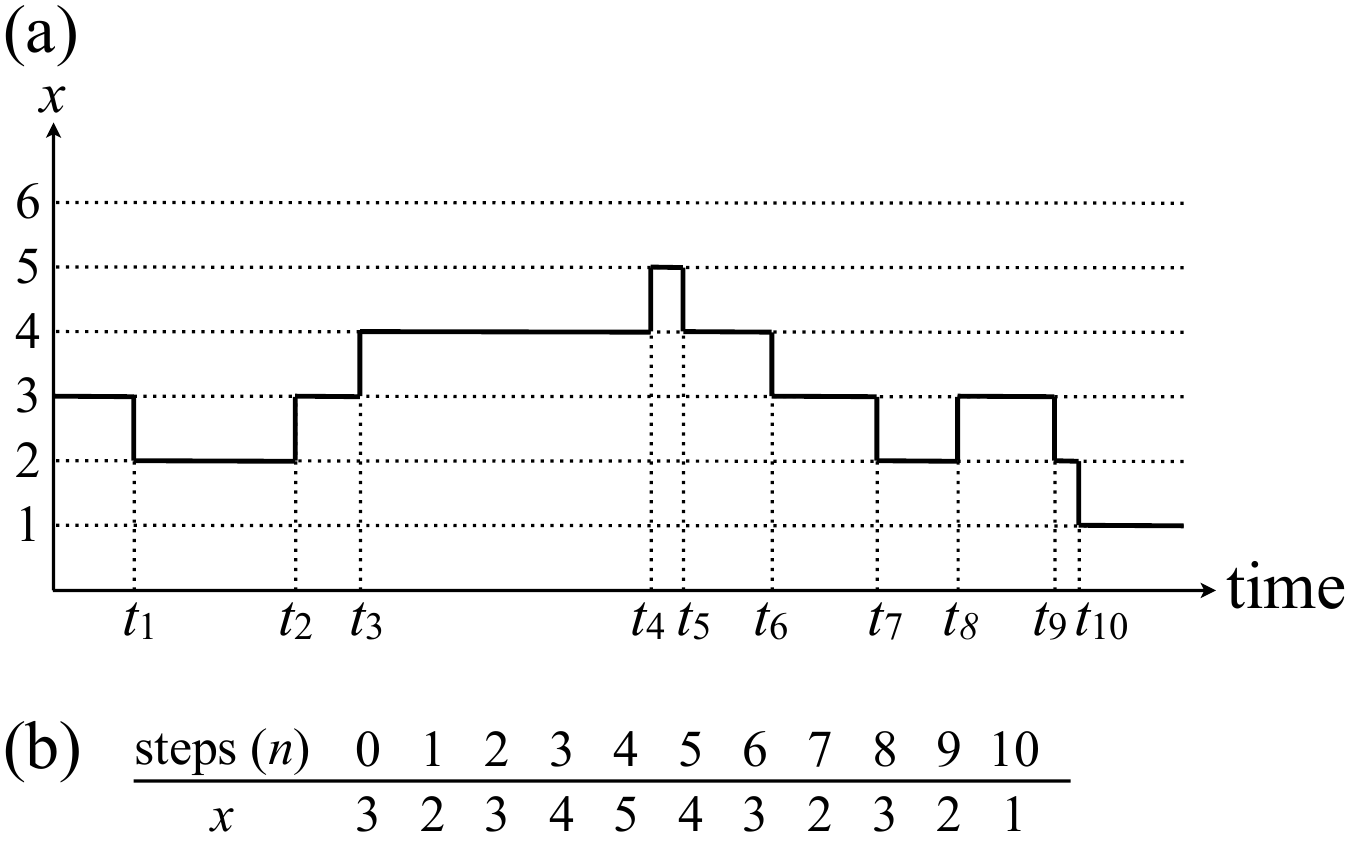}
\caption{Schematic of the standard continuous-time random walk (CTRW) on a one-dimensional lattice. (a) The position $x$ of the walker in physical time $t$ is described by $p(x; t)$. Note that $t_n$ represents the time of the $n$th move. (b) The position of the walker after $n$ moves is described by $p(x; n)$.}
\label{fig:p(x;t) vs p(x;n)}
\end{center}
\end{figure}

Let $t_n$ denote the time of the $n$th move. By definition, $t_n = \sum_{i=1}^n \tau_i$, where each $\tau_i$ is independent and identically distributed (i.i.d.) and drawn from some distribution $\psi(\tau)$.
Additionally, we can write
\begin{equation}
	p(x; t) = \sum_{n=0}^{\infty} p(x; n) p(n,t)\,,
\label{eq:p(x;t) with free n}
\end{equation}
where $p(x; t)$ is the probability that the walker is located at $x$ at time $t$, the quantity $p(x; n)$ is the probability that the walker is located at $x$ after $n$ steps, and $p(n, t)$ is the probability density that the walker has moved $n$ times at time $t$.
Note that it is crucial to distinguish $p(x; t)$ and $p(x; n)$, and we illustrate the difference between these probabilities with a schematic in Fig.~\ref{fig:p(x;t) vs p(x;n)}. Equation~\eqref{eq:p(x;t) with free n} reflects the fact that a walker can visit $x$ at time $t$ after some number $n$ of steps.

The probability $p(x; n)$ is given by the same solution, Eq.~\eqref{eq:inverse FT RW}, as for the DTRW. To obtain $p(x; t)$ from Eq.~\eqref{eq:p(x;t) with free n}, we need to examine $p(n, t)$, and we thus need to consider a renewal process generated by $\psi(\tau)$. According to the elementary renewal theorem \cite{Cox1962book}, the mean of $n$ at time $t$ is
\begin{equation}
	\langle n\rangle = \frac{t}{ \langle \tau\rangle}\,.
\label{eq:elementary renewal theorem}
\end{equation}
Equation~\eqref{eq:elementary renewal theorem} indicates that $n(t)$ grows linearly with time on average, irrespective of the details of the distribution $\psi(\tau)$. However, realized values of $n$ are random, inducing heterogeneity in the length of the RW ``trajectory'' (i.e., the walk measured in terms of the number of moves) observed at a given time $t$.

When the CTRW is driven by a Poisson process, $\psi(\tau)$ is the exponential distribution (i.e., $\psi(\tau) = \beta e^{-\beta \tau}$). In this case, $n$ obeys the Poisson distribution with mean $\beta t$. That is,
\begin{equation}
	p(n, t) = \frac{(\beta t)^n}{n!} e^{-\beta t}\,.
\end{equation}

It requires some effort to derive $p(n, t)$ when $\psi(\tau)$ is a general distribution. To calculate the time of the $n$th event or the number of events in a given time interval, we need to sum i.i.d. variables that obey $\psi(\tau)$. The duration $\tau\ge 0$ is nonnegative, so we take a Laplace transform
\begin{equation}
	\hat{\psi}(s) =  \int_{0}^{\infty} \psi(\tau) e^{- s \tau} {\rm d}\tau \equiv \langle e^{- s \tau} \rangle\,.
\label{eq:Laplace renewal}
\end{equation}
The Taylor expansion of Eq.~\eqref{eq:Laplace renewal} is given by
\begin{equation}
	\hat{\psi}(s) =  \sum_{n=0}^{\infty} (-1)^n \frac{\langle \tau^n\rangle s^n}{n!}
\label{eq:hatpsi(s) Taylor}
\end{equation}
and implies that $\hat{\psi}(s)$ generates the moments of $\psi(\tau)$ if they exist.
One computes the inverse Laplace transform by integrating in the complex plane:
\begin{equation}
	\psi(\tau) = \frac{1}{2 \pi i} \int_{c - i \infty}^{c + i \infty} \hat{\psi}(s) e^{s \tau} {\rm d}s\,,
\end{equation}
where  $c$ is a real constant that is larger than the real part of all singularities of $\hat{\psi}(s)$.

The probability that no event has occurred up to time $t$ is
\begin{equation}
	p(0,t) = \int_t^\infty \psi(t^{\prime}) {\rm d}t^{\prime}\,,
\end{equation}
whose Laplace transform is
\begin{equation}
	\hat{p}(0,s) = \frac{1 - \hat{\psi}(s)}{s}\,.
\label{eq:hatp(0,s)}
\end{equation}
The probability that one event occurs in $[0, t]$ is
\begin{equation}
	p(1,t) = \int_0^t \psi(t^{\prime}) p(0,t-t^{\prime}) {\rm d}t^{\prime}\,.
\label{eq:p(1,t)}
\end{equation}
By Laplace-transforming Eq.~\eqref{eq:p(1,t)} and applying Eq.~\eqref{eq:hatp(0,s)}, we obtain
\begin{equation}
	\hat{p}(1,s) = \hat{\psi}(s) \frac{1 - \hat{\psi}(s)}{s}\,.
\end{equation}
By the same arguments, the probability density that $n$ events occur at times $t_1$, $t_2$, $\ldots$, $t_n$ but at no other times in $[0, t]$ is given by $\psi(t_1)\psi(t_2-t_1)\cdots \psi(t_n-t_{n-1}) p(0,t-t_n)$. This yields \cite{Cox1962book,Grigolini2001Fractals}
\begin{equation}
	\hat{p}(n,s) =\left[\hat{\psi}(s)\right]^n \frac{1 - \hat{\psi}(s)}{s}\,.
\label{eq:hatp(n,s)}
\end{equation}
In the analysis of RWs, Eq.~\eqref{eq:hatp(n,s)} relates two ways to count time: one is in terms of the number of moves ($n$), and the other is in terms of the physical time ($t$).

For a CTRW driven by a Poisson process, we obtain
\begin{equation}
	\hat{\psi}(s) = \int_0^{\infty} \beta e^{-\beta \tau} e^{-s\tau} {\rm d}\tau = \frac{\beta}{s+\beta}\,.
\label{eq:hatpsi(s) Poisson}
\end{equation}
Substituting Eq.~\eqref{eq:hatpsi(s) Poisson} into Eq.~\eqref{eq:hatp(n,s)} yields
\begin{equation}
	\hat{p}(n,s) =\left( \frac{\beta}{s + \beta} \right)^n \frac{1}{s + \beta}\,.
\end{equation}

By taking the Fourier transform of Eq.~\eqref{eq:p(x;t) with free n} with respect to $x$ and the Laplace transform of Eq.~\eqref{eq:p(x;t) with free n} with respect to $t$ and then using Eqs.~\eqref{four} and \eqref{eq:hatp(n,s)}, we obtain
\begin{align}
	\hat{p}(k; s) &= \hat{p}(k; n) \hat{p}(n, s)\\
	&= \frac{1 - \hat{\psi}(s)}{s} \sum_{n=0}^{\infty} \hat{f}(k)^n \hat{\psi}(s)^n \notag\\
	&= \frac{1 - \hat{\psi}(s)}{s} \frac{1}{1 - \hat{f}(k) \hat{\psi}(s)}\,.
\label{eq:central CTRW}
\end{align}
This result is central to the theory of CTRWs \cite{Montroll1965JMathPhys}, and we will extend it to the case of general networks in Section~\ref{sub:CTRW nets}. Taking the inverse transform of Eq.~\eqref{eq:central CTRW} with respect to both time and space yields $p(x; t)$, and we can examine the behavior of the RW for large $t$ by expanding $\hat{p}(k; s)$ or $\hat{p}(x; s)$ for small $s$.


\section{Random walks on networks}\label{sec:nets}

\subsection{Notation} \label{sub:notations}

For our discussions, we assume that our networks are finite. However, to estimate how certain quantities scale with the number $N$ of nodes, we sometimes examine the $N \rightarrow \infty$ limit.
We allow our networks to have self-edges and multi-edges.
We assume that the edge weights are nonnegative, so our networks are unsigned.
For now, we assume that our networks are ordinary graphs (i.e., the best-studied types of networks), but we will consider multilayer networks in Section \ref{sec:multilayer} and temporal networks in Section \ref{sub:temporal networks}.
Because introducing edge weights does not usually complicate RW problems, we assume that our networks are weighted unless we state otherwise, and we consider unweighted networks to be a special case of weighted networks. We also assume that our networks are directed unless we state otherwise.
We summarize our main notation in Table~\ref{tab:notation}.

\begin{table*}[t]
\caption{Main notation.}
\begin{tabular}{|c|p{14cm}|}\hline
$N$ & number of nodes \\ \hline
$M$ & number of edges \\ \hline
$v_i$ & the $i$th node (where $i \in \{1, \dots, N\}$) \\ \hline
$A$ & The $N\times N$ weighted adjacency matrix of the network; the matrix component $A_{ij} \ge 0$ represents the weight of the edge from node $v_i$ to node $v_j$. In an undirected network, $A_{ij} = A_{ji}$ (where $i,j \in \{1, \dots, N\}$). In an unweighted network, $A_{ij}\in \{0, 1\}$ (again with $i,j \in \{1, \dots, N\}$).\\ \hline
$L$ & combinatorial Laplacian matrix\\ \hline
$L^{\prime}$ & RW normalized Laplacian matrix\\ \hline
$s_i$ & The strength of node $v_i$ in an undirected network; it is defined by
$s_i \equiv \sum_{j=1}^N A_{ij} = \sum_{j=1}^N A_{ji}$. In an undirected and unweighted network, $s_i$ is equal to the degree of $v_i$, which we denote by $k_i$.\\ \hline
$s_i^{\rm in}$ & In-strength of $v_i$; it is defined by $s_i^{\rm in} = \sum_{j=1}^N A_{ji}$. In an unweighted network, $s_i^{\rm in}$ is equal to the in-degree of $v_i$, which we denote by $k_i^{\rm in}$.\\ \hline
$s_i^{\rm out}$ & Out-strength of $v_i$; it is defined by $s_i^{\rm out} = \sum_{j=1}^N A_{ij}$. In an unweighted network, $s_i^{\rm out}$ is equal to the out-degree of $v_i$, which we denote by $k_i^{\rm out}$.\\ \hline
$\langle k\rangle$ & mean degree, which is given by
$\langle k\rangle = \sum_k k p(k)$ and indicates the sample mean of the degree for a network \\ \hline
$D$ & The $N\times N$ diagonal matrix whose $(i, i)$th element is equal to $s_i^{\rm out}$ (where $i \in \{1, \dots, N\}$). In an undirected network, the $(i, i)$th element of $D$ is equal to $s_i$.\\ \hline
$n$ & discrete time\\ \hline
$t$ & continuous time\\ \hline
$p_i$ & probability that a random walker visits $v_i$\\ \hline
$p_i^*$ & stationary density of a random walker at $v_i$\\ \hline
$\approx$ & approximately equal to \\ \hline
$\propto$ & proportional to \\ \hline
\end{tabular}
\label{tab:notation}
\end{table*}

%

An undirected network is called ``regular'' if all nodes have the same degree.
Notably, many mathematical results for RWs on networks are restricted to regular graphs \cite{Lovasz1993Boyal,Aldous2002book,Hoory2006BullAmMathSoc}. In this review, we are interested in networks with heterogeneous degree distributions, which tend to be the norm rather than the exception in empirical networks in numerous domains \cite{Clauset2009SiamRev}.

In our discussions, we assume that undirected networks are connected networks and that directed networks are ``weakly connected'' (i.e., that they are connected when one ignores the directions of the edges). It is clear (in the absence of jumps such as ``teleportation'' \cite{Gleich2015SiamRev}
 to augment the RW) that a random walker is confined in the component in which it starts, and the analysis of RWs is then reduced to analysis within each component. See \cite{Newman2010book} for extensive discussions of components and weakly connected components.


\subsection{Discrete time}

\subsubsection{Definition and temporal evolution}

Consider a DTRW on a directed network. We suppose that there is a single walker, which moves during each time step. When the walker is located at $v_i$, it moves to the out-neighbor $v_j$ with a probability proportional to $A_{ij}$. The transition-probability matrix $T$ has elements $T_{ij}$, which give the probability that the walker moves from $v_i$ to $v_j$, of
\begin{equation}
	T_{ij}=\frac{A_{ij}}{s_i^{\rm out}}\,,
\label{eq:T_ij random walk}
\end{equation}
where we assume that $s_i^{\rm out}>0$.
Other choices of $T$, informed by the adjacency matrix $A$, are also possible.
One example is a ``degree-biased RW'' in unweighted (and usually undirected) networks \cite{Eisler2005PhysRevE,WangWangYin2006PhysRevE,Fronczak2009PhysRevE,LeeYook2009EurPhysJB,Baronchelli2010PhysRevE,Bonaventura2014PhysRevE};
in this case, $T_{ij} \propto k_j^{\alpha}$, where $\alpha$ is a constant.
%
%
If $A_{ij}=A_{ji} = (k_i k_j)^{\alpha}$, then $T$ given by Eq.~\eqref{eq:T_ij random walk} gives this degree-biased RW. Another example of a biased transition-probability matrix $T$ is a ``maximum entropy RW'' \cite{Demetrius2005PhysicaA,Gomezgardenes2008PhysRevE,Burda2009PhysRevLett,Delvenne2011PhysRevE,Sinatra2011PhysRevE}.

Because a random walker must go somewhere --- including perhaps the current node --- in a given move, the following conservation condition holds:
\begin{equation}
	\sum_{j=1}^N T_{ij} = 1\,.
\label{eq:Markov chain conservation}
\end{equation}

A DTRW on a finite network is a Markov chain on $N$ states. There is a huge literature (both pedagogical and more advanced) on Markov chains in general and for RWs in particular.
This is especially true for finite state spaces (corresponding to finite networks) and for stationary Markov chains in which the transition probability does not depend on discrete time $n$ \cite{Kemeny1960-1976book,Papoulis1965-2002book,Iosifescu1980book,Stewart1994book,Norris1997book,TaylorKarlin1984-1998book,Aldous2002book,LevinPeresWilmer2009book,Blanchard2011book,Privault2013book}. We draw from this literature to explain several properties of DTRWs in the rest of this section.


Let $p_i(t)$ denote the probability that node $v_i$ is visited at discrete time $n$. This probability evolves according to
\begin{equation}
	p_j(n+1) =  \sum_{i=1}^N  p_i(n) T_{ij} \quad (j \in \{1, \dots, N\})\,.
\label{eq:Markov chain transition}
\end{equation}
Additionally,
\begin{equation}\label{eq:sum p_i(n) = 1}
	\sum_{i=1}^N p_i(n) = 1
\end{equation}
for any $n$ if Eq.~\eqref{eq:sum p_i(n) = 1} holds for $n=0$.
Equation~\eqref{eq:Markov chain transition} is equivalent to
\begin{equation}
	\bm p(n+1) = \bm p(n)T\,,
\label{eq:Markov chain transition vector}
\end{equation}
where $\bm p(t) = (p_1(n)\;, \ldots\; ,p_N(n))$.
From Eq.~\eqref{eq:Markov chain transition vector}, we see that
\begin{equation}
	\bm p(n) = \bm p(0) T^n\,.
\label{eq:m1}
\end{equation}


\subsubsection{Stationary density\label{sub:stationary density DTRW}}

Consider the stationary density (i.e., the so-called ``occupation probability'') ${\bm p}^* = (p_1^*, \ldots, p_N^*)$, where $p_i^* = \lim_{n\to\infty}p_i(n)$ (with $i \in \{1, \dots, N\}$). Substituting $p_i(n) = p_i(n+1) = p_i^*$
into Eq.~\eqref{eq:Markov chain transition vector} yields
\begin{equation}
	{\bm p}^* = {\bm p}^* T\,.
\label{eq:Markov chain stationary density}
\end{equation}
Therefore, the stationary density is the left eigenvector of $T$ with eigenvalue $1$. The corresponding right eigenvector is $(1\;, \ldots , \; 1)^{\top}$, where $\top$ represents transposition.

For a directed network that is ``strongly connected'' (i.e., a walker can travel from any node $v_i$ to any other node $v_j$ along directed edges \cite{Newman2010book}), ${\bm p}^*$ is unique. In undirected networks, one just needs a network to be connected, which we have assumed.

In undirected networks, we obtain the central result
\begin{equation}
	p_i^* = \frac{s_i}{\sum_{\ell=1}^N s_{\ell}} \quad (i \in \{1, \dots, N\})\,,
\label{eq:p_i^* propto k_i}
\end{equation}
which one can verify by substituting Eq.~\eqref{eq:p_i^* propto k_i} into Eq.~\eqref{eq:Markov chain stationary density}. For unweighted networks, Eq.~\eqref{eq:p_i^* propto k_i} reduces to $p_i^* = k_i/2M$. Regardless of other structural properties of a network, the stationary density is determined solely by strength (and thus by degree for unweighted networks). Equation~\eqref{eq:p_i^* propto k_i} also holds for directed networks that satisfy $s_i\equiv s_i^{\rm in} = s_i^{\rm out}$ (where $i \in \{1, \dots, N\}$). Such directed networks are sometimes called ``balanced'' \cite{Aldous2002book}.

In undirected networks,
\begin{equation}
	p_i^* T_{ij} = p_j^* T_{ji}\,.
\label{eq:detailed balance basic}
\end{equation}
In other words, for each edge, the flow of probability in each direction must equal each other at equilibrium. This property, called ``detailed balance'' in statistical physics \cite{sethnabook} and ``time reversibility'' in mathematics \cite{Lovasz1993Boyal,Aldous2002book}, does not generally hold for directed networks.

\begin{figure}[tb]
\begin{center}
\includegraphics[scale=0.4]{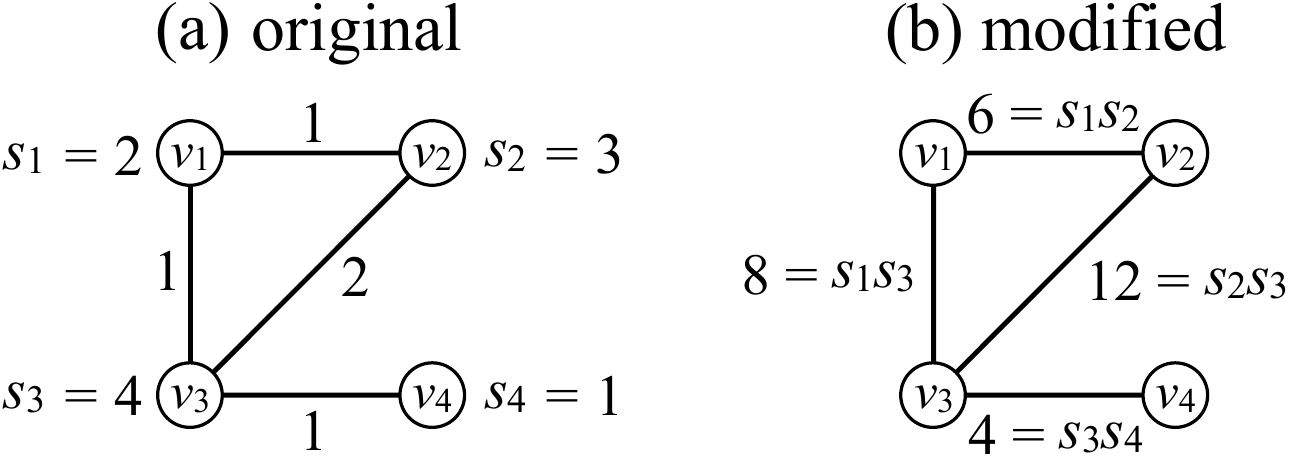}
\caption{Strength-biased RW. (a) An original undirected network, whose weighted adjacency matrix is given by $A$. (b) The modified undirected network, whose weighted adjacency matrix is given by $A^{\prime}$. The numbers attached to the edges represent the edge weight. We set $\alpha=1$.}
\label{fig:strength biased RW}
\end{center}
\end{figure}

Let's consider a generalization of the degree-biased RW to weighted networks (i.e., a strength-biased RW) in which the probability that a random walker located at node $v_i$ or $v_j$ traverses the edge ($v_i$, $v_j$) is proportional to $(s_i s_j)^{\alpha}$. It follows that
\begin{equation}
	T_{ij} = \frac{(s_i s_j)^{\alpha}}
{\sum_{\ell; v_{\ell}\in {\mathcal N}_i} (s_i s_{\ell})^{\alpha}}
	= \frac{s_j^{\alpha}}{\sum_{\ell; v_{\ell}\in {\mathcal N}_i} s_{\ell}^{\alpha}}\,,
\end{equation}
where ${\mathcal N}_i$ is the neighborhood of $v_i$.
A strength-biased RW is equivalent to an RW on a modified undirected network whose weighted adjacency matrix is given by $A^{\prime}_{ij} = (s_i s_j)^{\alpha}$ (see Fig.~\ref{fig:strength biased RW} for an example). The strength of node $v_i$ in this modified network is given by $s_i^{\prime} = \sum_{j=1}^N A^{\prime}_{ij} = s_i^{\alpha} \sum_{j; v_j \in {\mathcal N}_i}^N s_j^{\alpha}$. By substituting $s_i^{\prime}$ into Eq.~\eqref{eq:p_i^* propto k_i} in place of $s_i$, we obtain the stationary density
\begin{equation}
	p_i^* = \frac{s_i^{\alpha} \sum_{v_j \in {\mathcal N}_i} s_j^{\alpha}}
{\sum_{i^{\prime}=1}^N s_{i^{\prime}}^{\alpha} \sum_{v_{j^{\prime}} \in {\mathcal N}_{i^{\prime}}} s_{j^{\prime}}^{\alpha}}\,.
\end{equation}
For an unweighted network constructed using a ``configuration model'' \cite{Fosdick2016},
%
%
a standard model of random networks, we obtain
$p_i^* \approx k_i^{\alpha+1}/\sum_{\ell=1}k_{\ell}^{\alpha+1}$ \cite{Colizza2008JTheorBiol,LinZhang2013PhysRevE,ZhangShanChen2013PhysRevE}.
In particular, we obtain $p_i^* = 1/N$ for all nodes when $\alpha=-1$. Therefore, in general,
we expect that a node with a large strength tends to have a large $p_i^*$ when $\alpha > -1$ (including for the unweighted case $\alpha=0$) and that the same node tends to have a small $p_i^*$ when $\alpha<-1$. For nodes with a large strength, we expect $p_i^*$ to increase as $\alpha$ increases.

For directed networks in general, one can write a first-order approximation to the stationary density from Eq.~\eqref{eq:Markov chain stationary density}. We assume that we do not possess any information about the neighbors of $v_i$, so we replace $p_j^*$ and $s_j^{\rm out}$ by their mean values:
\begin{equation}
	p_i^* = \sum_{j=1}^N p_j^* \frac{A_{ji}}{s_j^{\rm out}}
\approx (\text{const}) \times \sum_{j=1}^N A_{ji} \propto s_i^{\rm in}\,.
\label{eq:in-degree approximation}
\end{equation}
On both synthetic and empirical networks, Eq.~\eqref{eq:in-degree approximation} is reasonably accurate in some cases but not in others
\cite{Donato2004EurPhysJB,Fortunato2006PNAS,Restrepo2006PhysRevLett,Davis2008JAmerSocInfoSciTech,Fortunato2008LNCS,Fersht2009PNAS,MasudaOhtsuki2009NewJPhys,Ghoshal2011NatComm}.


\subsubsection{Relaxation time\label{sub:relaxation time}}

To determine the relaxation time to the stationary state, it is instructive to project the solution, Eq.~\eqref{eq:m1}, onto an appropriate basis of vectors and to represent it in terms of its modes.
The procedure, which is analogous to taking a Fourier transform [see Eq.~\eqref{eq:def FT}], is sometimes called a ``graph Fourier transform'' \cite{Sandryhaila2013IEEETransSignalProc,Tremblay2014IEEESigProc} and will be explained in this section [see Eqs.~\eqref{eq:p_i(t) decomposed}--\eqref{eq:graph FT 2}].

For simplicity, we consider undirected networks. In general, the transition probability matrix $T$ is asymmetric even for undirected networks, except for regular graphs. However, one can derive its eigenvalues and eigenvectors from those of the symmetric matrix
\begin{equation}
	\tilde{A}_{ij}= \frac{A_{ij}}{\sqrt{s_i s_j}}\,,
\label{eq:def tilde A}
\end{equation}
which we can decompose as follows:
\begin{equation}
	\tilde{A}_{ij} = \sum_{\ell=1}^N \lambda_{\ell} \bm u_{\ell} \bm u_{\ell}^{\top}\,,
\end{equation}
where $\lambda_{\ell}$ is the $\ell$th eigenvalue of $\tilde{A}$ and $\bm u_{\ell}$ is the corresponding normalized eigenvector (so that $\langle \bm u_{\ell}, \bm u_{\ell^{\prime}}\rangle = \delta_{\ell \ell^{\prime}}$, where $\langle \, , \rangle$ is the inner product), and $\delta$ is the Kronecker delta.
Because $\tilde{A}$ is symmetric, each eigenvalue $\lambda_{\ell}$ is real.

Because $T_{ij} = \sqrt{s_j} \tilde{A}_{ij} / \sqrt{s_i}$, we have the following similarity relationship between $T$ and $A$ \cite{Aldous2002book,Samukhin2008PhysRevE}:
\begin{equation}
	T=D^{-1/2}\tilde{A}D^{1/2}\,,
\label{eq:T from tildeA}
\end{equation}
where we defined $D$ (a matrix whose nonzero entries lie only on the diagonal) in Section~\ref{sub:notations}. Equation~\eqref{eq:T from tildeA} implies that $T$ and $\tilde{A}$ have the same eigenvalues.
In particular, all eigenvalues of $T$ are real-valued, because that is the case for $\tilde{A}$. The left and right eigenvectors of $T$ corresponding to the eigenvalue $\lambda_{\ell}$ are, respectively,
\begin{align}
	\bm u_{\ell}^{\rm L} =& \bm u_{\ell}^{\top}D^{1/2} = \left((u_{\ell})_1\sqrt{s_1},\; \ldots\; ,(u_{\ell})_N\sqrt{s_N}\right)
\label{eq:u_ell^L}
\end{align}
and
\begin{align}
	\bm u_{\ell}^{\rm R} =& D^{-1/2}\bm u_{\ell} = \left((u_{\ell})_1/\sqrt{s_1},\; \ldots\; , (u_{\ell})_N/\sqrt{s_N}\right)^{\top}\,.
\label{eq:u_ell^R}
\end{align}
One can verify Eqs.~\eqref{eq:u_ell^L} and \eqref{eq:u_ell^R} using Eq.~\eqref{eq:T from tildeA} and the relation $\tilde{A} \bm u_{\ell} = \lambda_{\ell} \bm u_{\ell}$.

Using
\begin{align}
	T^n &= D^{-1/2}\tilde{A}^n D^{1/2}\notag\\
	&= D^{-1/2} \sum_{\ell=1}^N \lambda_{\ell}^n \bm u_{\ell} \bm u_{\ell}^{\top} D^{1/2}\notag\\
	&= \sum_{\ell=1}^N \lambda_{\ell}^n \bm u_{\ell}^{\rm R} \bm u_{\ell}^{\rm L}\,,
\label{eq:T^n eigen expansion}
\end{align}
we obtain the following mode expansion of the solution of the RW:
\begin{equation}
	\bm p(n) = \bm p(0) T^n = \sum_{\ell=1}^N \lambda^n_{\ell} \bm u_{\ell}^{\rm L} \langle \bm p(0), \bm u_{\ell}^{\rm R}\rangle\,.
\label{eq:p(t) RW discrete time final}
\end{equation}
That is,
\begin{equation}
	p_i(n) = \sum_{\ell=1}^N a_\ell(n) (u_{\ell}^{\rm L})_i\,,
\label{eq:p_i(t) decomposed}
\end{equation}
where
\begin{align}
	a_\ell(n) &= \lambda_{\ell}^n a_{\ell}(0) \,,
	\label{eq:graph FT 1}\\
	a_{\ell}(0) &\equiv \langle \bm p(0), \bm u_{\ell}^{\rm R}\rangle\,,
	\label{eq:graph FT 2}
\end{align}
and $a_\ell(n)$ is the projection onto the $\ell$th eigenmode. Equations~\eqref{eq:p_i(t) decomposed}--\eqref{eq:graph FT 2} map the state vector $\bm p(n)$, which is defined on the nodes, to a vector $(a_1(n),\; \ldots\; ,a_N(n))$ of eigenvector amplitudes (i.e., their coefficients). This transform, called the ``graph Fourier transform'', generalizes the standard Fourier transform of an RW [see Eqs.~\eqref{eq:inverse FT} and \eqref{eq:inverse FT RW}], and the eigenvectors of the transition-probability matrix $T$ play the role of the Fourier modes $e^{ikx}$.


For the matrix $T$ and $\tilde{A}$, the eigenvalues $\lambda_\ell$ each satisfy $-1\le \lambda_\ell \le 1$ \cite{Lovasz1993Boyal,Aldous2002book}. Except in the special cases of multipartite graphs, the strict inequality $\lambda_{\ell} > -1$ also holds. In this case, the mode with $\lambda_{\ell}=1$ corresponds to the stationary density, and we thus write $\bm u_{\ell}^{\rm L}=\bm p^*$. The right eigenvector that corresponds to this mode is $\bm u_{\ell}^{\rm R} \propto (1,\; \ldots\; , 1)^{\top}$. All modes for which $-1 < \lambda_\ell <1$ decay to $0$. The eigenvalue $\lambda_{\ell}=1$ is the largest-magnitude eigenvalue, and the Perron--Frobenius theorem guarantees that all elements of $\bm u_{\ell}^{\rm L}$ and $\bm u_{\ell}^{\rm R}$ are positive. Similar results hold for directed networks, although we cannot take advantage of the symmetric structure of the matrix $\tilde{A}$ in general. In directed networks, the eigenvalues satisfy $\left| \lambda_{\ell}\right|\le 1$. When $\left| \lambda_{\ell}\right| < 1$ holds for all but one eigenvalue, which is the case except for directed variants of multipartite graphs with an even number of components,
the mode with $\lambda_{\ell}=1$ corresponds to the stationary density.
In this case, we obtain $\bm u_{\ell}^{\rm L}=\bm p^*$ and $\bm u_{\ell}^{\rm R} \propto (1,\; \ldots\; , 1)^{\top}$. Again, the Perron--Frobenius theorem guarantees that all elements of $\bm u_{\ell}^{\rm L}$ are positive.

By letting $n\to\infty$ in Eq.~\eqref{eq:p(t) RW discrete time final}, we obtain
$\bm p^* = \bm u_{\max}^{\rm L} \langle \bm p(0), \bm u_{\max}^{\rm R}\rangle$,
where the subscript ``$\max$'' indicates the mode corresponding to the dominant eigenvalue (which is equal to $1$). Because $\bm u_{\max}^{\rm R} \propto (1,\; \ldots\; , 1)^{\top}$, it follows that $\langle \bm p(0), \bm u_{\max}^{\rm R}\rangle = 1$ regardless of the initial condition $\bm p(0)$. This is consistent with the fact that $\bm u_{\max}^{\rm L}$ gives the stationary density. By letting $n$ be large but finite, we obtain
\begin{equation}
	\bm p(n) \approx \bm u_{\max}^{\rm L} \langle \bm p(0), \bm u_{\max}^{\rm R}\rangle + \lambda^n_{\text{2}} \bm u_{\text{2}}^{\rm L} \langle \bm p(0), \bm u_{\text{2}}^{\rm R}\rangle\,,
\label{eq:up to 2nd}
\end{equation}
where $\lambda_{\rm 2}$ is the second-largest (in magnitude) eigenvalue of $T$. In deriving Eq.~\eqref{eq:up to 2nd}, we only kept two terms, because $|\lambda_{\ell}|^n \ll |\lambda_{\text{2}}|^n$ for all eigenvalues $\lambda_{\ell}$ with $\ell > 2$, assuming that $|\lambda_{\ell}| < |\lambda_2|$ (where $\ell \in \{3, \ldots, N\}$).
Equation~\eqref{eq:up to 2nd} indicates that the second-largest eigenvalue of $T$  governs the relaxation time. More generally, the relaxation speed is determined by the ratio
between $\left| \lambda_{\text{2}} \right|$ and $\lambda_{\max}=1$. The difference $1-\lambda_{\text{2}}$ is often called the ``spectral gap''. A large spectral gap (i.e., a small-magnitude for $\lambda_{\text{2}}$) entails fast relaxation.

The ``Cheeger inequality'' gives useful bounds on $\lambda_{\rm 2}$ \cite{Chung1997book-spectral}. The ``Cheeger constant'', which is also called ``conductance'', is defined by
\begin{equation}
	h = \min_{S} \left\{\frac{(\text{number of edges that connect } S \text{ and } \overline{S})}
{\min \{ {\rm vol}(S), {\rm vol}(\overline{S})\}}\right\}\,,
\label{eq:Cheeger constant}
\end{equation}
where $S$ is a set of nodes in a network, $\overline{S}$ is the complementary set of the nodes (i.e., $S\cap \overline{S} = \emptyset$ and $S\cup \overline{S}$ is the complete set of the $N$ nodes), and ${\rm vol}(S) \equiv \sum_{i=1; v_i \in S}^{N} s_i$.
In the minimization in Eq.~\eqref{eq:Cheeger constant}, we seek a bipartition of a network such that the two parts are the most sparsely connected. (In other words, we want a minimum cut.) The denominator in the right-hand side of Eq.~\eqref{eq:Cheeger constant} prevents the selection of a very uneven bipartition, which would easily yield a small value for the numerator. The Cheeger inequality is
\begin{equation}
	\frac{h^2}{2} < 1 - \left|\lambda_2\right| \le 2h\,,
\end{equation}
so a small Cheeger constant $h$ implies a small spectral gap $1- \left| \lambda_2 \right|$ and hence slower relaxation. This result is intuitive, because one can partition a network with a small value of $h$ into two well-separated communities such that it is difficult for random walkers to cross from one community to the other. Note that there are various versions of Cheeger constants and inequalities. They give qualitatively similar --- but quantitatively different --- results \cite{Lovasz1993Boyal,Aldous2002book,Donetti2006JStatMech,Arenas2008PhysRep,Cvetkovic2010book,piet-book}. As discussed in Ref.~\cite{Jeub2015PhysRevE} and references therein, such results are important considerations for community detection.

A fact related to the relaxation time is that the power method is a practical method to calculate the stationary density of an RW in a directed network \cite{Golub1996book}. Suppose that we start with an arbitrary initial vector $\bm p(0)$, excluding one that is orthogonal to $\bm p^*$,
and repeatedly left-multiply it by $T$.
After many iterations, we obtain an accurate estimate of $\bm p^*$. Because any $\bm p(0)$ that is orthogonal to $\bm p^*$ includes a negative entry, one can start iterations with any probability vector $\bm p(0)$.
In practice, one may have to normalize $\bm p(n)$ after each iteration (or after some number of iterations) to avoid the elements of $\bm p(n)$ becoming too large or small.


\subsubsection{Exit probability}

One is often interested in the probability that a random walker terminates at a particular node, which is then called an ``absorbing state''. Upon reaching an absorbing state, a stochastic process cannot escape from it. A node $v_i$ is ``absorbing'' if and only if $T_{ii}=1$, which implies that $T_{ij}=0$ (for $j\neq i$). A set of nodes is an ``ergodic'' set if (1) it is possible to go from $v_i$ to $v_j$ for any nodes in the set and (2) the process does not leave the set once it has been reached. An absorbing node is an ergodic set that consists of a single node. A state in a Markov chain is said to be a ``transient state'' if it does not belong to an ergodic set.


When an RW is composed of $N_1$ transient-state nodes and $N_2$ absorbing-state nodes, there are $N_1+N_2=N$ nodes in total. Without loss of generality, we relabel the nodes such that $v_1$, $\ldots$, $v_{N_1}$ are transient and $v_{N_1+1}$, $\ldots$, $v_N$ are absorbing. The transition-probability matrix $T$ then has the following form:
\begin{equation}
	T =  \begin{pmatrix}
		Q & R\\
		0 & I
	\end{pmatrix}\,,
\label{forme}
\end{equation}
where $Q$ is an $N_1 \times N_1$ matrix that describes transitions between transient-state nodes, $R$ is an $N_1 \times N_2$ matrix that describes transitions from transient-state nodes to absorbing-state nodes, and $I$ is the $N_2\times N_2$ identity matrix that corresponds to individual absorbing-state nodes. Taking powers of Eq.~\eqref{forme} yields
\begin{equation}
	T^n=  \begin{pmatrix}
		Q^n & R(I + Q + \cdots + Q^{n-1})\\
		0 & I
		\end{pmatrix}\,.
\end{equation}

Suppose that we start from transient-state node $v_i$ and want to calculate the mean number of visits to transient-state node $v_j$ before reaching an absorbing-state node. This number of visits is equal to the ($i$, $j$)th element of the matrix
\begin{equation}
	W = \sum_{n=0}^{\infty} Q^n = \left(I - Q\right)^{-1}\,,
\label{eq:num transient visit 1}
\end{equation}
because the $(i, j)$th element of $Q^n$ is equal to the probability that a random walker starting from $v_i$ visits $v_j$ at discrete time $n$.
The matrix $W$ is called the ``fundamental matrix'' associated with $Q$. The matrix on the right-hand side of Eq.~\eqref{eq:num transient visit 1} is called the ``resolvent'' of $Q$. Similar considerations arise in the study of ``central'' (i.e., important) nodes in networks \cite{Est12comm}.

The ``exit probability'' (i.e., the ``first-passage-time probability'') is defined as the probability $U_{ij}$ that the walker terminates at an absorbing state $v_j$ when it starts from a transient state $v_i$. When there are multiple absorbing-state nodes, it is nontrivial to determine the exit probability. The probability that the walker reaches $v_j$ after exactly $n$ steps is given by the $(i, j)$th element of $Q^{n-1} R$. Therefore, we obtain the exit probability in matrix form as follows:
\begin{equation}
	U = \sum_{n=1}^{\infty} Q^{n-1} R = W R\,.
\end{equation}


\subsubsection{Mean first-passage and recurrence times\label{sub:MFPT}}

When does a random walker starting from a certain source node arrive at a target node for the first time? The answer to this question is known as the ``first-passage time'' (or ``first-hitting time'') if the source and target nodes are different and is known as the ``recurrence time'' (or the ``first-return time'') when the source and target nodes are identical. Let $m_{ij}$ (with $i\neq j$) denote the mean first-passage time (MFPT) from node $v_i$ to node $v_j$. The mean recurrence time is $m_{ii}$. For directed networks, we assume strongly connected networks throughout this section to guarantee that $m_{ij}<\infty$ (for $i,j \in \{1, \dots, N\}$). For reviews on first-passage problems on networks and other media, see \cite{Redner2001book,Benichou2014PhysRep}.

\vspace{.1 in}

\emph{General networks}: Let's first consider some general results.
The following identity holds \cite{Kemeny1960-1976book,Papoulis1965-2002book,Stewart1994book,Aldous2002book}:
\begin{equation}
	m_{ij} = 1 + \sum_{\ell=1; \ell\neq j}^N T_{i\ell} m_{\ell j}\,.
\label{eq:MFPT self-consistent}
\end{equation}
In its first step, a random walker moves from node $v_i$ to node $v_{\ell}$, which produces the $1$ on the right-hand side of Eq.~\eqref{eq:MFPT self-consistent}. If $\ell=j$, then the walk terminates at $v_{\ell}$, resulting in a first-passage time of $1$. Otherwise, we seek the first-passage from node $v_{\ell}$ (with $\ell\neq j$) to node $v_j$. This produces the second term on the right-hand side. Note that Eq.~\eqref{eq:MFPT self-consistent} is also valid when $i=j$.

In matrix notation, we write Eq.~\eqref{eq:MFPT self-consistent} as
\begin{equation}
	M = J + T (M-M_{\rm dg})\,,
\label{eq:MFPT self-consistent matrix}
\end{equation}
where $M=(m_{ij})$, all of the elements of the matrix $J$ are equal to $1$, and $M_{\rm dg}$ is the diagonal matrix whose diagonal elements are equal to $m_{ii}$.
By left-multiplying Eq.~\eqref{eq:MFPT self-consistent matrix} by $\bm p^*$ and using $\bm p^* J = (1,\; \ldots\; ,1)$ and $\bm p^* T = \bm p^*$, we obtain the mean recurrence time
\begin{equation}
	m_{ii} = \frac{1}{p_i^*}\,.
\label{eq:Kac}
\end{equation}
Equation~\eqref{eq:Kac} is called ``Kac's formula'' \cite{Aldous2002book,LevinPeresWilmer2009book,Blanchard2011book}.


There are several different ways to evaluate the MFPT $m_{ij}$ (with $i\neq j$), and it is insightful to discuss different approaches.

One method is simply to iterate Eq.~\eqref{eq:MFPT self-consistent} \cite{Stewart1994book}.

A second method to calculate the MFPT, for a given $j$, is to rewrite Eq.~\eqref{eq:MFPT self-consistent} as
\begin{equation}
	{\overline{\bm m}^{(j)} = \bm 1 + \overline{T}^{(j)}} \overline{\bm m}^{(j)}\,,
\label{eq:MFPT self-consistent 2}
\end{equation}
where $\overline{\bm m}^{(j)} = (m_{1j},\; \ldots\; ,m_{j-1,j}\; ,m_{j+1,j}\; \ldots\; ,m_{Nj})^{\top}$ and $\bm 1 = (1,\; \ldots\; ,1)^{\top}$ are $(N-1)$-dimensional column vectors and $\overline{T}^{(j)}$ is the $(N-1)\times (N-1)$ submatrix of $T$ that excludes the $j$th row and $j$th column \cite{LinZhang2013PhysRevE}. The formal solution of Eq.~\eqref{eq:MFPT self-consistent 2} is
\begin{equation}
	\overline{\bm m}^{(j)} = \left(\overline{L}^{(j)}\right)^{-1} \overline{D}^{(j)} \bm 1\,,
\label{eq:m_ij final with reduced Laplacian}
\end{equation}
where $\overline{D}^{(j)}$ is the submatrix of $D$ that excludes the $j$th row and $j$th column and $\overline{L}^{(j)} = \overline{D}^{(j)} - \overline{A}^{(j)}$, where $\overline{A}^{(j)}$ is the submatrix of $A$ that excludes the $j$th row and $j$th column. The matrix $\overline{L}^{(j)}$ is sometimes called a ``grounded Laplacian matrix'' \cite{Miekkala1993Bit} (although it is not a Laplacian matrix), and it is invertible because we assumed strongly connected networks.
One can derive and solve Eq.~\eqref{eq:m_ij final with reduced Laplacian} separately for each $j$.

A third method to calculate the MFPT is to take advantage of relaxation properties of RWs \cite{Noh2004PhysRevLett}.
Let $p_{ij}(n)$ denote the probability that a walker starting at node $v_i$ visits node $v_j$ after $n$ moves. The master equation is
\begin{equation}
	p_{ij}(n+1)=\sum_{\ell=1}^N p_{i\ell}(n)T_{\ell j}\,.
\label{eq:master eq Noh}
\end{equation}
Let $F_{ij}(n)$ denote the probability that the walker starting from $v_i$ arrives at $v_j$ for the first time after $n$ moves. We obtain
\begin{equation}
	p_{ij}(n) = \delta_{n0}\delta_{ij}+\sum^n_{n^{\prime}=0}
F_{ij}(n^{\prime}) p_{jj}(n-n^{\prime})\,.
\label{eq:P F first-passage}
\end{equation}
Using a discrete-time Laplace transform (see, e.g., \cite{wilf2005} for an extensive discussion of such generating functions), defined by
\begin{equation}
	\hat{p}_{ij}(s)\equiv \sum^{\infty}_{n=0}e^{-sn}p_{ij}(n)
\end{equation}
and
\begin{equation}
	\hat{F}_{ij}(s)\equiv \sum^{\infty}_{n=0}e^{-sn}F_{ij}(n)\,,
\end{equation}
we transform Eq.~\eqref{eq:P F first-passage} to
\begin{equation}
	\hat{p}_{ij}(s)=\delta_{ij}+\hat{F}_{ij}(s)\hat{p}_{jj}(s)
\end{equation}
and thereby obtain
\begin{equation}
	\hat{F}_{ij}(s)=\frac{\hat{p}_{ij}(s)-\delta_{ij}}{\hat{p}_{jj}(s)}\,.
\label{eq:P F first-passage Laplace}
\end{equation}
Using Eq.~\eqref{eq:P F first-passage Laplace} then yields
\begin{align}
	m_{ij} &= \sum^{\infty}_{n=0}n F_{ij}(n)=
-\hat{F}_{ij}^{\prime}(0) \notag \\
	&= \frac{-\hat{p}_{ij}^{\prime}(0)\hat{p}_{jj}(0)+\hat{p}_{jj}^{\prime}(0)
\left[\hat{p}_{ij}(0)-\delta_{ij}\right]}
	{\hat{p}_{jj}(0)^2}\,.
\label{eq:<T_ij> 1}
\end{align}
To evaluate Eq.~\eqref{eq:<T_ij> 1}, we define
\begin{equation}
	R_{ij}^{(m)}\equiv \sum^{\infty}_{n=0} n^m\left[
p_{ij}(n)-p_j^* \right]\,.
\label{eq:R_ij^n def}
\end{equation}
Equation~\eqref{eq:R_ij^n def} quantifies the relaxation speed at which $p_{ij}(n)$ approaches the stationary density. To write the Laplace transform, we multiply both sides of Eq.~\eqref{eq:R_ij^n def} by $(-1)^m s^m/m!$ and sum over $m$. We thereby obtain
\begin{align}
	\sum^{\infty}_{m=0}R_{ij}^{(m)}(-1)^m\frac{s^m}{m!}
	&= \sum^{\infty}_{m=0}\sum^{\infty}_{n=0}n^m(-1)^m\frac{s^m}{m!}\left[
p_{ij}(n)-p_j^*\right]\notag\\
	&= \sum^{\infty}_{n=0}e^{-sn}
\left[p_{ij}(n)-p_j^*\right]\notag\\
	&= \hat{p}_{ij}(s) - \frac{p_j^*}{1-e^{-s}}\,.
\label{eq:Noh derivation 1}
\end{align}
Substituting Eq.~\eqref{eq:Noh derivation 1} into Eq.~\eqref{eq:P F first-passage Laplace} then yields
\begin{align}
	\hat{F}_{ij}(s) &=
\frac{\frac{p_j^*}{s+o(s)}+
\sum^{\infty}_{m=0}R_{ij}^{(m)}(-1)^m\frac{s^m}{m!}-\delta_{ij}}
{\frac{p_j^*}{s+o(s)}+
\sum^{\infty}_{m=0}R_{jj}^{(m)}(-1)^m\frac{s^m}{m!}}\notag\\
	&= \frac{p_j^*+R_{ij}^{(0)}s-\delta_{ij}s+o(s)}
{p_j^*+R_{jj}^{(0)}s+o(s)}\notag\\
	&= 1+\frac{R_{ij}^{(0)}-R_{jj}^{(0)}-\delta_{ij}}{p_j^*}s
+o(s)\,,
\end{align}
where $o(s)$ represents a quantity that is much smaller than $s$ in the relevant asymptotic limit ($s\to 0$ in the present case).
Consequently,
\begin{equation}
	m_{ij} = -\hat{F}_{ij}^{\prime}(0) =
\begin{cases}
\frac{1}{p_j^*} & (j=i)\,,\\
\frac{R_{jj}^{(0)}-R_{ij}^{(0)}}{p_j^*} & (j\neq i)\,,
	\end{cases}
\label{eq:T_ij general}
\end{equation}
which is consistent with Kac's formula [see Eq.~\eqref{eq:Kac}]. For undirected networks, substituting $p_j^*=s_j/\sum_{\ell=1}^N s_{\ell}$ into Eq.~\eqref{eq:T_ij general} yields
\begin{equation}
	m_{ij} = \begin{cases}
\frac{\sum_{\ell=1}^N s_{\ell}}{s_j} & (j=i)\,,\\
\frac{\sum_{\ell=1}^N s_{\ell}}{s_j}\left(R_{jj}^{(0)}-R_{ij}^{(0)}\right) & (j\neq i)\,.
	\end{cases}
\label{eq:T_ij general undirected}
\end{equation}

A fourth method to examine the MFPT is to estimate $m_{ij}$ using a mean-field approximation \cite{Kittas2008EPL,Perra2012PhysRevLett,Starnini2012PhysRevE}. Regardless of the source node $v_i$, the target node $v_j$ is reached with an approximate probability of $p_j^*$ in each time step. Therefore,
\begin{equation}
	m_{ij} \approx \sum_{n=1}^{\infty} n p_j^* (1-p_j^*)^{n-1} = \frac{1}{p_j^*} = m_{jj}\,.
\label{eq:m_ij meanfield}
\end{equation}
Equation~\eqref{eq:m_ij meanfield} is a rather coarse approximation, and $m_{ij}$ can deviate considerably from $m_{jj} = 1/p_j^*$. More sophisticated mean-field approaches can likely do better, especially for networks with structures that are well-suited to the employed approximation.




There have been many studies of MFPTs for various network models using both analytical and numerical approaches \cite{Redner2001book,Almaas2003PhysRevE,Masuda2004PhysRevE-rw,Hwang2012PhysRevE,Hwang2012PhysRevE,Hwang2013PhysRevE,PengAgliariZhang2015Chaos}. We will discuss some examples of undirected and unweighted networks. We focus mainly on the MFPT between different nodes, although it is of course also interesting to calculate recurrence times.


\emph{Regular networks}: For a complete graph, $m_{ij}$ (with $i\neq j$) is independent of $i$ and $j$ because of the symmetry of the network. Therefore, Eq.~\eqref{eq:MFPT self-consistent} reduces to
\begin{equation}
	m_{ij} = \frac{1}{N-1} + \frac{N-2}{N-1}(1+m_{ij})\,,
\end{equation}
which yields $m_{ij} = N-1$ for $i \neq j$. Kac's formula [see Eq.~\eqref{eq:Kac}] implies that $m_{ii}=N$.

For regular lattices ${\mathbb Z}^d$ of any dimension $d$, Eq.~\eqref{eq:Kac} implies that $m_{ii}\propto N$ because $p_i^*\propto k_i = 2d$ for any $i$. Define $m_{\bullet j}$ to be the MFPT averaged over all source nodes $v_i$ ($i \neq j$) \cite{Montroll1969JMathPhys}. For {${\mathbb Z}^d$}, it satisfies the scalings $m_{\bullet j}\propto N^2$ for $d=1$, $m_{\bullet j} \propto N\ln N$ for $d=2$, and $m_{\bullet j}\propto N$ for $d=3$. %

\emph{Erd\H{o}s--R\'{e}nyi (ER) random graphs}: Consider an ER random graph $G(N,p)$, where $p$ denotes the (independent) probability that each node pair has an edge. Assuming that the mean degree $\langle k\rangle$ is kept constant (i.e., $p=\langle k\rangle/(N-1)\propto 1/N$), we obtain $m_{ii}\propto N$ and $m_{ij}\propto N^{3/2}$ (with $i \neq j$) as $N\to\infty$ \cite{Bollt2005NewJPhys}
for the ``giant component'' (i.e., a largest connected component that scales linearly with the number $N$ of network nodes as $N \rightarrow \infty$ \cite{Newman2010book}). Now suppose that we assume instead that $p > \ln N/N$, so that all nodes belong to a single component (in the $N\to\infty$ limit) and thus $m_{ij}$ (for $i,j \in \{1, \dots, N\}$) is well-defined. It then follows that $m_{ij}$ averaged over all source and target nodes is equal to $N-1$, independently of $p$ \cite{Sood2005JPhysA,Lowe2014StatProbLett}. In other words, for a sufficiently dense ER random graph, the MFPT is the same as that for the complete graph.
The MFPT is much longer for directed ER graphs than for undirected ones, because random walkers do not backtrack on directed networks \cite{TishbyBiham2017arxiv}.

\emph{Other network models with random features}: Much effort in studying RWs on networks has considered first-passage times on Watts--Strogatz (WS) small-world networks
\cite{Jespersen2000PhysRevE,PanditAmritkar2001PhysRevE,Lahtinen2001PhysRevE,Jasch2001PhysRevE,Almaas2003PhysRevE,Parris2005PhysRevE,YangSJ2005PhysRevE}. As expected, given that WS networks interpolate between regular lattices and ER networks \footnote{Technically, it is a variant
of WS networks with edge rewiring (rather than edge addition) that interpolates between regular lattices and ER networks \cite{smallworld-scholarpedia}.}, these studies have found that the behavior of an RW on WS networks lies somewhere between that on a regular lattice and that on ER graphs.

Equation~\eqref{eq:T_ij general undirected} has also been elaborated further for ``scale-free'' networks, which are defined as networks with a power-law degree distribution $p(k)\propto k^{-\gamma}$, where $p(k)$ is the degree distribution.
Let's consider scale-free networks that are generated by a ``configuration model'' \cite{Fosdick2016}, so there are no degree--degree correlations. We examine the mean
of the MFPT $m_{ij}$ over the position of the source node $v_i$ (with $i \neq j$), which we select according to the stationary density. We use $\tilde{m}_{\bullet j}$ to denote this weighted mean of the MFPT over $i$. This mean is distinct from the unweighted mean $m_{\bullet j}$.
For scale-free networks constructed using a configuration model, we obtain for large $N$ that
\cite{Hwang2012PhysRevLett}
\begin{equation}
	\tilde{m}_{\bullet j} \propto
\begin{cases}
N^{2/d_{\rm s}} & (d_{\rm s}<2)\,,\\
N k_j^{(1-2/d_{\rm s})(\gamma-1)} & (2 < d_{\rm s} < 2(\gamma-1)/(\gamma-2))\,,\\
N k_j^{-1} & (d_{\rm s} > 2(\gamma-1)/(\gamma-2))\,,
	\end{cases}
\label{eq:Hwang2012PhysRevLett}
\end{equation}
where $d_{\rm s} \equiv 2 d_{\rm f}/d_{\rm w}$ is the ``spectral dimension'' of the network; the ``fractal dimension'' $d_{\rm f}$ is defined as the exponent of the scaling relation $N_r\propto r^{d_{\rm f}}$, where $N_r$ is the number of nodes within distance $r$ from a source node; and the ``walk dimension'' $d_{\rm w}$ is defined from the scaling relation $\langle r^2 \rangle \propto t^{2/d_{\rm w}}$, where $r$ is the distance between the current position of the walker and the source node \cite{Rammal1984JStatPhys,Benavraham2000book}.
In practice, one calculates the walk dimension as the scaling exponent for the time $t_{\rm exit}$ for a random walker to exit from a sphere of radius $r$ from the source node (so that $t_{\rm exit}\propto r^{d_{\rm w}}$) \cite{Condamin2007Nature}.
For regular lattices, $d_{\rm w}=2$, and the diffusion is thus called ``normal''. If $d_{\rm w} \neq 2$, the diffusion is called ``anomalous'' \cite{Benavraham2000book}.
For the ``compact exploration'' case of $d_{\rm s}<2$, Eq.~\eqref{eq:Hwang2012PhysRevLett} suggests that the asymptotic scaling of $\tilde{m}_{\bullet j}$ with $N$ does not depend on the target node at leading order.
However, if $d_{\rm s}>2$ (the second and the third cases in Eq.~\eqref{eq:Hwang2012PhysRevLett}), nodes with higher degrees are reached faster.
In particular, for networks that satisfy the ``small-world property'' (i.e., the mean path length between nodes scales proportionally to $\ln N$ or even more slowly) \cite{smallworld-scholarpedia}, including popular scale-free network models (such as ones generated by a configuration model), one obtains $d_{\rm s}=\infty$ (and $d_{\rm s}$ is very large for many empirical networks).
Therefore, the third case in Eq.~\eqref{eq:Hwang2012PhysRevLett} applies.


\emph{Fractal and pseudo-fractal networks}: There are various deterministic mechanisms to grow networks in a recursive manner. Depending on the mode, these algorithms yield ``pseudo-fractal'' scale-free networks \cite{Dorogovtsev2002PhysRevE_pseudofractal} (also called ``hierarchical networks'' \cite{Ravasz2002Science,Ravasz2003PhysRevE} or ``transfractals'' \cite{Rozenfeld2007NewJPhys}; see  Table~\ref{table:hierarchical} for different meanings of the term ``hierarchical network'' that exist in the literature),
which have a highly symmetric structure and satisfy the small-world property; fractal networks that do not satisfy the small-world property \cite{Song2005Nature,Song2006NatPhys,Rozenfeld2007NewJPhys}; or classical fractals \cite{Benavraham2000book}. These objects are defined and studied in the limit $N \to \infty$.
For such models, it is often possible to exploit their deterministic and recursive nature to exactly calculate the MFPT, and generating functions again can be helpful.

\begin{table}[tb]
\begin{center}
\caption{
The term ``hierarchical network'' has been used (sometimes in a misleading way) to describe various network structures. To help readers, we provide a short summary of three common uses.}\vspace{.2cm}
\begin{tabular}{|p{4cm}|p{10cm}|} \hline
Hierarchical modularity & A hierarchical network can indicate the presence of ``hierarchical modularity'', in which dense modules are themselves composed of dense submodules in the recursive manner of a ``Russian doll'' \cite{Simon1962ProcAmPhilosSoc}.\\ \hline
Status theory & One can also understand a hierarchy in the context of ``status theory'', in which certain nodes have a higher status than others, and a directed edge indicates a difference of status \cite{Leskovec2010CHI}. This notion leads naturally to trees that are dominated by a root and, more generally, to acyclic networks \cite{Karrer2009PhysRevE}.\\ \hline
Pseudo-fractal networks & Some models of pseudo-fractal networks are sometimes called hierarchical networks. Ravasz and Barab\'{a}si proposed to characterize such ``hierarchical'' structure by examining a scaling relation between clustering coefficient and node degree \cite{Ravasz2002Science,Ravasz2003PhysRevE}. \\ \hline
\end{tabular}
\label{table:hierarchical}
\end{center}
\end{table}

Let's start by looking at fractals that do not have a heavy-tailed degree distribution. In a recursive process of generating a fractal structure from a model of a fractal, we stop the process in each iteration and regard any intersection with more than one edges as a node. In this way, we define a network corresponding to each iteration. The recursive process generates a series of networks, where the number $N$ of nodes becomes larger as one iterates further. We are interested in how the MFPT scales in such networks as a function of $N$. For example, consider a network constructed from the Sierpinski gasket \cite{falconer2013}. When the target node is located at the apex of the gasket, the MFPT averaged over a uniform distribution of the source node is $m_{\bullet j} \propto N^{\ln 5/\ln 3} \approx N^{1.46}$ \cite{Benavraham2000book,Kozak2002PhysRevE,Bollt2005NewJPhys}.
Another example is the so-called ``T-graph'', which is produced by the initial condition of two nodes connected by an edge and recursive replacement of each edge by a star composed of four nodes to produce a fractal \cite{HavlinWeissman1986JPhysA,Kahng1989JPhysA}. For the T-graph, the MFPT
when the target is the unique central node and the source node is distributed uniformly over the $N-1$ remaining nodes is $m_{\bullet j} \propto N^{\ln 6/\ln 3} \approx N^{1.63}$ \cite{Agliari2008PhysRevE}.
Yet another example are so-called ``Vicsek fractals'', which are produced by the initial condition of a star having $f+1$ nodes and recursive addition of $f$ replicas of the current network, such that each replica network is connected to the current network by one edge between leaves (i.e., between a node with degree 1 in a replica and a node with degree 1 in the current network) \cite{Vicsek1983JPhysA,BlumenJurjiu2003PhysRevE}.
For Vicsek fractals, the MFPT averaged over all pairs of source and target nodes, chosen from all possible pairs and denoted by $m_{\bullet\bullet}$, scales as $m_{\bullet \bullet}\propto N^{\ln (3f+3)/\ln(f+1)}$ \cite{ZhangWuZhang2010PhysRevE}. Similar scaling results have also been studied in other deterministic and stochastic fractals and heterogeneous media \cite{Kahng1989JPhysA,Matan1989PhysRevA,Benavraham2000book,Redner2001book}.

Now let's consider fractal networks that have a power-law degree distribution. One generates a so-called ``$(u, v)$-flower'', where $u$ and $v$ are integers, by starting with two nodes connected by an edge and replacing each edge by two parallel paths of length $u$ and $v$ in each generation. This model produces fractal and scale-free networks for $u, v\ge 2$ \cite{Berker1979JPhysC,Rozenfeld2007NewJPhys}. The degree distribution of a $(u, v)$-flower is
$p(k) = k^{-\gamma}$, where $\gamma= 1+ \ln (uv) / \ln 2$.
For this network, the MFPT between so-called ``hubs'' (which, in this context, are defined as nodes that are present in the same finite generation and whose degree thus becomes infinite as $N\to\infty$) scales as $m_{ij} \propto N^{\frac{\ln (uv)}{\ln (u+v)}}$ \cite{Rozenfeld2007NewJPhys}.
%
Consistent with this result, when $u=v$, the MFPT, averaged over source-node position (which is distributed according to the stationary density), to the node with the largest degree (i.e., one of the two nodes that exist initially) is given by $\tilde{m}_{\bullet j} \propto N^{2\ln u / \ln (2u)}$ \cite{Tejedor2009PhysRevE}.
A tree-like network model, called the ``$(u, v)$-tree'', is produced if, in each generation, one replaces every edge by a path of $u$ edges and add two new paths of $v/2$ edges that start from each end point of the already-added path of $u$ edges and have a loose end. (If $v$ is odd, one adds two paths of $(v\pm 1)/2$ edges.) When $u \ge 2$, the $(u, v)$-tree model produces fractal and scale-free networks with $\gamma=1+\ln(u+v)/\ln 2$ \cite{Song2006NatPhys,Rozenfeld2007NewJPhys}. For such networks, the MFPT between hubs (which here too are defined as nodes that are present in the same finite generation) scales as $m_{ij}\propto N^{\frac{\ln [u(u+v)]}{\ln (u+v)}}$ \cite{Rozenfeld2007NewJPhys,ZhangLinMa2011JPhysA}.

All of the above results on fractals and fractal scale-free networks are consistent with a known scaling law for the MFPT: it scales proportionally to $N^{2/d_{\rm s}} = N^{d_{\rm w}/d_{\rm f}}$ \cite{Bollt2005NewJPhys}. There are known analytical expressions for $d_{\rm f}$ and $d_{\rm w}$ for the fractals and fractal scale-free networks whose MFPT we discussed above. The spectral dimension is $d_{\rm s} =  \ln 9/\ln 5\approx 1.37$ for the Sierpinski gasket \cite{Havlin1987AdvPhys}, $d_{\rm s} = \ln 9/\ln 6\approx 1.23$ for the T-graph \cite{HavlinWeissman1986JPhysA}, $d_{\rm s}= 2\ln (f+1)/\ln (3f+3)$ for the Vicsek fractals \cite{BlumenJurjiu2003PhysRevE}, $d_{\rm s}=2\ln(u+v)/\ln(uv)$ for the fractal $(u, v)$-flowers \cite{Rozenfeld2007NewJPhys,YunKahngKim2009NewJPhys}, and $d_{\rm s} = 2\ln(u+v)/\ln u(u+v)$ for the fractal $(u, v)$-trees \cite{Rozenfeld2007NewJPhys,YunKahngKim2009NewJPhys}.

As we mentioned in the beginning of this section, there are also scale-free network models that are constructed deterministically and recursively. The resulting networks are not fractals \cite{Barabasi2001PhysicaA,Dorogovtsev2002PhysRevE_pseudofractal,JungKimKahng2002PhysRevE,Ravasz2002Science,Ravasz2003PhysRevE,Andrade2005PhysRevLett,Doye2005PhysRevE,Rozenfeld2007NewJPhys} and are sometimes called ``pseudo-fractals'' \cite{Dorogovtsev2002PhysRevE_pseudofractal}. In the literature, fractal and pseudo-fractal networks are usually distinguished as follows. By definition, pseudo-fractal networks satisfy the small-world property, as they have a small mean path length (which scales as $\log N$ or smaller \cite{smallworld-scholarpedia}) between pairs of nodes, possibly due to the creation of shortcuts during the generation of the network. In contrast, the fractal network models discussed above, as well as conventional fractals, have large worlds, as the mean path length scales as a power of $N$ \cite{Song2005Nature}. Similar to the case of fractal networks, it is possible to exactly calculate the MFPT for a variety of pseudo-fractals by exploiting the recursive nature of their definitions.

Before general $(u, v)$-flowers were proposed in Ref.~\cite{Rozenfeld2007NewJPhys}, the special case with $u=1$ and $v=2$ had already been studied \cite{Dorogovtsev2002PhysRevE_pseudofractal}.
A $(1, 2)$-flower has degree distribution
$p(k)\propto k^{-\gamma}$, where $\gamma = 1 + \ln 3/\ln 2 \approx 2.59$
 \cite{Dorogovtsev2002PhysRevE_pseudofractal}. A $(u, v)$-flower has a small mean path length and is non-fractal when $u$ or $v$ is equal to $1$ \cite{Rozenfeld2007NewJPhys}.
%
In a $(1,2)$-flower, the MFPT for an arbitrary pair of nodes (present in a particular finite generation of the network) scales as $m_{ij}\propto N$ \cite{Bollt2005NewJPhys}. For the same network, $m_{ij}$ averaged over a uniformly distributed location of the source node scales as $m_{\bullet j}\propto N^{\ln 2/ \ln 3} \approx N^{0.63}$ when the target node $v_j$ is the largest hub (whose degree $k\approx N^{\ln 2/\ln 3}$) \cite{ZhangQi2009PhysRevE}.
For a $(1,v)$-flower for general $v$, the MFPT between hubs
(i.e., nodes that are present in the same finite generation, so their degree becomes infinite as $N\to\infty$) scales as $m_{ij}\propto N^{\ln v / \ln (v+1)}$, which is consistent with the results in Ref.~\cite{ZhangQi2009PhysRevE} that we explained above.
For a $(1,v)$-tree for general $v$, which produces non-fractal scale-free networks \cite{Rozenfeld2007NewJPhys}, the MFPT between hubs (i.e., nodes present in the same finite generation) scales as $m_{ij} \propto N$ and that between non-hub nodes (i.e., nodes of finite degree) scales as $m_{ij}\propto N \ln N$ \cite{Rozenfeld2007NewJPhys}.
The MFPT to the most connected hub $v_j$ (i.e., the node that is present initially) averaged over the position of the uniformly distributed source node $v_i$ (with $i \neq j$) scales as $m_{\bullet j}\propto N$ \cite{ZhangLinMa2011JPhysA}.
Consider a different scale-free tree model, in which, in each generation, $m$ new nodes are connected to each of the already existing nodes. This model produces a power-law degree distribution with $\gamma= 1 + \ln (2m+1) / \ln (m+1)$ \cite{JungKimKahng2002PhysRevE}. For this network model, the MFPT averaged over all pairs of source and target nodes selected uniformly at random scales as
$m_{\bullet\bullet} \propto N\ln N$ \cite{ZhangQiZhou2010PhysRevE}. The MFPT when the target node is selected from the stationary density of an RW is also proportional to $N\ln N$ as $N\rightarrow \infty$ for an arbitrary source node \cite{ZhangGuoLin2014PhysRevE}.
%
%
Similar results have also been derived for pseudo-fractal scale-free networks that include loops.
In one such network model, one starts from a single node and, in each generation, adds two replicas of the present network and connects some nodes in each replica to the initially-present single node. This model produces scale-free networks with loops and with $\gamma= \ln 3 / \ln 2 \approx 1.59$ \cite{Barabasi2001PhysicaA}.
For this model, the MFPT from the largest-degree hub (i.e., the initially-existing node) to a low-degree node created in the latest generation in the growth (and the corresponding MFPT in the reverse direction) scales as $m_{ij}\propto N^{1-\ln 2/\ln 3} \approx N^{0.37}$ \cite{Agliari2009PhysRevE}. The MFPT to the largest-degree hub starting from a uniformly distributed source node (where the position of the source node is selected with the equal probability from the $N-1$ nodes excluding the target hub node) also scales as $m_{\bullet j} \propto N^{1-\ln 2/\ln 3}$ \cite{Agliari2009PhysRevE}.
One obtains a related pseudo-fractal scale-free network model by starting the recursive growth process of a network from an $N_{\rm init}$-node connected network in which one root node is specified
\cite{Ravasz2002Science,Ravasz2003PhysRevE}.
In each generation, one adds $N_{\rm init}-1$ replicas $N_{\rm init}\ge 3$) and connects them to the root node by some edges. This model produces a scale-free network with
$\gamma= 1 + \ln N_{\rm init}/\ln (N_{\rm init}-1)$. For this network model, the MFPT to the root node, which has the largest degree, starting from a source node, selected with equal probability from all nodes but the root, scales as $m_{\bullet j} \propto N^{1-\ln (N_{\rm init}-1)/\ln N_{\rm init}}$
\cite{ZhangLinGao2009PhysRevE}. Because $N_{\rm init}\ge 3$, the MFPT scales no faster than
$N^{1-\ln 2/\ln 3}\approx N^{0.37}$.
Finally, a so-called ``Apollonian network'' is defined through an Apollonian packing (i.e., a space-filling packing of spheres)
and produces a power-law degree distribution with $\gamma = 1 + \ln 3 / \ln 2 \approx 2.58$ \cite{Andrade2005PhysRevLett,Doye2005PhysRevE}. For Apollonian networks,
the MFPT to the node with the largest degree, where the source node is selected with the equal probability from all but the target node, is given by $m_{\bullet j}\propto N^{2-\ln 5/\ln 3} \approx N^{0.54}$ \cite{ZhangGuan2009EPL}.

In the results in the above paragraph for pseudo-fractal scale-free (but non-fractal) networks, the MFPT scales at most proportional to $N\ln N$ and mostly scales sublinearly in $N$.
The MFPT is smaller than for fractals and fractal scale-free networks for which $m_{ij}$ (or its mean over source or target nodes) scales superlinearly (i.e., in proportion to
$N^{2/d_{\rm s}}$, where $d_{\rm s}<2$).
Because $d_{\rm s}=\infty$ for the aforementioned pseudo-fractal scale-free networks, which satisfy the small-world property, the MFPT does not scale in proportion to
$N^{2/d_{\rm s}}$. These results are consistent qualitatively with the third case in Eq.~\eqref{eq:Hwang2012PhysRevLett}, although Eq.~\eqref{eq:Hwang2012PhysRevLett} was derived for a source node whose location satisfies the stationary density, and many of the aforementioned theoretical results were derived for specific source --- target pairs or a source node selected with equal probability from all nodes (excluding the target node).
Note that the largest degree in the aforementioned pseudo-fractal scale-free networks (including the $(1,v)$-flowers and $(1,v)$-trees) scales as a sublinear power of $N$ \cite{Barabasi2001PhysicaA,Dorogovtsev2002PhysRevE_pseudofractal,JungKimKahng2002PhysRevE,Ravasz2002Science,Ravasz2003PhysRevE,Andrade2005PhysRevLett,Doye2005PhysRevE,Rozenfeld2007NewJPhys}. Therefore, the third line of Eq.~\eqref{eq:Hwang2012PhysRevLett} suggests sublinear power-law scaling of the MFPT with respect to $N$ for these networks.

Unsurprisingly, the MFPT can depend on the distance between source and target nodes.
The results in Ref.~\cite{Noh2004PhysRevLett} have been extended to the case of
networks such as fractal and pseudo-fractal networks in a way that takes into account the distance between the source and target \cite{Condamin2007Nature,Reuveni2010PhysRevE}. The MFPT is
\begin{align}
	m_{ij} \propto
	\begin{cases}
		N(A + B r^{d_{\rm w} - d_{\rm f}})\qquad\quad (d_{\rm f} < d_{\rm w}\,; \text{ i.e., } d_{\rm s}<2)\,,\\
		N(A + B \ln r)\qquad\quad (d_{\rm w} = d_{\rm f}\,; \text{ i.e., } d_{\rm s}=2)\,,\\
		N(A-Br^{d_{\rm w} - d_{\rm f}})\qquad\quad (d_{\rm w} > d_{\rm f}\,; \text{ i.e., } d_{\rm s}> 2)\,,
	\end{cases}
\label{eq:Condamin 2007 main result}
\end{align}
where $r$ is the distance between nodes $v_i$ and $v_j$, and $A$ and $B$ are constants. For example, the Sierpinski gasket has $d_{\rm f}=\ln 3/\ln 2$ and $d_{\rm w}=\ln 5/\ln 2$. Therefore, Eq.~\eqref{eq:Condamin 2007 main result} implies that $m_{ij}\propto N r^{(\ln 5-\ln 3)/\ln 2}$.
The pseudo-fractal scale-free networks that we discussed above satisfy the small-world property, so $d_{\rm f}=\infty$ because the number $N_r$ of nodes within radius $r$ grows exponentially in $r$ \cite{Song2005Nature}.
Additionally, Eq.~\eqref{eq:Condamin 2007 main result} still holds if we replace $d_{\rm f}$ by the box-counting dimension $d_{\rm B}$. The box-counting dimension is defined by the scaling relation $N_{\rm B}/N \propto \ell_{\rm B}^{-d_{\rm B}}$, where $N_{\rm B}$ is the number of non-overlapping boxes of linear size $\ell_{\rm B}$ (e.g., the length of a side for a square) that are necessary to cover an entire fractal (and, in the present context, an entire network). For fractals without a heavy-tailed degree distribution, $d_{\rm B} = d_{\rm f}$ \cite{Song2005Nature}.

For discussion of scaling theory based on renormalization theory for first-passage time and other quantities on networks, see Refs.~\cite{Gallos2007PNAS,Hwang2013PhysRevE}.
For other approaches to first-passage times and return times on networks, see Refs.~\cite{Masuda2004PhysRevE-rw,Baronchelli2006PhysRevE,LauSzeto2010EPL}.


\subsubsection{Cover time}

``Cover time'' is defined as the time required for a random walker to visit all nodes \cite{Lovasz1993Boyal,Aldous2002book}. 
It has been proven that the expected cover time $c$, maximized with respect to the source node, scales approximately as $c \ln \left[c/(c-1)\right] N\ln N$
in an Erd\H{o}s--R\'{e}nyi random graph in which each pair of nodes is adjacent with a probability of approximately $c (\ln N)/N$ \cite{Cooper2007RandStructAlgor}. For a Barab\'{a}si--Albert scale-free network, the expected cover time scales as $2m/(m-1) N\ln N$, where $m$ is the number of edges in each new node  \cite{Cooper2007JCombThSerB}.
These results hold with high probability in the limit of infinite network size (i.e., with probability tending to $1$ as $N\to\infty$). For arbitrary networks, researchers have developed a universal form of the distribution of cover times \cite{Chupeau2015NatPhys} and a method for accurately calculating the mean cover time for networks on which RWs relax rapidly \cite{Maier2017arxiv}.

In practice, exactly covering all nodes tends to be a rather strong requirement.
In contrast to the above and other rigorous mathematical results on exact cover time, physicists have tended to instead examine ``coverage'' $C(n)$ in terms of the number of distinct nodes visited at least once within $n$ steps \cite{Montroll1965JMathPhys,Rammal1984JStatPhys,Barnes1993STOC,Almaas2003PhysRevE,Gallos2004PhysRevE,Stauffer2005PhysRevE,Dafontouracosta2007PhysRevE,Baronchelli2008PhysRevE,Baronchelli2010PhysRevE}.
For a complete graph, one can calculate that
\begin{equation}\label{mean}
	 C(n) =  \sum_{i=1}^N \left[ 1 - \left(1 - p_i^*\right)^n \right]\,.
\end{equation}
because each node is visited with probability $p_i^* = 1/N$ in a single step. In some situations, one can also expect Eq.~\eqref{mean} to hold approximately as a mean-field calculation.
The ``edge coverage'' (i.e., the number of distinct edges visited at least once within $n$ steps) has also been examined for various networks \cite{Barnes1993STOC,Asztalos2010EPL}.




\subsection{Continuous-time random walks (CTRWs)}\label{sub:CTRW nets}


Similar to the case of RWs on a line, CTRWs on networks have two main components: the statistics of a walker's trajectory in terms of the number of steps and the statistics of the times at which events take place. By combining these two components, one can specify the probability that a random walker visits a specified node at a specified time. For RWs on networks, the dynamics of a walker are affected not only by the statistical properties of temporal events, but also by the type of network unit in which a temporal process is defined. First, we distinguish between node-centric CTRWs and edge-centric CTRWs \cite{Aldous2002book,Samukhin2008PhysRevE,Hoffmann2012PhysRevE,Speidel2015PhysRevE}. For dynamical processes in general, there are often substantial differences between node-based dynamics and edge-based dynamics \cite{Porter2016book}, so it is crucial to distinguish between these situations. A second delineation is between active and passive CTRWs, depending on whether a walker passively follows edges when available or actively initializes them as it travels. This second distinction becomes crucial for temporal process other than Poisson process. One can combine the above components to consider various types of walks (e.g., node-centric active CTRWs).

\begin{figure}[tb]
\begin{center}
\includegraphics[scale=0.315]{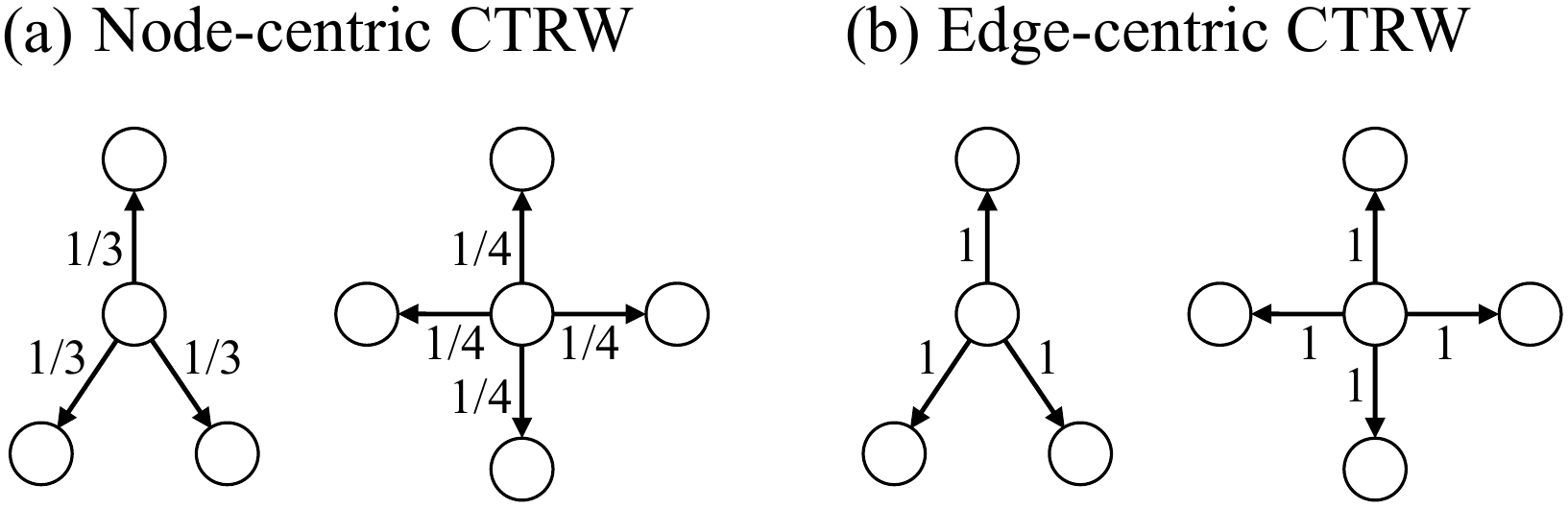}
\caption{Schematic of two types of continuous-time random walks (CTRWs) on networks: (a) a node-centric CTRW and (b) an edge-centric CTRW. In each case, a walker is visiting either a degree-3 node or a degree-4 node in a network, which we assume is unweighted for simplicity. We show the transition rates for each edge. In panel (a), the walker travels at a unit rate and moves to one of its out-neighbors with equal probability for each choice. Therefore, the transition rate for each edge is the reciprocal of the out-degree of the node that the walker is visiting. In panel (b), however, the transition rate on each edge is equal to $1$. Therefore, on average, a walker visiting the node with out-degree $4$ leaves the node earlier than a walker visiting the node with out-degree $3$.
}
\label{fig:node-centric and link-centric CTRW}
\end{center}
\end{figure}


\subsubsection{Node-centric versus edge-centric random walks} \label{sub:node-centric vs edge-centric}

In a CTRW, a walker waits until the next move for a time $\tau$, where $\tau$ is a random variable.
For the sake of simplicity, let's start with a scenario in which moves occur as independent Poisson processes. In other words, $\tau$ is distributed according to the exponential distribution with parameter $\lambda$. We can safely normalize $\lambda$ to $1$, because $\lambda$ only sets the time scale. In a node-centric CTRW, a walker moves from node $v_i$ when it becomes active, and it selects one of the out-neighbors, which we denote by $v_j$, as the destination with a probability proportional to $A_{ij}$ [see Fig.~\ref{fig:node-centric and link-centric CTRW}(a)]. This assumption is the same as that for a DTRW.

The master equation for the Poissonian node-centric CTRW on a network is
\begin{equation}
	\frac{{\rm d}{\bm p}(t)}{{\rm d}t} = \bm p(t) (-I+T) = - \bm p(t) D^{-1} L\,,
\label{eq:master normalized CTRW}
\end{equation}
where
\begin{equation}
	L\equiv D - A
\label{eq:def L}
\end{equation}
is the (``combinatorial'') ``Laplacian matrix'' of the network. The process is driven by the ``random-walk normalized Laplacian''
\begin{equation}
	L^{\prime} \equiv D^{-1} L = I - T\,.
\label{eq:RW normalised Laplacian}
\end{equation}
That is, $(L^{\prime})_{ij} = \delta_{ij} - (A_{ij}/s_i^{\rm out})$.
If we examine the node-centric CTRW in terms of the number $n$ of moves, the trajectories are statistically the same as those of the DTRW in Eq.~\eqref{eq:Markov chain transition}. Consistent with this observation, node-centric CTRWs are also called the ``continuization'' of the DTRW~\cite{Aldous2002book}. In particular, the stationary density of the node-centric CTRW is the same as that of the DTRW. By setting the left-hand side of Eq.~\eqref{eq:master normalized CTRW} to $0$, we obtain $\bm p^* (-I+T)= 0$, so that $\bm p^* = \bm p^* T$. If the network is undirected, $p_i^* = s_i/\sum_{\ell=1}^N s_{\ell}$.
Node-centric CTRWs have been used in, for example, some empirical-data-driven metapopulation disease-spreading models
\cite{Colizza2007NatPhys,Vespignani2011NatPhys}. In those models, a network consists of subpopulations of individuals, and individuals move from one subpopulation to another through a mobility rule. The simplest mobility rule, which has been used widely, is that individuals move according to a Poissonian node-centric CTRW. (For a discussion of mobility models, see Ref.~\cite{MasudaLambiotte2016book}.)

Another type of CTRW is an edge-centric CTRW, in which each edge (rather than a node) is activated independently according to a renewal process [see Fig.~\ref{fig:node-centric and link-centric CTRW}(b)].
By definition, once an edge is activated, it becomes available, and a random walker can use it to move to the associated adjacent node.
This RW model has also been called the ``fluid model'' \cite{Aldous2002book}.

When a Poisson process with a rate proportional to the edge weight is assigned independently to each edge, the master equation is
\begin{equation}
	\frac{{\rm d}{\bm p}(t)}{{\rm d}t} = \bm p(t) (-D+A) = - \bm p(t) L\,.
\label{eq:master unnormalized CTRW}
\end{equation}
The Poissonian edge-centric CTRW is associated with
the unnormalized (i.e., combinatorial) Laplacian $L$. Equation~\eqref{eq:master unnormalized CTRW} implies that the transition rate at node $v_i$ is equal to $s_i^{\rm out}$. A walker leaves a node with a large out-strength (such a node may be a network ``hub'') more quickly than a node with a small out-strength.  This situation contrasts with the aforementioned node-centric CTRW, for which the transition rate of a walker is the same for all nodes.

The stationary density for Eq.~\eqref{eq:master unnormalized CTRW} is
\begin{equation}
	\bm p^* L = 0\,.
\label{eq:p^* CTRW unnormalized}
\end{equation}
Equation~\eqref{eq:p^* CTRW unnormalized} is equivalent to $p_i^* s_i^{\rm out} - \sum_{j=1}^N p_j^* A_{ji} = 0$ (for $i\in \{1, \ldots, N\}$), which indicates that the in-flow of the probability (i.e., $\sum_{j=1}^N p_j^* A_{ji}$) and the out-flow of the probability (i.e., $p_i^* s_i^{\rm out}$) are balanced at each node.
Equation~\eqref{eq:p^* CTRW unnormalized} also indicates that $\bm p^*$ is a left eigenvector of $L$ with eigenvalue $0$. In connected undirected networks, the $0$ eigenvalue, which we denote by $\lambda_1=0$, is an isolated eigenvalue. Its associated eigenvector is
\begin{equation}
	\bm p^* = \frac{1}{N}(1\;, \ldots ,\; 1)\,.
\end{equation}
For a directed network, the right eigenvector corresponding to $\lambda_1=0$ is still given by $(1,\; \ldots\;, 1)^{\top}/N$, but the left eigenvector (i.e., $\bm p^*$) is different in general. Equation~\eqref{eq:p^* CTRW unnormalized} is equivalent to $\bm p^* D = \left(\bm p^* D\right)\left(D^{-1} A\right) = \bm p^* D T$, where (as usual) $T$ is the transition-probability matrix of the DTRW. Therefore, $\bm p^* D$ is the stationary density for the DTRW (and hence for the above node-centric CTRW) in general directed networks. In other words, for the edge-centric CTRW, $p_i^*$ is given by the expression for $p_i^*$ for the node-centric CTRW divided by $s_i^{\rm out}$ and properly normalized.
Using this relationship, we divide
Eq.~\eqref{eq:in-degree approximation} by $s_i^{\rm out}$ to derive the first-order approximation~\cite{MasudaOhtsuki2009NewJPhys,Scardal2016PhysRevE}:
\begin{equation}
	p_i^* \approx (\text{const}) \times \frac{s_i^{\rm in}}{s_i^{\rm out}}\,.
\label{eq:first-order kin/kout}
\end{equation}

For Poissonian node-centric CTRWs and Poissonian edge-centric CTRWs (and also for DTRWs), one can express the stationary density for directed networks by enumerating spanning trees.
We present this technique now because it is easier to understand this approach using $L$ rather than $L^{\prime}$. The ``$(i,j)$ cofactor'' of $L$ is defined by
\begin{equation}
	{\rm Co}\left( i, j \right)
\equiv (-1)^{i+j} \det \overline{L}^{(i, j)}\,,
\label{eq:adjoint}
\end{equation}
where $\overline{L}^{(i,j)}$ is the
$(N-1)\times (N-1)$ matrix obtained by deleting
the $i$th row and the $j$th column of $L$. (Previously, we used $\overline{L}^{(i)}$ to denote the $(N-1)\times (N-1)$ matrix obtained by deleting the $i$th row and column from $L$ (see Section~\ref{sub:MFPT}), and here we use the notation $\overline{L}^{(i,j)}$ without ambiguity. Taking $i = j$ yields $\overline{L}^{(i,i)} \equiv \overline{L}^{(i)}$.)
Because $\sum_{j=1}^N L_{ij} = 0$ (with $i \in \{1, \dots, N\}$),
the value of ${\rm Co}\left( i, j \right)$ is independent of $j$.
Using Eq.~\eqref{eq:adjoint} and the fact that
$L$ is singular because of the $0$ eigenvalue, we obtain
\begin{align}
  \sum_{i=1}^N {\rm Co}(i,i) L_{ij}
  &= \sum_{i=1}^N {\rm Co}(i,j)L_{ij}\nonumber\\
  &= \det L = 0
\end{align}
for any $j$. This yields
\begin{equation}
	p_i^* \propto {\rm Co}\left( i,i\right)
=\det \overline{L}^{(i,i)}\,.
\label{eq:p_i^* spanning tree}
\end{equation}
From the matrix--tree theorem (i.e., Kirchhoff's theorem), $\det \overline{L}^{(i,i)}$ is equal to the sum of the weights of all possible directed spanning trees rooted at $v_i$ (called ``arbolescence'')~\cite{Biggs1997BullLondMathSoc,Agaev2000AutomRemCont}. One thereby obtains $p_i^*$ from weighted spanning trees in a formula called the ``Markov-chain tree formula'' \cite{Aldous2002book}. The ``weight'' of a spanning tree is defined as the product of the weight of the $N-1$ edges that form the tree. For unweighted networks, the weight of a spanning tree is $1$, and $\det \overline{L}^{(i,i)}$ is equal to the number of spanning trees rooted at $v_i$.
When we apply Eq.~\eqref{eq:p_i^* spanning tree} to a node-centric CTRW (or
to a DTRW), we replace $L$ by $L^{\prime}$. In doing this, we must be aware of the weight of spanning trees even for unweighted networks because $L^{\prime}$ is the combinatorial Laplacian for the weighted adjacency matrix $D^{-1}A$, where $A$ is a binary (i.e., unweighted) adjacency matrix.

Equation~\eqref{eq:p_i^* spanning tree} is useful for exacting calculating $p_i^*$ for some directed networks, including a variant of Watts--Strogatz small-world networks and multipartite networks \cite{MasudaKawamuraKori2009PhysRevE}, and for approximately calculating $p_i^*$ for some types of directed networks with community structure \cite{MasudaKawamuraKori2009NewJPhys}.

Although the stationary density differs for node-centric and edge-centric CTRWs, their trajectories (and also those of the DTRW) are statistically the same and are determined by the transition-probability matrix $T$ [see Eq.~\eqref{eq:T_ij random walk}] for Poisson processes. For edge-centric CTRWs, this is true because the probability that a Poisson process on the edge ($v_i$, $v_j$) occurs first among the Poisson processes on all edges ($v_i$, $v_{\ell}$) (where $\ell \in \{1, \ldots, N\}$) is proportional to the rate of the process on the edge ($v_i$, $v_j$) (i.e., it is proportional to $A_{ij}$).
Let $\bm p(n) = (p_1(n),\;  \ldots \;, p_N(n))$ denote the distribution of the random walker, where $p_i(n)$ is the probability that the walker visits $v_i$ after exactly $n$ moves. In the Poissonian case, the master equations for the DTRW, the node-centric CTRW, and the edge-centric CTRW in terms of $n$ are each given by Eq.~\eqref{eq:Markov chain transition vector}. However, the temporal properties along these trajectories are in general different for the two Poissonian CTRWs.
In the Poissonian node-centric CTRW, moves are triggered by a Poisson process at a constant rate, so the probability $p(n,t)$ of having performed $n$ steps at time $t$ is given by a Poisson distribution. In the Poissonian edge-centric CTRW, however, $p(n,t)$ depends on a walker's trajectory. When a walker is at a node $v_i$, the time to the next event is drawn from the exponential distribution with mean $1/s_i^{\rm out}$. If a trajectory includes many nodes with large out-strengths, the number $n$ of moves at a given time $t$ tends to be larger than for trajectories that traverse many nodes with small out-strengths.

The combinatorial Laplacian $L$ of a connected, undirected network includes exactly one $0$ eigenvalue, so $0 = \lambda_1 < \lambda_2 \le \cdots \le \lambda_N$, where $\lambda_{\ell}$ is its $\ell$th smallest eigenvalue. The combinatorial Laplacian of a directed network satisfies an analogous relationship, $0 = \lambda_1 < \mathrm{Re} (\lambda_2) \le  \cdots \le \mathrm{Re} (\lambda_N)$, provided the network is strongly connected or has just one strongly connected component from which all other nodes can be reached by a directed path \cite{Ermentrout1992SiamJApplMath,Agaev2000AutomRemCont,Arenas2008PhysRep}. In the latter case, we call such a strongly connected component the ``root component'' (including the case of a single node, which is then a ``root node''). If there are multiple components in an undirected network or multiple root components, then there are multiple $0$ eigenvalues in $L$, although we do not consider such situations in the present article. The spectral gap (and thus $\lambda_2$) governs the relaxation time. The corresponding eigenvector $\bm u_2$ is called the ``Fiedler vector''. For details of spectral properties of networks, see Refs.~\cite{Merris1994LinAlgeItsAppl,Chung1997book-spectral,Mohar1997chapter,Godsil2001book,Arenas2008PhysRep,Cvetkovic2010book,CohenHavlin2010book,Newman2010book,piet-book}.

When a network is undirected, one can also construct Eq.~\eqref{eq:master unnormalized CTRW} as a type of deterministic, linear synchronization or coordination dynamics in which $p_i(t)$ is the state of node $v_i$ and nodes $v_i$ and $v_j$ attract each other with a coupling strength of $A_{ij}$ \cite{Arenas2008PhysRep}. The only difference between CTRW dynamics and linearized synchronization dynamics is that $p_i(t)$ is confined between $0$ and $1$ and normalized in CTRWs, whereas it is not in synchronization dynamics. Therefore, various theoretical results on linear synchronization dynamics on networks are applicable to edge-centric CTRWs. In particular, methods to estimate the relaxation time via the spectral gap of $L$ are useful for understanding relaxation properties of RWs \cite{Arenas2008PhysRep,pecora-review,Motter2007NewJPhys}.


\subsubsection{Active versus passive random walks} \label{sub:active vs passive}

In Section~\ref{sub:node-centric vs edge-centric}, we assumed that temporal events are determined from Poisson processes.
In that case, it was not necessary to specify if temporal events are defined on the walker or on the
network. However, for non-Poisson processes, it is crucial to specify these properties. In this section,
we assume that temporal events are generated by renewal processes with arbitrary distributions of inter-event times. Various empirical data sets related to human activity support heavy-tailed (and hence non-exponential) distributions \cite{VazquezA2006PhysRevE-burst,HolmeSaramaki2012PhysRep}. See Ref.~\cite{kivela-PRE2015} for a discussion of how to estimate such distributions from empirical data.

One type of model arises when a renewal process describes the timings of the moves of a random walker. In other words, the walker carries its own clock and re-initializes it after each move. The CTRW is then said to be active, which may be appropriate components of models of human or animal trajectories.

A second model consists of assuming that it is the timings at which nodes or edges become active that are generated by a renewal process. In such scenarios, the node or the edge (rather than a walker) carries a clock, and the arrival of a walker does not modify it. The random walker is thus a passive entity that follows edges when they become available \cite{Hoffmann2012PhysRevE,Speidel2015PhysRevE}. Passive RWs are often used in models of spreading of a virus on a time-dependent contact network or in the spreading of information on a communication network.\footnote{However, spreading processes are typically non-conservative, so one needs to be careful about using RWs in these situations.}
Active and passive walks model different types of situations. One can interpret active walks as a continuous-time process that can take place on a fixed network architecture. One can then construe the resulting flickering of edges induced by a walker as components of a temporal network. In contrast, passive walks are event-driven processes that take place on a temporal network, which has its own intrinsic dynamics. As we will see, the two types of walks have radically different mathematical properties.


{\em Node-centric active CTRWs.} When the inter-event time between two moves obeys a distribution $\psi(\tau)$ that is not exponential, the RW dynamics are non-Markovian. In a non-Markovian setting, the rate at which a walker moves depends on the time since the last move. To analyze this scenario, we consider the extension of Eq.~\eqref{eq:p(x;t) with free n} to the case of general networks and write
\begin{equation}
	\bm p(t) = \sum_{n=0}^{\infty} \bm p(n) p(n,t)\,,
\label{eq:p(t) with p(n)}
\end{equation}
where we recall that $p(n, t)$ is the probability that a walker has moved $n$ times at time $t$. By taking the Laplace transform of Eq.~\eqref{eq:p(t) with p(n)} and using Eqs.~\eqref{eq:hatp(n,s)}, we obtain
\begin{equation}
	\hat{\bm p}(s)  = \frac{1 - \hat{\psi}(s)}{s} \sum_{n=0}^{\infty} \bm p(n) \hat{\psi}(s)^n\,.
\label{eq:Montroll 0}
\end{equation}
We then substitute $\bm p(n) = \bm p(0) T^n$ [see Eq.~\eqref{eq:m1}] into Eq.~\eqref{eq:Montroll 0}, where $T$ is the transition-probability matrix of the DTRW, to obtain
\begin{equation}
	\hat{\bm p}(s) = \frac{1 - \hat{\psi}(s)}{s} \bm p(0) \left[ I - T\hat{\psi}(s) \right]^{-1}\,.
\label{eq:Montroll}
\end{equation}
Equation~\eqref{eq:Montroll} is a generalization to arbitrary networks of results by Montroll and Weiss \cite{Montroll1965JMathPhys}. We have implicitly taken a node-centric perspective, as the waiting time (i.e., the time to the next event) of the walker does not depend on the node degree; when the walker is ready for a move, it chooses one of the node's edges uniformly at random and traverses it.
 The inverse Laplace transform of Eq.~\eqref{eq:Montroll} gives the probability $p_i(t)$ that the walker visits $v_i$ at time $t$.

For a Poisson process (i.e., when $\psi(\tau) = \beta e^{-  \beta \tau}$), substituting $\hat{\psi}(s) = \beta/(s+ \beta)$ [see Eq.~\eqref{eq:hatpsi(s) Poisson}] in Eq.~\eqref{eq:Montroll} yields
\begin{equation}
	s \hat{\bm p}(s) - \bm p(0) =  \beta \hat{\bm p}(s) (-I+T)
\label{eq:p(s) Poisson general net}
\end{equation}
after some calculations. Because the inverse Laplace transform of $s \hat{\bm p}(s) - \bm p(0)$ is equal to $\frac{{\rm d}{\bm p}}{{\rm d}t}(t)$, Eq.~\eqref{eq:p(s) Poisson general net} leads to Eq.~\eqref{eq:master normalized CTRW} up to a multiplicative constant $\beta$.


To understand how the form of $\psi(\tau)$ affects diffusive processes, let's work in the graph-Fourier domain. That is, we work in terms of the amplitude of the eigenmodes, and we examine how the relaxation of different eigenmodes deviates from the situation for Poisson processes \cite{Delvenne2015NatComm}. Combining Eqs.~\eqref{eq:p_i(t) decomposed}--\eqref{eq:graph FT 2}
and \eqref{eq:Montroll 0} yields
\begin{equation}
	\hat{\bm p}(s)  = \frac{1 - \hat{\psi}(s)}{s} \sum_{\ell=1}^N \frac{a_{\ell}(0)}{1-\lambda_{\ell}\hat{\psi}(s)}
\bm u_{\ell}^{\rm L}\,,
\label{eq:hatp(s) with eigenmodes}
\end{equation}
where $\lambda_{\ell}$ is an eigenvalue of $T$ and $\bm u_{\ell}^{\rm L}$ is the corresponding left eigenvector. By taking the inner product of both sides of Eq.~\eqref{eq:hatp(s) with eigenmodes} with the right eigenvector $\bm u_{\ell}^{\rm R}$ of $T$ for a particular value $\ell$, we obtain
\begin{equation}
	\hat{a}_{\ell}(s) = \frac{1 - \hat{\psi}(s)}{s \left[1 -  \lambda_{\ell} \hat{\psi}(s)\right]} a_{\ell}(0)\,.
\label{eq:a(s) network}
\end{equation}

For CTRWs driven by Poisson processes, an eigenmode relaxes exponentially in time. However, relaxation dynamics can be rather different when $\psi(t)$ is not an exponential distribution. For simplicity, we assume that $\psi(t)$ has finite mean and finite variance. (When these moments are not defined, one can examine dynamical processes using the framework of fractional calculus \cite{Denigris2016EurPhysJB}.) We substitute a small-$s$ expansion
\begin{equation}
	\hat{\psi}(s) = 1 - \langle\tau \rangle s + \frac{1}{2} \langle \tau^2 \rangle s^2 + o(s^2)
\end{equation}
into Eq.~\eqref{eq:a(s) network}. For the $\ell$th mode, where $\lambda_{\ell} \neq 1$, one can calculate that
\begin{equation}
	a_{\ell}(s)
	      = \frac{\langle \tau \rangle}{1-\lambda_{\ell}} \left[ 1 - s \left( \frac{\lambda_{\ell} \langle \tau \rangle}{1-\lambda_{\ell}}
+ \frac{\langle \tau^2 \rangle}{2 \langle \tau \rangle} \right) \right]\,.
\label{eq:a_ell(s) with Tauberian}
\end{equation}
This leads to a characteristic time $t_{\rm cha}$ of
\begin{align}
	t_{\rm cha} &= \frac{\lambda_{\ell} \langle \tau \rangle}{1-\lambda_{\ell}}
+ \frac{\langle \tau^2 \rangle}{2 \langle \tau \rangle}\notag\\
	&= \langle \tau \rangle \left(\frac{1}{\epsilon_{\ell}} + \beta_{\text{burst}} \right)\,,
\label{JCBurst}
\end{align}
where $\epsilon_{\ell} = 1 - \lambda_{\ell}$ is the eigenvalue of the random-walk normalized Laplacian $L^{\prime}$ and
\begin{equation}
	\beta_{\text{burst}} =\frac{\sigma_{\tau}^2-\langle \tau\rangle^2}{2 \langle \tau\rangle^2}\,,
\label{eq:beta}
\end{equation}
where $\sigma_{\tau}^2 = \langle \tau^2\rangle - \langle \tau\rangle^2$ is the variance of $\tau$. The quantity $\beta_{\text{burst}} \in [-1/2,\infty)$ is a measure of burstiness. Poisson processes have $\beta_{\text{burst}}=0$, and $\beta_{\text{burst}}=-1/2$ when $\psi(\tau)$ is distributed as a delta function. A heavy-tailed distribution, implying bursty activity of nodes, generates a large value of $\beta_{\text{burst}}$.

Let's consider the slowest-decaying mode associated with the spectral gap $\epsilon_{\ell}$ (i.e., the smallest nonzero eigenvalue of $L^{\prime}$). The corresponding characteristic decay time $t_{\rm cha}$ indicates the relaxation time of the
CTRW towards equilibrium. Equation~\eqref{JCBurst} includes competition between two factors. When the spectral gap is small relative to $1/\beta_{\rm burst}$, the first term on the right-hand side of Eq.~\eqref{JCBurst} is dominant. In this case, $t_{\rm cha}$ is determined primarily by structural bottlenecks in a network (e.g., through the existence of sets of densely-connected nodes called ``communities'' (see Section~\ref{sub:community}), which are connected weakly to each other) \cite{Chung1997book-spectral,piet-book,Jeub2015PhysRevE}.
When the spectral gap is larger or when an event sequence is bursty (in the sense of a large variation in inter-event times), the second term dominates the right-hand side of Eq.~\eqref{JCBurst}. In this case, $t_{\rm cha}$ is determined primarily by the properties of $\psi(\tau)$ rather than by network structure.

Because the inter-event time and the number of moves in a RW are statistically independent, the stationary density of the node-centric CTRW with a general $\psi(\tau)$ is the same as those for a DTRW or a Poissonian node-centric CTRW. One can thus calculate the recurrence time and first-passage time of a node-centric CTRW by multiplying the corresponding results for the DTRW (see Section~\ref{sub:MFPT}) by $\langle \tau \rangle$.

{\em Edge-centric active CTRWs.} One can define other types of active RWs
that have qualitatively different behaviors of the stationary density and first-passage times. For instance, consider the following edge-centric active RW: when a walker arrives at a node, it considers each edge and takes the first edge available for transport. The time at which each edge appears is independently drawn from the same distribution $\psi(\tau)$ where, as before, the clock on each edge is re-initialized upon the arrival of a walker at an incident node.
Because only the first edge to appear is taken by the walker, there is a competition between different edges. The probability density that a random walker moves from node $v_i$ to node $v_j$ at time $\tau$ since the walker arrived at $v_i$ is
\begin{equation}
\label{eq:f(t;j <- i)}
	f(\tau;j \gets i) = \psi(\tau) \left[\int_{\tau}^{\infty} \psi(\tau^{\prime}) {\rm d}\tau^{\prime}\right]^{k_i-1}\,.
\end{equation}
Some calculations yield
\begin{equation}
	p_i^* = \frac{\langle \min_{\ell=1,\dots,k_i} \tau_{\ell} \rangle k_i}{\sum_{j=1}^N \langle \min_{\ell=1,\ldots,k_j} \tau_{\ell} \rangle k_j}\,,
\label{eq:p_i^* active RW STN}
\end{equation}
where the factors of $\tau_\ell$ are independent copies of inter-event times that are drawn from the distribution $\psi(\tau)$.
Because
\begin{equation}
	\left\langle \min_{\ell=1,\dots,k_i} \tau_{\ell} \right\rangle = \int_0^\infty \left[ \int_{\tau^{\prime}}^\infty \psi({\tau^{\prime}}) {\rm d}{\tau^{\prime}} \right]^{k_i} {\rm d}{\tau^{\prime}}
\label{eq:actmin RW STN}
\end{equation}
depends only on $k_i$, Eqs.~\eqref{eq:p_i^* active RW STN} and \eqref{eq:actmin RW STN} imply that $p_i^*$ depends only on $k_i$. Note that the stationary density for the active RW is not proportional to $k_i$ unless $\tau$ is constant, which reduces the model to the DTRW. The mean recurrence time for node $v_i$ is
\begin{equation}
	m_{ii} =  \frac{\sum_{j=1}^N \left\langle \min_{\ell=1,\ldots,k_j} \tau_{\ell} \right\rangle k_j}{k_i} \propto \frac{1}{k_i}\,.
\label{eq:T_{i|i} active identical RW STN}
\end{equation}
Equations~\eqref{eq:p_i^* active RW STN} and \eqref{eq:T_{i|i} active identical RW STN} indicate that Kac's formula [see Eq.~\eqref{eq:Kac}] is not satisfied unless the network is regular.

\begin{figure}[tb]
\begin{center}
\includegraphics[scale=0.45]{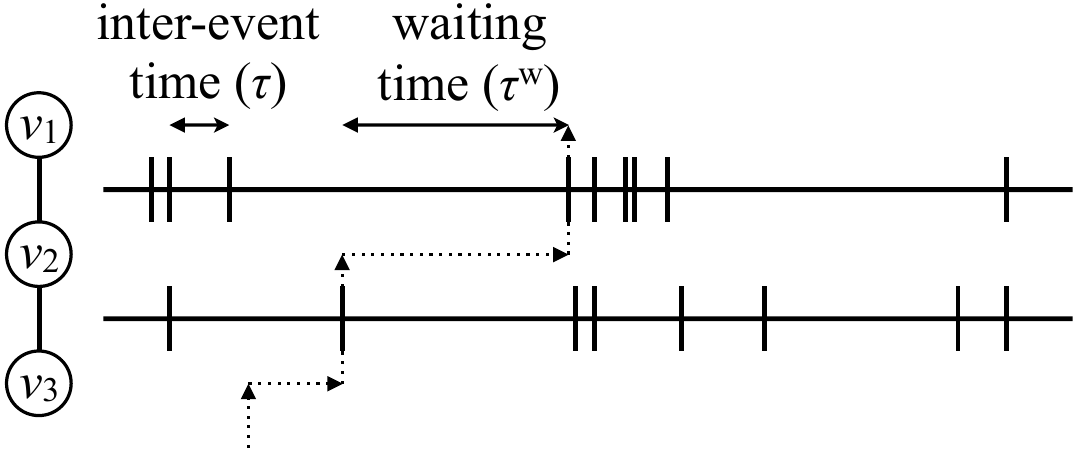}
\caption{Schematic illustrating
the concept of waiting time. We show a trajectory of a random walker using dotted arrows. The walker moves from node $v_3$ to node $v_2$, and it then moves to node $v_1$. This example corresponds to $j=1$, $i=2$, and $\ell=3$ in the main text. (See the $j\neq \ell$ case in Eq.~\eqref{eq:approx T RW STN}.)
}
\label{fig:waiting time}
\end{center}
\end{figure}

{\em Edge-centric passive CTRWs.} Passive RWs differ from active ones in that properties of a network (rather than a random walker) evolves as a renewal process. We start with edge-centric passive RWs, which
have attracted considerable attention because of their many applications (e.g., diffusion on temporal networks). We thus assume that each edge is governed by an independent renewal process, which we assume for simplicity is the same distribution $\psi(\tau)$ for each edge.
A first important difference from active walks arises from the ``waiting-time paradox'' (which is also called the ``bus paradox'') \cite{Feller1971book2,Allen1990book}.
In this paradox, a walker arrives at node $v_i$ from node $v_{\ell}$. The waiting time before edge $(v_i, v_j)$ (with $j \neq \ell$) is activated is typically longer than the naive expected value $\langle \tau\rangle/2$. Let $\psi^{\rm w}(\tau^{\rm w})$ denote the distribution of waiting times $\tau^{\rm w}$ on edge ($v_i$, $v_j$) after a walker has arrived at node $v_i$ from node $v_{\ell}$ (where $\ell \neq j$). See Fig.~\ref{fig:waiting time} for a schematic. One can calculate $\psi^{\rm w}(\tau^{\rm w})$ from $\psi(\tau)$ when the arrival of a walker to $v_i$ and the activation of edge ($v_i$, $v_j$) are statistically independent processes.
In that situation, the probability density
for the time at which a walker moves from $v_{\ell}$ to $v_i$ lies in an interval of length $\tau$ satisfies
\begin{equation}
	f(\tau) = \frac{\tau \psi(\tau)}{\int_0^{\infty} \tau^{\prime} \psi(\tau^{\prime}){\rm d}\tau^{\prime}} = \frac{\tau \psi(\tau)}{\langle\tau\rangle}\,.
\label{eq:fall within}
\end{equation}
Conditioned on the walker's arrival time to $v_i$ lying in an interval of length $\tau$, the probability density for the waiting time to be equal to $\tau^{\rm w}$ is
\begin{equation}
	g(\tau^{\rm w}|\tau) = \begin{cases}
1/\tau & (0\le \tau^{\rm w}\le \tau)\,,\\
0 & (\tau > \tau^{\rm w})\,.
	\end{cases}
\label{eq:waiting time pdf conditional}
\end{equation}
Equations~\eqref{eq:fall within} and \eqref{eq:waiting time pdf conditional} yield
\begin{equation}
	\psi^{\rm w}(\tau^{\rm w})= \int_{\tau^{\rm w}}^{\infty} f(\tau)g(\tau^{\rm w}|\tau) {\rm d}\tau=
\frac{1}{{\langle\tau\rangle}} \int_{\tau^{\rm w}}^\infty \psi(\tau) {\rm d}\tau\,.
\label{eq:waiting-time distribution}
\end{equation}
In particular, the mean waiting time is given by
$\int_0^{\infty} \tau^{\rm w} \psi^{\rm w}(\tau^{\rm w}) {\rm d}\tau^{\rm w}
=  \langle\tau^2\rangle/\left(2\langle\tau\rangle\right)$.
If $\psi(\tau)$ is heavy-tailed, $\langle\tau^2\rangle$ is much larger than $\langle\tau\rangle$, so a typical waiting time is very long. For example, if $\psi(\tau)\propto \tau^{-\gamma}$, with $\gamma \in (2,3]$, the mean inter-event time is finite, whereas the mean waiting time diverges because $\langle\tau^2\rangle$ diverges.

A second difference is that one can only derive approximate master equations for edge-centric passive CTRWs, whereas they are exact for active CTRWs. When a random walker moves from node $v_{\ell}$ to node $v_i$ at time $t$, the waiting time (i.e., the time to the next event) on edge ($v_i$, $v_j$), where $j\neq \ell$ (we will consider the case $j=\ell$ in the next paragraph), is
estimated by the distribution $\psi^{\rm w}$
%
However, if a random walker has already traversed edge ($v_i$, $v_j$) in the past --- let's suppose that the last traversal time occurred at $t^{\prime}$ --- the independence assumption that is required to derive Eq.~\eqref{eq:waiting-time distribution} is not satisfied, and the waiting time on ($v_i$, $v_j$) is not given exactly by the distribution $\psi^{\rm w}$, unless the process is Poissonian and $\psi$ is an exponential distribution. The deviation between the waiting-time distribution and $\psi^{\rm w}$ increases when $t^{\prime}$ approaches $t$. In the remainder of the present section, we ignore any modification of the distribution of the subsequent waiting time caused by past events on ($v_i$, $v_j$); this corresponds to assuming that $t^{\prime}=-\infty$. To our knowledge, the impact of such a memory effect (i.e., finite $t^{\prime}$)
has not been considered in detail in the literature.


A third difference stems from the possibility of non-Markovian trajectories for random walkers. To explain this point, consider the case of backtracking moves (i.e., $v_\ell \to v_i \to v_\ell$). For such backtracking moves, the waiting time on the edge ($v_i$, $v_{\ell}$) is distributed according to $\psi$, rather than $\psi^{\rm w}$, as the waiting-time paradox does not apply. The existence of different waiting times for backtracking and non-backtracking moves has impacts the motion of a walker.
For a walker to move to node $v_j$ at time $\tau^{\rm w}$ since the walker moved from node $v_{\ell}$ to node $v_i$, there cannot be any events on any edges emanating from $v_i$ in $[0, \tau^{\rm w}]$, and then an event must occur on the edge ($v_i$, $v_j$) at time $\tau^{\rm w}$.
Let $f(\tau^{\rm w};j \gets i | i \gets \ell)$ denote the probability density of the event that a walker that has moved from $v_{\ell}$ to $v_i$ moves to node $v_j$ at time $\tau^{\rm w}$.
We obtain
\begin{align}
	f(\tau^{\rm w};j \gets i | i \gets \ell) \approx
\begin{cases}
	\psi(\tau^{\rm w}) \left[ \int_{\tau^{\rm w}}^{\infty} \psi^{\rm w}(\tau^{\prime}) {\rm d}\tau^{\prime} \right]^{k_i-1} & (j = \ell)\,,\\
\psi^{\rm w}(\tau^{\rm w}) \left[ \int_{\tau^{\rm w}}^{\infty} \psi^{\rm w}(\tau^{\prime}) {\rm d} \tau^{\prime}\right]^{k_i-2} \int_{\tau^{\rm w}}^{\infty} \psi(\tau^{\prime}) {\rm d}\tau^{\prime} & (j \neq \ell)\,.
	\end{cases}
\label{eq:approx T RW STN}
\end{align}
Equation~\eqref{eq:approx T RW STN} indicates that where a walker moves 
depends not only on its current position but also on the edge that it used to arrive to that position.
For trajectories of RWs, one can construe this situation as a special case of the ``memory networks'' that we will discuss in Section~\ref{sec:memory net}.

Unless $\psi$ is an exponential distribution, $f(\tau^{\rm w};\ell \gets i | i \gets \ell)$ is not equal to $f(\tau^{\rm w};j \gets i | i \gets \ell)$ (with $j\neq \ell$) in general, so the trajectory of an RW (i.e., the walk measured in terms of the number of moves) is non-Markovian.
In particular, if $\psi$ is a heavy-tailed distribution, the mean waiting time is larger than the mean inter-event time. Therefore, a walker tends to backtrack (i.e., there are sequences of moves of the form $v_{\ell} \to v_i \to v_{\ell}$), and diffusion dynamics are slowed down. This slowing down is caused entirely by the modification of trajectories in non-exponential distributions, and, in particular, it does not arise from a competition between structural and temporal factors (in contrast to Eq.~\eqref{eq:beta}).
If $\psi$ has lighter tails than an exponential distribution, a walker tends to avoid backtracking. (We briefly discuss non-backtracking RWs in Section~\ref{sec:outlook}.) When $\psi$ is not an exponential distribution, trajectories of the edge-centric passive CTRW are different from those of active CTRWs or DTRWs.

We now evaluate the stationary density and recurrence time of non-Poissonian edge-centric passive CTRWs \cite{Speidel2015PhysRevE}. Let $q_{j \gets i}(t)$ denote the rate at which a random walker moves from node $v_i$ to node $v_j$ at time $t$. This quantity satisfies the following approximate self-consistency equation:
\begin{align}
	q_{j \gets i}(t) &\approx \sum_{\ell\in {\mathcal N}_i} \left[ \int_0^t f(t-t^{\prime}; j \gets i | i \gets \ell) q_{i \gets \ell}(t^{\prime}) {\rm d}t^{\prime} \right] \notag \\
		&\qquad + p_{j \gets i}(0) \delta(t)\,,
\label{eq:qpas RW STN}
\end{align}
where we recall that ${\mathcal N}_i$ is set of the neighbors of $v_i$. The initial condition satisfies
\begin{equation}
\label{eq:initial RW STN}
	\sum_{j\in {\mathcal N}_i} p_{i \gets j}(0) = p_i(0)\,.
\end{equation}
Equation~\eqref{eq:initial RW STN} implies that one needs to specify an initial condition that includes not only the current position of the walker but also its previous location. More generally, the transition probability of a move depends on the previous move.
The master equation is given by
\begin{equation}
	\frac{d}{dt}p_i(t) = \sum_{j\in {\mathcal N}_i} \left[ q_{i \gets j}(t) - q_{j \gets i}(t) \right]\,.
\label{eq:master RW STN}
\end{equation}

To derive the stationary density, we work in terms of $q_{i \gets j}(t)$ rather than $p_i(t)$. We take the Laplace transform of Eq.~\eqref{eq:qpas RW STN} to obtain
\begin{equation}
	\hat{q}_{j \gets i}(s) \approx \sum_{\ell\in {\mathcal N}_i} \left[ \hat{f}(s; j \gets i | i \gets \ell) \hat{q}_{i \gets \ell}(s) \right] + p_{j \gets i}(0)\,.
\label{eq:hatq_ji(s) RW STN}
\end{equation}
Note that $\hat{q}_{j \gets i}(s) \neq \hat{q}_{i \gets j}(s)$ in general even for undirected networks. Equation~\eqref{eq:hatq_ji(s) RW STN} is a set of linear equations with $2M$ unknowns. We solve $\hat{q}_{j \gets i}(s)$ and then calculate the stationary value of $q_{j \gets i}(t)$ (i.e., $q_{j\gets i}^*\equiv \lim_{t\to\infty}q_{j \gets i}(t)$ as $\hat{q}_{j\gets i}(0)$). We thereby obtain $p_i^*$ as a weighted sum of $q_{i\gets j}^*$ terms, where $j\in {\mathcal N}_i$. In fact, $q_{j\gets i}^*$ does not depend on $i$ or $j$, and the final result is
\begin{equation}
	p_i^* = \frac{1}{N} \quad (i \in \{1, \dots ,N\})\,.
\label{eq:p_i^* link-centric CTRW renewal}
\end{equation}
Therefore, the stationary density is the uniform density, independent of the network structure and the form of $\psi(\tau)$. The mean recurrence time is
\begin{equation}
	m_{ii} \approx \frac{N \langle \tau \rangle }{k_i}\,.
\label{eq:T_{i|i} passive}
\end{equation}
Equation~\eqref{eq:T_{i|i} passive} indicates that the mean recurrence time is essentially independent of $\psi(\tau)$, as it depends only on the mean $\langle \tau\rangle$, which gives the trivial normalization of time. Equations~\eqref{eq:p_i^* link-centric CTRW renewal} and \eqref{eq:T_{i|i} passive} imply that Kac's formula [see Eq.~\eqref{eq:Kac}] is not satisfied by any edge-centric passive CTRW except in regular networks.


{\it Node-centric passive CTRWs.} To conclude our taxonomy of CTRWs on networks, we mention a fourth combination: passive node-centric RWs. We are not aware of studies of node-centric passive RWs,
though they may be relevant for situations in which the activity of a temporal network is driven by node dynamics more than by interactions between nodes.
Node-centric passive CTRWs are also subject to the bus paradox, but they are substantially simpler mathematically than edge-centric active walks, because non-Markovian trajectories do not arise when the renewal processes on the nodes are independent.




\section{Random walks on generalized networks}

\subsection{Multilayer networks} \label{sec:multilayer}

A multilayer network includes different ``layers'' and allows one to explicitly incorporate different types of subsystems and/or different types of ties between edges \cite{Kivela2014JCompNetw,Boccaletti2014PhysRep}. The latter case, which is often called a ``multiplex'' network, occurs when there are different types of interactions between individuals, different modes of transportation, and so on. If there are $\ell_{\max}$ layers, one can represent a multilayer network as an ordinary (i.e., ``monolayer'') network with  $\ell_{\max}N$ nodes, where there are $\ell_{\max}$ replicates of each node if each entity (represented by a node) exists on every layer.
How strongly different layers are connected to each other (and which interlayer edges are present) has an enormous effect on diffusive dynamics in multilayer networks \cite{Kivela2014JCompNetw,naturephysicsspreading,deford2015}. It thereby affects anything else, such as various community-detection methods, that are based on RWs (see Section~\ref{sub:community}) \cite{Mucha2010Science,Dedomenico2015PhysRevX,Jeub2017NetwSci}.

\begin{figure}[tb]
\centering
\includegraphics[scale=0.45]{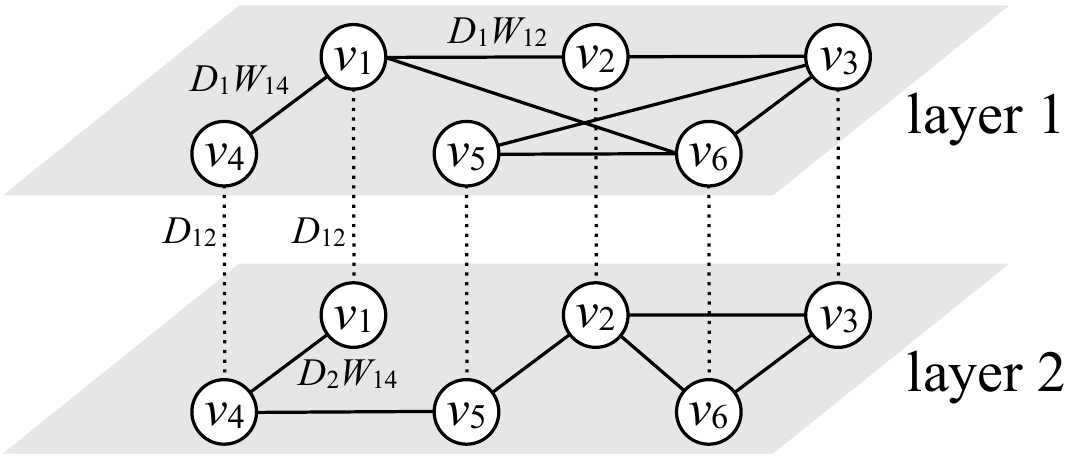}
\caption{Schematic of a Poissonian edge-centric CTRW on a multilayer network with $\ell_{\max}=2$ layers. The values on the edges represent edge weights.}
\label{fig:multilayer Gomez}
\end{figure}

Let's consider Poissonian edge-centric CTRWs. For simplicity, we also assume undirected multilayer networks in which each intra-layer network is a connected network \cite{Gomez2013PhysRevLett,Soleribalta2013PhysRevE,Radicchi2013NatPhys}
and each node is present on every layer (though of course this need not be true in general).
We also assume that inter-layer edges occur only between the same entity in different layers (i.e., so-called ``diagonal'' coupling) and that there is only a single type (i.e., ``aspect'') of layering \cite{Kivela2014JCompNetw}. (For example, a single-aspect multilayer network can be a multiplex network, but it cannot be both multiplex and time-dependent.)
%
Let $A^{\alpha} = (A^{\alpha}_{ij})$ denote the adjacency matrix for the $\alpha$th layer. One needs to think about both diffusion within layers and diffusion between layers (see Fig.~\ref{fig:multilayer Gomez}). Let $D_{\alpha}$ denote the intra-layer diffusion constant in the $\alpha$th layer, and let $D_{\alpha\beta}$ (with $\alpha,\beta \in \{1,\dots, \ell_{\max} \}$) denote the inter-layer diffusion constant between the $\alpha$th and $\beta$th layers. Such constants set the edge weights between pairs of nodes that represent the same entity in different layers, and the corresponding nodes in the $\alpha$th and $\beta$th layers are connected by an edge on which there is a Poisson process with rate $D_{\alpha\beta}$. The master equation is given by
\begin{equation}
	\frac{{\rm d}p_i^{\alpha}(t)}{{\rm d}t} =
D_{\alpha} \sum_{j=1}^N A_{ij}^{\alpha} \left[p_j^{\alpha}(t) - p_i^{\alpha}(t)\right]
+ \sum_{\beta=1}^{\ell_{\max}} D_{\alpha \beta}\left[ p_i^{\beta}(t) - p_i^{\alpha}(t)\right]\,,
\label{eq:master Gomez2013}
\end{equation}
where $p_i^{\alpha}(t)$ is the probability that a random walker visits the $i$th node in the $\alpha$th layer. The normalization is given by
$\sum_{\alpha=1}^{\ell_{\max}} \sum_{i=1}^N p_i^{\alpha}(t)=1$.

Consider the case of two layers and $D_{\rm x} \equiv D_{12} = D_{21}$ \cite{Gomez2013PhysRevLett,Radicchi2013NatPhys}. Equation~\eqref{eq:master Gomez2013} is written concisely as
\begin{equation}
	\frac{{\rm d}\bm p(t)}{{\rm d}t} = - \bm p(t) \mathcal{L}\,,
\end{equation}
where $\bm p(t) =$ $(p_1^1(t),\;$ $p_2^1(t),\;$ $\ldots\;,$ $p_N^1(t),\;$ $p_1^2(t),\;$ $p_2^2(t),\;$ $\ldots\;,$ $p_N^2(t))$, and
\begin{equation}
	\mathcal{L}  =
\begin{pmatrix}
D_1 L_1 + D_{\rm x} I & -D_{\rm x} I\\
	-D_{\rm x} I & D_2 L_2 + D_{\rm x} I
\end{pmatrix}
\end{equation}
is the (combinatorial) ``supra-Laplacian'', where $L_1$ and $L_2$ are the (combinatorial) Laplacian matrices for the intra-layer network. Because this RW is an edge-centric CTRW on an undirected network, the stationary density is $\left(p_i^{\alpha}\right)^* = 1/(2N)$ (with $i \in \{1, \dots , N\}$ and $\alpha \in \{1,2\}$).

The supra-Laplacian matrix $\mathcal{L}$ has a $0$ eigenvalue that corresponds to the stationary density. The relaxation time is governed by the smallest positive eigenvalue (i.e., the spectral gap) $\lambda_2$ of $\mathcal{L}$. One of the nonzero eigenvalues is $2D_{\rm x}$ and has a corresponding eigenvector of $(1,\; \ldots\;, 1,\; -1, \;  \ldots\;, -1)$. If the inter-layer diffusion constant $D_{\rm x}$ is small, then $\lambda_2 = 2D_{\rm x}$, so the inter-layer hopping is a bottleneck for diffusion in the entire multilayer network. In the opposite limit ($D_{\rm x}\gg 1$), one can examine diffusion properties using a perturbative analysis \cite{Gomez2013PhysRevLett}. The quantity $2D_{\rm x}$ is still an eigenvalue, but it diverges to infinity in the limit $D_{\rm x}\to\infty$, and there are $N$ copies of the same eigenvalue in this limit. Another important quantity is $\lambda_{\rm s}/2$, the eigenvalue of $(L_1+L_2)/2$; and there are also $N$ copies of this eigenvalue. Therefore, $\lambda_2 = \lambda_{\rm s}/2$. Note that $L_1+L_2$ is the (combinatorial) Laplacian for the monolayer network obtained by adding the intra-layer edge weights for each intra-layer edge and ignoring the inter-layer edges. We obtain
\begin{equation}
	\frac{\lambda_{\rm s}}{2} \ge \frac{\lambda_2^{\alpha=1}+\lambda_2^{\alpha = 2}}{2}
\ge \min (\lambda_2^{\alpha=1}, \lambda_2^{\alpha=2})\,,
\label{eq:bound Gomez2013}
\end{equation}
where $\lambda_2^{\alpha}$ is the second-smallest eigenvalue (i.e., the spectral gap) of $L_{\alpha}$, so it specifies the speed at which an RW on the network consisting only of the $\alpha$th layer (so there are no inter-layer edges) relaxes to the stationary density. Equation~\eqref{eq:bound Gomez2013} implies that above diffusion in the two-layer network is faster than diffusion in the slower layer. For some multilayer networks, however, diffusion can occur faster than in each layer considered individually
\cite{Gomez2013PhysRevLett,Soleribalta2013PhysRevE}.

The small-$D_{\rm x}$ and $D_{\rm x} \gg 1$ regimes are connected by a discontinuous
(i.e., ``first-order'') phase transition \cite{Radicchi2013NatPhys}. More precisely, there exists a threshold value $D_{\rm x}^*$ of $D_{\rm x}$, such that $\lambda_2 = 2 D_{\rm x}$ for $D_{\rm x} \le D_{\rm x}^*$ and
$\lambda_2 \le \lambda_{\rm s}/2$ for $D_{\rm x} \ge D_{\rm x}^*$. Note that $D_{\rm x} \to \lambda_{\rm s}/2$ as $D_{\rm x}\to\infty$. The first derivative of $\lambda_2$ with respect to $D_{\rm x}$ is discontinuous at $D_{\rm x} = D_{\rm x}^*$. The transition point has an upper bound given by $D_{\rm x}^* \le \lambda_{\rm s} /4$.

Reference \cite{Dedomenico2014PNAS} investigated the so-called ``coverage'' time of different types of CTRWs in multilayer networks by calculating the mean fraction of distinct nodes that are visited at least once (in any layer) in some time period by a walk (which can start from any node in a network).
Reference~\cite{Dedomenico2014PNAS} then examined coverage as a function of time when some nodes are deleted and used it to consider the resilience of multilayer networks to random node failures. In their paper, node failure is defined with respect to the removal of nodes in individual layers (rather than, e.g., removal from all layers), such as a failure of a station in a single transportation mode (i.e., a single layer) in a transportation network.


See Refs.~\cite{Kivela2014JCompNetw,naturephysicsspreading,Boccaletti2014PhysRep} and references therein for further discussion of diffusion processes in multilayer networks. For example, RWs have been employed to estimate the number of layers in multilayer networks \cite{Lacasa2017arxiv}. The investigation of RWs in multilayer networks is a very active area of research.


\subsection{Temporal networks} \label{sub:temporal networks}

Many empirical networks vary over time, and one can describe them as temporal networks \cite{HolmeSaramaki2012PhysRep,Holme2015EurPhysJB}. CTRWs with non-exponential distributions of inter-event times (see Section~\ref{sub:CTRW nets}) are often discussed in the context of temporal networks, because non-Poissonian distributions of inter-event times are a fundamental property of most empirical temporal networks \cite{VazquezA2006PhysRevE-burst,HolmeSaramaki2012PhysRep}.

In this section, we discuss some situations in which a temporal network is given in the form of a sequence of static networks (which are called ``snapshots'' in \cite{MasudaLambiotte2016book})\footnote{There are also other types of temporal networks \cite{HolmeSaramaki2012PhysRep,Holme2015EurPhysJB}, 
and it is important to consider the time scales of both network evolution and the evolution of dynamical processes on a network to determine appropriate frameworks for network analysis
\cite{Porter2016book}.}. In this type of example, one time-independent network corresponds to a single observation (with a time stamp) of a temporal network, whose time resolution may correspond to that imposed by a recording period (e.g., every 20 secs). One can then consider an RW on a (temporal) sequence of adjacency matrices:
\begin{equation}
	\mathcal{A} = \{A(1), A(2), \ldots, A(n_{\max})\}\,,
\end{equation}
where $(A(n))_{ij}$ encodes the activation of edge ($v_i$, $v_j$) at discrete time $n$ (with $n \in \{1, \dots, n_{\max}\}$).
See the review \cite{Holme2015EurPhysJB} for a discussion of several models of RWs on temporal networks in addition to the ones that we will discuss in the following sections.


\subsubsection{Activity-driven model}

RWs on temporal networks have been examined both analytically and computationally. One useful approach is to examine RWs on an ``activity-driven model'' of temporal networks \cite{Perra2012PhysRevLett}.

The simplest type of activity-driven model generates a sequence of uncorrelated time-independent networks \cite{Perra2012SciRep}.
First, we associate each node $v_i$ (with $i \in \{1 \;, \ldots,\; N\}$) with a random variable $a_i$, called the ``activity potential'', drawn from a given distribution $F(a)$ (with $a\ge 0$).
Second, at each discretized time $t$, each node $v_i$ is independently active with probability $a_i\Delta t < 1$ and inactive with probability $1-a_i\Delta t$, where $\Delta t$ is the time difference (which we assume to be homogeneous) between two consecutive time points. Third, at each $t$, each activated node generates $m$ undirected edges that connect to $m$ other nodes uniformly at random. When nodes $v_i$ and $v_j$ are both active and each connects to the other with an edge at time $t$, we suppose that there is exactly one unweighted edge ($v_i$, $v_j$) at $t$. In practice, we suppose that $a_i\Delta t$ is sufficiently small to prevent such mutual edge creation to occur too often. We regard the network at each $t$ as an undirected and unweighted network, and we repeat this procedure independently to generate a time-independent network for the time interval $\Delta t$.

Consider the aggregation of a temporal network into a time-independent network, which we construct by summing the edge weights across some time window for each edge. The aggregated network neglects any temporal information contained in the temporal network during that window. If we aggregate observed time-independent networks over some time --- which cannot be too long, or else the aggregated network might be a complete weighted graph --- the aggregated (and sometimes called ``annealed'') adjacency matrix is given by
\begin{equation}
	A_{ij}^* \approx \frac{m\left(a_i+a_j\right)}{N}\,,
\label{eq:A_ij^* activity driven}
\end{equation}
where we neglect $o(1/N)$ terms.
The degree distribution of the aggregated network is
\begin{equation}
	p(k^*) \approx \frac{1}{m} F\left(\frac{k}{m}-\langle a \rangle\right)\,,
\end{equation}
where $\langle a\rangle = \int a F(a) {\rm d}a$ is the ensemble average of $a$. Therefore, a heterogeneous distribution $F(a)$ yields a comparably heterogeneous degree distribution in the aggregated network.

When we observe a temporal network with a fine temporal resolution, the network at each time point is very sparse\footnote{We use the term ``sparse'' to indicate the presence of an extremely small number of edges rather than in a conventional graph-theoretic sense, in which a sparse network still typically has a large number of edges (but with an edge density that scales sufficiently slowly as the number $N$ of nodes becomes large) \cite{Newman2010book,Bollobas2001book}.}. 
This also occurs for the above activity-driven model if $a_i\Delta t$ and $m$ are sufficiently small. A walker has to remain at a node if the node is isolated at the present time $t$, and this fact has a substantial effect on RW dynamics. In the above activity-driven model, there are two ways for a walker located at node $v_i$ to move to node $v_j$ in a network at time $t$ \cite{Perra2012PhysRevLett,Ribeiro2013SciRep}. The first way is to combine the following three independent events: (i) $v_i$ is activated with probability $a_i\Delta t$, (ii) node $v_i$ is connected to $v_j$ with probability $m/N$, and (iii) the edge ($v_i$, $v_j$) is traversed with probability $1/(m+m\langle a\rangle \Delta t)$. Note that the mean degree of $v_i$ in a time-independent network at an arbitrary time $t$ when $v_i$ is activated is equal to $m+m\langle a\rangle \Delta t$, because $v_i$ has $m\langle a\rangle \Delta t$ edges from the activation of other nodes. The second way is to combine the following four independent events: (i) node $v_i$ is not activated with probability $1-a_i\Delta t$, (ii) node $v_j$ is activated with probability $a_j\Delta t$, (iii) $v_j$ is connected to $v_i$ with probability $m/N$, and (iv) the edge ($v_i$, $v_j$) is traversed with probability $1/(1+m\langle a\rangle \Delta t)$. By adding these contributions and assuming that $\Delta t$ is small, we obtain a transition-probability matrix $T$ with elements
\begin{align}
	T_{ij} &\approx a_i\Delta t \frac{m}{N} \frac{1}{m+m\langle a\rangle \Delta t}
+ (1-a_i\Delta t) a_j\Delta t \frac{m}{N}\frac{1}{1+m\langle a\rangle \Delta t}\notag\\
	&\approx \frac{\Delta t}{N}\left(a_i+m a_j\right)\quad (j\neq i)\,.
\label{eq:T_ij activity driven}
\end{align}
Note that $T_{ii} = 1 - \sum_{j=1; j\neq i}^N T_{ij}$.

We aggregate all nodes with the same value of $a$ into one group, and we regard $a$ as continuous. Let $p_a(t)$ denote the probability that a single node with activity potential $a$ is visited at time $t$. The normalization is
$\int p_a(t) F(a) {\rm d}a = 1$, and the master equation in the $\Delta t\to 0$ limit is
\begin{equation}
	\frac{{\rm d}p_a(t)}{{\rm d}t} = \int a^{\prime} p_{a^{\prime}}(t) F(a^{\prime}){\rm d}a^{\prime}
-a p_a(t) + ma \frac{1}{N} - m\langle a\rangle p_a(t)\,.
\label{eq:master RW on activity driven}
\end{equation}
The first and second terms on the right-hand side of Eq.~\eqref{eq:master RW on activity driven} account, respectively, for the in-flows and out-flows of probability driven by $(\Delta t/N)a_i$ on the right-hand side of Eq.~\eqref{eq:T_ij activity driven}. The third and fourth terms account, respectively, for the in-flows and out-flows driven by $(\Delta t/N)m a_j$ in Eq.~\eqref{eq:T_ij activity driven}. This RW is a Poissonian node-centric CTRW 
whose general master equation is given by Eq.~\eqref{eq:master normalized CTRW}.

The stationary density of Eq.~\eqref{eq:master RW on activity driven} is
\begin{equation}
	p_a^* = \frac{\frac{ma}{N} + \phi}{a+m\langle a\rangle}\,,
\label{eq:stationary density activity driven}
\end{equation}
where
\begin{equation}
	\phi = \int a p_a^* F(a) {\rm d}a
\label{eq:def phi}
\end{equation}
is the mean probability flow from active nodes at equilibrium. By combining Eqs.~\eqref{eq:stationary density activity driven} and \eqref{eq:def phi}, we obtain the following self-consistency equation:
\begin{equation}
	\phi = \int a \frac{\frac{ma}{N} + \phi}{a+m\langle a\rangle} F(a) {\rm d}a\,.
\label{eq:phi solution}
\end{equation}
Because we are considering a Poissonian node-centric CTRW in an undirected network, the stationary density for a time-independent, aggregated network has components that are proportional to node degree. Equation~\eqref{eq:A_ij^* activity driven} implies that $p_a^*$ for the aggregated network is proportional to $m(a+\langle a\rangle)$. However, the stationary density for the CTRW on the activity-driven temporal network model, obtained by numerically solving Eq.~\eqref{eq:phi solution} for a given heterogeneous $F(a)$, is rather different from the time-independent case \cite{Perra2012PhysRevLett}. In particular, in the activity-driven model, $p_a^*$ saturates as the degree (or, equivalently, $a$) increases.

The MFPT is also different in the temporal and
aggregated
networks. At equilibrium, the probability that a walker moves to node $v_j$ in each discrete
step of time $\Delta t$ is $\xi_j = \sum_{i=1; i\neq j}^N p_i^* T_{ij}$. The probability that the walker arrives at $v_j$ for the first time after $n$ steps is thus given by $\xi_j(1-\xi_j)^{n-1}$ under the mean-field approximation in Eq.~\eqref{eq:m_ij meanfield}. One can then calculate that the MFPT for the above activity-driven model is
\begin{equation}\label{120}
	m_{ij} \approx \sum_{n=1}^{\infty} \Delta t n \xi_j(1-\xi_j)^{n-1}
= \frac{\Delta t}{\xi_j} = \frac{N}{m a_j + \sum_{\ell=1}^N a_{\ell} p_{\ell}^*}\,.
\end{equation}
This result is different from the aggregated (time-independent) network case, in which $m_{ij} \approx 1/p_j^*$ under the mean-field approximation in Eq.~\eqref{eq:m_ij meanfield}. A crucial difference between RW dynamics in the temporal and aggregated cases is that a walker in the activity-driven model can be trapped for some time in an isolated node $v_i$ and is temporarily unable to travel to a different node. At a later time, $v_i$ becomes connected to another node, and the walker can then move away from $v_i$. This phenomenon never happens in a time-independent (i.e., aggregated) network, as edges are always present.
These results were recently extended to RWs on an extended activity-driven model in which each node is assigned an attractiveness value in addition to an activity potential \cite{Alessandretti2017PhysRevE}.


One can also define RWs on empirical temporal networks. For example, given a sequence of time-independent networks, one can use each time-independent network to induce one time step of a DTRW \cite{Starnini2012PhysRevE}. (Another approach is to construct a multilayer representation of such a temporal network, and examine an RW on the resulting multilayer network \cite{Mucha2010Science,Kivela2014JCompNetw}.) In Ref.~\cite{Starnini2012PhysRevE}, the authors compared properties of RWs on empirical temporal networks to those on randomized temporal networks, which included ones in which the times of activating edge ($v_i$, $v_j$) are redistributed uniformly over time while keeping the weight of each edge in the aggregated network the same as that in the original temporal network. In comparison to such randomized temporal networks, the numerical computations in Ref.~\cite{Starnini2012PhysRevE} suggest that empirical temporal networks tend to slow down RW processes, as the MFPT is large and the coverage at a given time is small. See Refs.~\cite{Masuda2013PhysRevLett,Lambiotte2015JCompNetw,Scholtes2014NatComm,Delvenne2015NatComm,Sousadamata2015EurPhysJB,Denigris2016EurPhysJB} for discussions of the effects of temporal networks on the speed of diffusion on networks.

\begin{figure}[tb]
\centering
\includegraphics[scale=0.45]{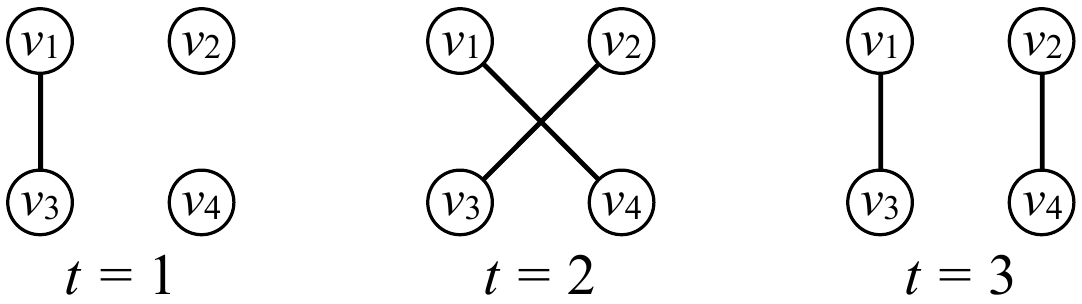}
\caption{A temporal network with three time points and $N=4$ nodes.}
\label{fig:deterministic walk}
\end{figure}

Note that if the time-independent network at each time point is sparse, the trajectory of a random walker may not be as random as the terminology RW might suggest.
For example, if the degree of $v_i$ equals $1$ at a certain time $t$, then the walker located at $v_i$ must move to its one neighbor. If $v_i$ is isolated at time $t$, then the walker does not move at $t$.
In the extreme case in which each node is adjacent to just one node or is isolated at all times, the trajectory of the ``random'' walk is deterministic. For example, in the temporal network on $N=4$ nodes in Fig.~\ref{fig:deterministic walk}, a walker starting from node $v_1$ always visits node $v_4$ after three time steps, so there is no randomness. In a CTRW, this situation always occurs in some sense: if $\psi(\tau)$ is a continuous distribution, then multiple events occur at the same time with probability $0$ because of the continuous-time nature of the stochastic dynamics. However, because the event times themselves are determined from a random process, we safely regard CTRWs as RWs. This situation is not shared by RWs on temporal networks when a network is given by a single realization of empirical or numerical data. Fortunately, there are at least two (imperfect) ways out of this conundrum. One solution is to aggregate a sequence of time-independent networks with a sufficiently large time window to make them sufficiently dense. Another solution is to allow walkers to wait at the current node with some probability even if an edge is available for it to move to another node.


\subsubsection{Memory networks\label{sec:memory net}}

By definition, a DTRW is a (stationary) Markov chain such that the transition probability does not depend on the past trajectory. Poissonian CTRWs and non-Poissonian active CTRWs (either node-centric or edge-centric) also share this property. However, many real temporal networks have correlations in edge activations \cite{HolmeSaramaki2012PhysRep,Holme2015EurPhysJB,MasudaLambiotte2016book}. Therefore, one does not expect a trajectory of RWs on an empirical temporal network to be a Markov chain, as certain trajectories are favored and others are discouraged or even forbidden. Such trajectories are poorly reproduced by the first-order Markov chains that we have considered thus far. In this situation, using higher-order Markov chains may be helpful \cite{Rosvall2014NatComm,Scholtes2014NatComm}, and it is also important to explore non-Markovian stochastic processes.

\begin{figure}[tb]
\begin{center}
\includegraphics[width=8.5cm]{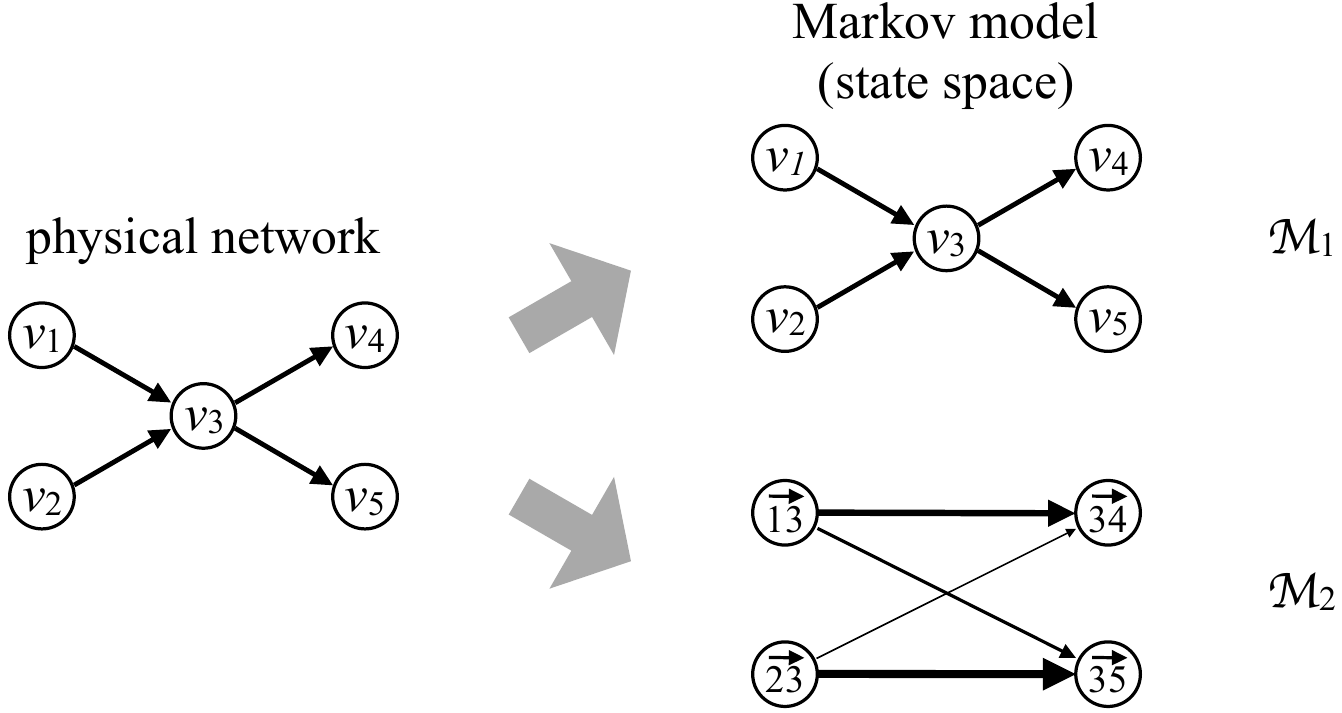}
\caption{Memory networks (of order 2). The network on the left shows a part of a directed network (a ``physical network'').
The width of each edge represents edge weight. In the present example, we assume for simplicity that the physical network is unweighted. In the first-order Markov chain $\mathcal{M}_1$, a state is a node of the physical network. In the second-order Markov chain $\mathcal{M}_2$ (of which we show a part), a state is a directed edge of the physical network. The state space is the directed line graph of the physical network. If the process that occurs on the physical network is Markovian, transitions in $\mathcal{M}_2$ are uniform in the following sense. Suppose, as indicated in the figure, that node $v_3$ has two in-edges and two out-edges in the physical network.
One then should be able to reach node $\protect\vv{34}$ with equal probability from nodes $\protect\vv{13}$ and $\protect\vv{23}$, yielding the same weight for edges $\protect\vv{13} \to \protect\vv{34}$ and $\protect\vv{23} \to \protect\vv{34}$. In the part of $\mathcal{M}_2$ (determined from, for example, a temporal network) that we show in this figure has edge weights that are different from the expectation of the first-order Markov chain $\mathcal{M}_1$. In other words, a move from node $v_3$ to node $v_4$ is more likely to occur when a walker arrives at $v_3$ from $v_1$ than from $v_2$. Therefore, the process represented by $\mathcal{M}_2$ network is not Markovian on the physical network.}
\label{fig:M2}
\end{center}
\end{figure}

To consider the above issue with empirical data in the context of temporal networks, we first map time series of edge activations in a sequence of time-independent networks to trajectories of walkers \cite{Rosvall2014NatComm}. We assume that a walker is located initially at a uniformly randomly selected node $v_i$. (The choice of initial condition can matter if RW trajectories simulated in the following are short.) A walker waits there until at least one edge is available for it to move. When at least one edge becomes available, the walker leaves the node with probability $1-q$ and does not move with probability $q$. As usual, the destination node $v_j$ is selected with probability $A_{ij}(t)/\sum_{\ell=1}^N A_{i\ell}(t)$. We repeat this procedure several times and thereby generate multiple trajectories starting at $n = 1$ and finishing at $n = n_{\max}$. When $q=0$, the walker always moves to a different node using the first available edge \cite{Starnini2012PhysRevE,Delvenne2015NatComm}. When $q \in (0,1)$, some randomness is introduced into the trajectories \cite{RochaMasuda2014NewJPhys}, preventing spurious effects such as a strong tendency for backtracking \cite{Saramaki2015EurPhysJB}.
However, for sufficiently large $q$, the effect of temporal correlations between edges at short time scales
becomes unimportant, which may dilute the impact of the temporality of the data.
If trajectories are statistically independent of the past locations of a walker, it is sufficient to use a first-order Markov chain. In this case, the transition-probability matrix $T=(T_{ij})$ constructed from an aggregated network, in which the weight of edge ($v_i$, $v_j$) is equal to the sum of $(A(t))_{ij}$ over time, is sufficient for describing the RWs. We denote a first-order Markov chain on an aggregated network by $\mathcal{M}_1$. See the top right panel of Fig.~\ref{fig:M2}.

In general, the probability that a random walker visits node $v_i$ after the ($n+1$)th step depends on the entire history of a stochastic process. To partially take into account temporal correlations between edge activations, one can use a second-order Markov chain. We define a process, which we denote by $\mathcal{M}_2$, using an expanded transition probability tensor, whose element $T_{i^{\prime}ij}$ represents the probability that a walker moves from node $v_i$ to node $v_j$ given that the previous position is node $v_{i^{\prime}}$. Another representation of the process $\mathcal M_2$ is to use a memoryless RW (i.e., a first-order Markov chain) between directed edges of the original network. In this representation, the probability that directed edge $\vv{v_i v_j}$  is visited depends on $\vv{v_{i^{\prime}} v_i}$ rather than only on node $v_i$, as in the first-order Markov chain $\mathcal{M}_1$. For simplicity, for the rest of the present discussion, we use the shorthand notation $\vv{ij}$ for a directed edge $\vv{v_i v_j}$.
For this representation, we regard the state space (i.e., the set of directed edges) as the nodes of a new network, which we call the ``$\mathcal{M}_2$ network'' or ``(second-order) memory network''. One construes the original network as a ``physical network'', and the state space of  $\mathcal M_2$ is the so-called ``directed line graph'' of the original network \cite{Harary1960Rendi}. The memory network has $2M$ nodes whether the original network is directed or undirected. We sometimes use the term ``memory nodes'' for the nodes of a memory network. Even for undirected networks, we must assign two memory nodes $\vv{ij}$ and $\vv{ji}$ to each pair of adjacent nodes $v_i$ and $v_j$ in the original network, because a memory node encodes the time ordering of visits. The number of edges in a memory network is proportional to $\langle k^2\rangle N$ \cite{EvansLambiotte2009PhysRevE}.

To improve accuracy, one can also examine memory networks in the form of higher-order Markov chains. For example, in a third-order Markov chain, the transition probability depends on the currently visited node $v_i$ and two previously visited nodes $v_j$ and $v_{\ell}$. A memory node is then specified by $\vv{v_{\ell} v_j v_i}$. However, going beyond second-order Markov chains is not always practical. First, a second-order memory network is conceptually simpler than higher-order counterparts, as the memory nodes are given by edges of the original network rather than by higher-order structures. Second, one may only obtain marginal gains by considering higher orders \cite{Rosvall2014NatComm}.
Third, higher-order memory networks require a lot of data, because the number of memory nodes and transition probabilities to be estimated increases exponentially with the order of the Markov chain.

One encodes the dynamics of a second-order Markov chain by a transition-probability matrix on the network with $2M$ nodes whose elements are given by $p(\vv{ij} \rightarrow \vv{jk})$ (see Fig.~\ref{fig:M2}). In practice, one estimates $p(\vv{ij} \rightarrow \vv{jk})$ with
\begin{equation}
	p(\vv{ij} \rightarrow \vv{jk}) =
\frac{(\text{number of transitions }\vv{ij} \to \vv{jk})}
{\sum_{\ell=1}^N (\text{number of transitions }\vv{ij} \to \vv{j\ell})}\,,
\label{eq:transition probability M2}
\end{equation}
where one counts the number of transitions in the RW trajectories generated by the sequence of time-independent networks.
One interprets the transitions as movements between directed edges. The normalization is given by $\sum_{\ell=1}^N p(\vv{ij} \rightarrow \vv{j\ell}) = 1$.
In situations in which one can measure RW trajectories in empirical data, they can be used directly to estimate Eq.~\eqref{eq:transition probability M2} \cite{Rosvall2014NatComm}.

In a first-order Markov chain $\mathcal{M}_1$ (i.e., a DTRW) on an unweighted network, we obtain
\begin{equation}
\label{mar}
	p(\vv{ij} \to \vv{j\ell})=
\begin{cases}
1/k_j      & (v_{\ell} \text{ is a neighbor of } v_j)\,,\\
0      & (\text{otherwise})\,.
	\end{cases}
\end{equation}
In general second-order Markov chains, the probability that a walker visits node $\vv{j\ell}$ after $n+1$ steps is given by
\begin{equation}
\label{general1}
	p(\vv{j\ell}; n+1) = \sum_{i=1}^{N} p(\vv{ij}; n) p(\vv{ij} \rightarrow \vv{j\ell})\,.
\end{equation}

Edge-centric passive CTRWs with a non-exponential distribution $\psi(\tau)$ of inter-event times are one example of a situation that is appropriate to model using a second-order Markov chain rather than a first-order chain. Equation~\eqref{eq:approx T RW STN} implies that $p(\vv{\ell i}\rightarrow \vv{ij})$ depends on whether $j=\ell$ or $j\neq \ell$. In particular, if $\psi(\tau)$ is a heavy-tailed distribution, then $p(\vv{\ell i}\rightarrow \vv{i\ell})$ (i.e., the probability to backtrack) is larger than is expected in a first-order Markov chain. All other $p(\vv{\ell i}\rightarrow \vv{ij})$ ($j\neq \ell$) values are the same. In contrast, if $\psi(\tau)$ is a lighter-tailed distribution than an exponential distribution, $p(\vv{\ell i}\rightarrow \vv{i\ell})$ is smaller than expected in a first-order Markov chain, and random walkers tend to avoid backtracking. The extreme case of the latter situation is a non-backtracking RW \cite{Rosvall2014NatComm,Scholtes2014NatComm,Salnikov2016SciRep,Xu2016SciAdv}.
 In such an RW, a walker performs an RW, except that it is not allowed to backtrack \cite{Alon2007CommContempMath,Fitzner2013JStatPhys}, so $p(\vv{ij} \rightarrow \vv{ji}) = 0$ and $p(\vv{ij} \rightarrow \vv{j\ell}) = 1/(s_j^{\rm out}-A_{ji})$ (with $\ell\neq i$).

A network's associated non-backtracking matrix, which is a $2M\times 2M$ adjacency matrix for the $\mathcal{M}_2$ network, has been used recently in several applications, including percolation \cite{Karrer2014PhysRevLett,Hamilton2014PhysRevLett}, network centralities \cite{martin2014PRE}, community detection \cite{Krzakala2013PNAS,Newman2013arxiv-backtracking,Bordevane2015Ieeeconf}, and efficient ``immunization'' algorithms \cite{Morone2015Nature}.
More generally, we also note that non-backtracking matrices help with ``message passing'' and ``belief propagation'' approaches to network analysis.

To quantify the difference between a first-order Markov chain $\mathcal{M}_1$ and a second-order Markov chain $\mathcal{M}_2$, we compare their entropy rates. ``Entropy rate'' quantifies the uncertainty of the next state given the current state, weighted by the stationary density. For $\mathcal{M}_1$, the entropy rate is
\begin{equation}
	H_1 = -\sum_{i,j=1}^N p_i^* T_{ij} \log{T_{ij}}\,.
\end{equation}
In $\mathcal{M}_2$, one calculates the entropy rate for a first-order Markov chain on the memory network and thereby obtains
\begin{equation}
	H_2 = -\sum_{i, j, \ell=1}^N p_{\vv{ij}}^* p(\vv{ij} \to \vv{j\ell}) \log p(\vv{ij} \to \vv{j\ell})\,,
\end{equation}
where $p_{\vv{ij}}^*$ is the stationary density at node $\vv{ij}$ in the memory network.
In many empirical temporal networks, $H_2$ is considerably smaller than $H_1$, implying that one cannot neglect memory effects \cite{Rosvall2014NatComm,Scholtes2014NatComm} (also see \cite{Takaguchi2011PhysRevX,Pfitzner2013PhysRevLett} for similar measurements). The first-order Markov chain $\mathcal{M}_1$ tends to overestimate the number of available neighbors around the current node of a random walker compared to its higher-order counterparts.

The observation that $H_2 < H_1$ can influence RW dynamics, other dynamical processes on networks, and how one wants to calculate certain structural features of networks. For example, communities of networks found by second-order Markov chains (see Section~\ref{sub:modularity}) tend to contain edges that are activated at the same time \cite{Salnikov2016SciRep}. Such communities are undetectable using first-order models (such as the usual RWs). Memory also affects the relaxation time of an RW or other Markov processes towards a stationary state \cite{Lambiotte2015JCompNetw}.

\begin{figure}[tb]
\begin{center}
\includegraphics[width=4cm]{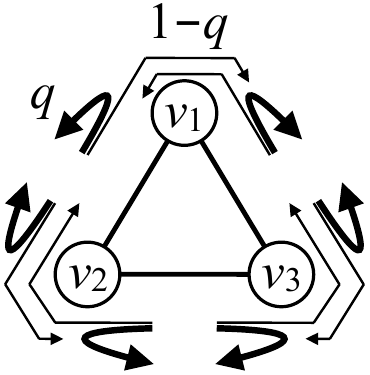}
\caption{A second-order Markov chain on a 3-clique. The widths of the arrows represent (schematically) the transition probabilities in a second-order Markov chain. For example, a walker that has moved from node $v_2$ to node $v_1$ moves back to $v_2$ with probability $q$ and moves to $v_3$ with probability $1-q$ in the next move. Because $q>1/2$ in this figure, random walkers tend to backtrack.}
\label{fig:chain N=3 memory net}
\end{center}
\end{figure}

The eigenvalue $\lambda_2$ of $T$ with the second-largest absolute value influences network community structure and determines the relaxation time of RWs \cite{Delvenne2015NatComm}. (See Section \ref{sub:community} for more discussions of community structure.) Temporal correlations can either increase or decrease $\lambda_2$, depending on how temporal correlations are introduced \cite{Lambiotte2015JCompNetw}. If memory increases $\left|\lambda_2\right|$, a random walker in a second-order Markov process tends to be confined in a certain part of the original network (i.e., the ${\mathcal M_1}$ network) than is suggested by network structure alone. In the corresponding $\mathcal M_2$ network, a random walker tends to be trapped in a community. In this case, memory has slowed down relaxation to a steady state. However, if memory decreases $\left|\lambda_2\right|$, a walker moves from one community to another faster than is suggested by the original network. In this case, memory accelerates relaxation to a steady state. Moreover, non-Markovian pathways in a network without community structure can still create community structure in the associated $\mathcal M_2$ network \cite{MasudaLambiotte2016book}. As a simple example (see Fig.~\ref{fig:chain N=3 memory net}), consider an undirected 3-clique (i.e., a triangle).

The transition-probability matrix of the usual DTRW (i.e., the $\mathcal M_1$ process) is
\begin{equation}
	T=\begin{pmatrix}
0 & \frac{1}{2} & \frac{1}{2}\\
\frac{1}{2} & 0 & \frac{1}{2}\\
\frac{1}{2} & \frac{1}{2} & 0
	\end{pmatrix}\,,
\end{equation}
which yields $\lambda_2 = -1/2$. On the triangle network, consider the second-order Markov chain process defined by
\begin{align}
	& p(\vv{12} \to \vv{21}) = p(\vv{21}\to \vv{12}) = p(\vv{13}\to \vv{31}) = p(\vv{31}\to \vv{13}) = p(\vv{23}\to \vv{32}) = p(\vv{32}\to \vv{23}) = q\,,\\
	& p(\vv{12}\to \vv{23}) = p(\vv{21}\to \vv{13}) = p(\vv{13}\to \vv{32}) = p(\vv{31}\to \vv{12}) = p(\vv{23}\to \vv{31}) = p(\vv{32}\to \vv{21}) = 1-q\,,
\end{align}
where $q \in [1/2,1)$ (see Fig.~\ref{fig:chain N=3 memory net}).
This RW backtracks the edge traversed in the previous step with probability $q$. If we order the nodes in the ${\mathcal M_2}$ network as $\vv{12}$, $\vv{21}$, $\vv{13}$, $\vv{31}$, $\vv{23}$, and $\vv{32}$, the transition-probability matrix is
\begin{equation}
	T = \begin{pmatrix}
0 & q & 0 & 0 & 1-q & 0\\
q & 0 & 1-q & 0 & 0 & 0\\
0 & 0 & 0 & q & 0 & 1-q\\
1-q & 0 & q & 0 & 0 & 0\\
0 & 0 & 0 & 1-q & 0 & q\\
0 & 1-q & 0 & 0 & q & 0
	\end{pmatrix}\,.
\end{equation}
The eigenvalues of $T$ are $1$, $1-2q$, and $\left[-1+q\pm \sqrt{(1-q)^2+4(2q-1)}\right]\big/2$. The last eigenvalues (for each of $\pm$) have multiplicity two. The relaxation time is governed by $\lambda_2=\left[-1+q- \sqrt{(1-q)^2+4(2q-1)}\right]\big/2 < 0$. When $q=1/2$, we obtain $\lambda_2=-1/2$, which is consistent with the memoryless case. When $q>1/2$, we see that $\lambda_2$ decreases monotonically towards $-1$, which one obtains in the limit $q\to 1$. A large value of $q$ makes $\left|\lambda_2\right|$ large and hence makes the spectral gap small, so a random walker tends to spend a long time in a community in the ${\mathcal M_2}$ network. In this situation, each of the three edges constitutes a community, and it is difficult for the walker to leave any edge.

Storing the stationary density of a second-order Markov chain (i.e., $p_{\vv{ij}}^*$) may be prohibitive, particularly for a network that is not sparse, because the $\mathcal{M}_2$ network has $2M$ nodes. A space-friendly alternative is to introduce an approximation $p_{\vv{ij}}^* \approx \hat{p}_i^* \hat{p}_j^*$ (with $i, j \in \{1, \ldots, N\}$) and estimate $\hat{p}_i^*$ \cite{LiNg2014LinMultilinAlgebra}. The estimated $\hat{p}_i^*$ is the stationary density of a modified second-order-like Markov chain called a ``spacey RW'' \cite{Benson2017SiamRev}. In a spacey RW, a walker visiting node $v_j$ forgets the last node $v_i$ that it has visited. The walker then draws the fictive last position $v_i$ uniformly at random from the list of the nodes visited in the past. (The probability that each node is selected is weighted by the number of past visits to the node.) The walker then moves to $v_{\ell}$ according to the probability $p(\vv{ij} \to \vv{j\ell})$. Spacey RWs are a type of ``reinforced RW'', in which nodes or edges (nodes in the present case) visited frequently in the past are also visited more frequently in subsequent steps \cite{Pemantle2007ProbSurv}. Spacey RWs have such a richer-get-richer mechanism embedded in the process to select the fictive last position $v_i$.

The formalism in this section allows one to examine how temporal 
correlations in a network affect spreading processes \cite{Salnikov2016SciRep,Pfitzner2013PhysRevLett}. It can 
also be used to directly exploit knowledge of the trajectories of 
diffusing entities (so-called ``trajectory data'') when they can be observed 
and collected. For instance, the trajectory of a traveler between 
different airports is rather different from a first-order Markov process, 
so it is important to consider higher-order Markov processes or even 
non-Markovian dynamics \cite{Rosvall2014NatComm}. Similar conclusions arise when studying 
animal movements \cite{Kareiva1983Oeco}, Website traffic \cite{Chierichetti2012WWW}, and other applications.

Although trajectory data are becoming increasingly available, it is difficult to measure trajectories for the vast majority of systems. Moreover, even when they can be measured, a high-order Markov model or non-Markovian model may be unnecessarily complicated to extract the most salient features of a system. Consequently, researchers have proposed 
simple models of second-order Markov dynamics based on the distinction between different types of transitions on networks. In practice, one can calibrate the model parameters in systems in which trajectories can be measured and then use these models to simulate trajectories in similar systems for which data on trajectories are not available.

In \cite{Rosvall2014NatComm}, Rosvall et al. enumerated three different types of transitions:
\begin{enumerate}
     \item A \textit{return step}, in which a walker coming from $\vv{ij}$ jumps to $\vv{ji}$. In other words: a walker coming from node $i$ to $j$ returns to node $i$.
      \item A \textit{triangular step}, in which a walker coming from $\vv{ij}$ moves to edge $\vv{j\ell}$, where $\ell\neq i$ is a neighbor of node $i$. 
    \item An \textit{exploratory step}, in which a walker moves from 
$\vv{ij}$ to an edge $\vv{j\ell^{\prime}}$ whose end point $\ell^{\prime}$ is neither node $i$ nor 
any of $i$'s neighbors.
\end{enumerate}
To each of above types of transition, one then assigns a positive weight (denoted $r_2$, $r_3$, and $r_{>3}$, respectively) to account for their relative contributions. One can recover several existing types of processes for specific choices of parameters. For example, $r_2=r_3=r_{>3}$ yields a first-order Markov process and $r_2=0$, $r_3=r_{>3}>0$ yields a non-backtracking RW.

\section{Applications} \label{sec:applications}

\subsection{Search on networks}

People are often interested in finding a resource, service, or piece of information that is available only at some nodes in a network \cite{Newman2010book}. If network structure is completely known to a user or a designer, a shortest path from the initially visited node to a destination node provides the most efficient way of searching, although it may be sensible to plan a detour if one expects congestion from traffic somewhere along a shortest path.

If a searcher has partial information about his/her destination (e.g., the geographic distance to it), one can of course use such information to inform search paths \cite{Kleinberg2006ProcICMath}. In contrast, if one does not have any information about network structure or has only local information (such as the degrees of neighbors), RWs provide a viable approach for searching in networks. One context in which this idea has been investigated and implemented are decentralized peer-to-peer networks \cite{Lv2002Ics,Bisnik2005Hotp2p}. A node that sends a query emits $N_{\rm rw}$ packets to neighbors selected uniformly at random. Each packet behaves as a random walker, which travels until it finds the item or reaches a prescribed lifetime $n_{\max}$, which is the maximum number of steps it is allowed to take before it is removed from the network.
Search overhead is determined by $N_{\rm rw} n_{\max}$, which is a measure of the number of walkers, averaged over time, that are wandering in a network. One expects larger $N_{\rm rw} n_{\max}$ to yield better search efficiency (i.e., a higher probability that an item is found). Therefore, there is a trade-off between search overhead and search efficiency.
RW search methods are comparable with flooding search methods in various networks and scenarios \cite{Lv2002Ics}.
In a flooding method, first used by Gnutella, a node with a query asks all of its neighbors, each of which in turn asks all of its unvisited neighbors, and so on \cite{Tanenbaum2011book}.

Most empirical networks are highly heterogeneous in node degree \cite{Newman2010book}.
If a node that is making or passing on a query knows the degrees of its neighbors, one can enhance search efficiency by sending the query to high-degree neighbors \cite{Adamic2001PhysRevE}. The main limitation of such an approach is that most queries are forwarded to hubs, potentially causing overloading at such nodes (depending on their capacity).


\subsection{Ranking} \label{sub:ranking}

In the study of networks, one often seeks to rank nodes, edges, or other structures based on their relative importances (i.e., ``centralities''). There are myriad ways to measure centralities in networks, especially for ranking nodes \cite{Newman2010book,vital2016}, and new ones are published at a very rapid pace. Many methods for computing node centralities are based on eigenvectors of matrices and are derived from various types of RWs or other walks. These include ``Katz centrality'' \cite{Katz1953Psychometrica} and related measures (such as ``communicability'') \cite{Est12comm}, ``eigenvector centrality'' \cite{bonacich1972}, ``PageRank'' \cite{Gleich2015SiamRev}, ``hubs'' and ``authorities'' \cite{kleinberg1999}, ``non-backtracking centrality'' \cite{martin2014PRE}, and many others. By considering RWs on multilayer and temporal networks, one can also generalize such notions of centrality \cite{Grindrod2011PhysRevE,Halu2013PlosOne,Dedomenico2014PNAS,RochaMasuda2014NewJPhys,Dedomenico2015NatComm-ranking,Soleribalta2016PhysicaD,Taylor2017MultiscaleModelSimul,Liao2017arxiv}.



\subsubsection{PageRank} \label{sub:PageRank}

The most famous centrality measure is probably ``PageRank'', which was introduced originally for ranking web pages. In this context, it was introduced by Brin and Page \cite{Brin1998conf} (see also \cite{pagerank}), although an equivalent formulation had already existed for two decades \cite{Pinski1976InfoProcMan}. (Brin and Page's discovery was independent of Ref.~\cite{Pinski1976InfoProcMan}.)

PageRank is discussed thoroughly in many review papers and monographs \cite{Langville2004InternetMath,Berkhin2005InternetMath,Langville2006book,Kamvar2010book,Gleich2015SiamRev,Ermann2016RevModPhys}, and it has been used (and generalized) for numerous applications --- including ranking of academic journals and papers, professional sports, disease-gene identification, discovery of correlated genes and proteins, systemic risk in financial networks, anomaly detection in distributed engineered systems, ordering of the most important functions in Linux, prediction of traffic flow and human movement, recommendation systems in online marketplaces, image search engines, identifying community structure in networks, and much more \cite{Gleich2015SiamRev}.
We indicate a few fascinating applications in passing. For example, seven new genes that predict the survival of patients in a type of pancreatic cancer were identified using PageRank \cite{Winter2012PlosComputBiol}.
PageRank has also been used to rank professional tennis players
\cite{Radicchi2011PlosOne}, and PageRank and other RW-based ranking methods have been used for ranking teams in U.S. college football
\cite{Callaghan2007AmMathMonthly,Park2005JStatMech} and ranking players in Major League Baseball \cite{Saavedra2010PhysicaA}.
PageRank and other eigenvector-based centrality measures have also been used to rank universities \cite{Lages2016EurPhysJB}, mathematics research programs \cite{myers2011,Taylor2017MultiscaleModelSimul}, baby names \cite{baby},
and many other things.

The PageRank vector is defined as the stationary density of a DTRW on a network that is a modification of an original network to guarantee that the stationary density always exists. For the original network, the temporal evolution of the probability $\bm p(n)$ that node $v_i$ (with $i \in \{1, \dots, N\}$) is visited at time $n$ is governed by Eq.~\eqref{eq:Markov chain transition} (or, equivalently, by Eq.~\eqref{eq:Markov chain transition vector}). The essential idea of PageRank is to use the stationary density in Eq.~\eqref{eq:Markov chain stationary density} as a centrality measure. Equation~\eqref{eq:Markov chain stationary density} implies that node $v_i$ is central if many edges enter node $v_i$ (i.e., it has a large in-degree), the source node of the edge that enters $v_i$ is a central node, and the source node $v_j$ of the edge that enters $v_i$ has a small out-degree. The last condition ensures that the total centrality of $v_j$ is shared among its out-neighbors. This recursive relationship (i.e., a node is central if it is adjacent to central nodes) leads to an eigenvalue problem. Other centrality measures --- including eigenvector centrality, Katz centrality, the hyperlink-induced topic search algorithm (which uses ``hubs'' and ``authorities''), and many others --- are based on the same basic idea \cite{Newman2010book}. In PageRank, the eigenvalue problem corresponds specifically to the stationary density of a DTRW.


In an empirical directed network, one cannot typically use a transition-probability matrix $T$ without modification to measure centralities, because such networks are not usually strongly connected. Consequently, there are transient nodes with stationary density equal to $0$, and the stationary density need not be unique, as it depends on the initial condition of an RW when there are multiple absorbing states. To overcome these problems, we allow walkers to ``teleport'' (e.g., uniformly at random) to other nodes to construct an effective network that is strongly connected. The master equation for the altered RW is
\begin{equation}
	p_i(t+1) = \alpha \sum_{j=1}^N p_j(t) T_{ji} + (1-\alpha) u_i\,,
\label{eq:preference}
\end{equation}
where the ``preference vector'' $(u_1,\; \ldots\; ,u_N)$, which satisfies the constraint $\sum_{i=1}^N u_i=1$, determines the conditional probability that a walker teleports to node $v_i$ when it teleports. At any node with at least one out-edge, a walker teleports with probability $1-\alpha$. To prevent the transition probability in Eq.~\eqref{eq:T_ij random walk} from being ill-defined, it is standard to ensure that a walker teleports with probability $1$ (rather than with probability $1-\alpha$) when it visits a so-called ``dangling node'' (which have no out-edges, so $s_i^{\rm out} = 0$ for a dangling node $v_i$). Mathematically, we set $T_{ij} = u_j$ (with $j \in \{1,\dots, N\}$) for any dangling node $v_i$. For web browsing, one interprets teleportation as a move to a new web page without following a hyperlink on the web page that is currently being visited. If $u_i > 0$ (with $i \in \{1,\dots, N\}$), any $\alpha \in (0,1)$ renders the altered RW ergodic, and Eq.~\eqref{eq:preference} thus converges to a unique stationary density. The PageRank vector is the stationary state of Eq.~\eqref{eq:preference}, and it is equal to the normalized eigenvector corresponding to the largest positive eigenvalue of the matrix $T^{\prime}$ with elements $T^{\prime}_{ij} = \alpha T_{ij} + (1-\alpha)u_j$.

Power iteration of $T^{\prime}$ converges rapidly if the spectral gap of $T^{\prime}$ is large (or, equivalently, if the second-largest eigenvalue of $T^{\prime}$ has small magnitude). The second-largest (in magnitude) eigenvalue of $T^{\prime}$ is equal to $\alpha \lambda_2$, where $\lambda_2$ is the second-largest (in magnitude) eigenvalue of $T$ \cite{Langville2004InternetMath}. Therefore, power iteration converges towards the PageRank vector at a rate that is proportional to $1/\alpha$ \cite{Gleich2015SiamRev}. However, a small value of $\alpha$, which corresponds to a large teleportation probability, dilutes the effect of the original network structure (which is encoded in the transition-probability matrix $T$). A rule of thumb is to set $\alpha$ near $1$ to suppress the effect of teleportation, but to also make sure that it is not too close. A popular choice is to let $\alpha=0.85$ and use a preference vector of $u_i=1/N$ (with $i \in \{1,\dots, N\}$) so that one teleports to nodes uniformly at random. An alternative choice is a ``personalized PageRank'' \cite{Brin1998BullIeee,Haveliwala2002ProcWWW,Jeh2003ProcWWW,Langville2004InternetMath,Berkhin2005InternetMath,Langville2006book,Kamvar2010book,Gleich2015SiamRev,XieBindel2015SIGKDD},
in which the preference vector is localized around one node or a small number of nodes (which can be helpful for applications to community detection \cite{Jeub2015PhysRevE}). One can also examine other teleportation strategies \cite{LambiotteRosvall2012PhysRevE}.

The stationary density of Eq.~\eqref{eq:preference} has components
\begin{equation} \label{formal}
	p_{i;\alpha}^* = (1-\alpha) \sum_{j=1}^N u_j \left[(I-\alpha T)^{-1}\right]_{ji}\,,
\end{equation}
and we note that we explicitly include the dependence on $\alpha$ in our notation. The Taylor expansion of Eq.~\eqref{formal} yields \cite{Boldi2005WWW,Brinkmeier2006}
\begin{equation}
\label{eq:formalB}
	p_{i;\alpha}^* \approx u_i + \sum_{\ell=1}^{\infty} \alpha^{\ell} \sum_{j=1}^N u_j \left( T^{\ell}_{ji} - T^{\ell-1}_{ji} \right)\,.
\end{equation}
Equation~\eqref{eq:formalB} includes terms for walks of all lengths $\ell$, and it thereby reveals the non-local nature of PageRank. When the value of $\alpha$ is large, a lot of credit is given to long walks. (See Ref.~\cite{Est12comm} for similar discussions in the context of centrality measures such as communicability.)
In fact, the stationary density can change drastically as a function of $\alpha$ \cite{Langville2004InternetMath}. Let's set $u_i=1/N$ (with $i \in \{1,\dots, N\}$) and rewrite Eq.~\eqref{eq:formalB} as
\begin{align}
\label{eq:formalK2}
	p_{i;\alpha}^* = \frac{1}{N} + \sum_{\ell=1}^{\infty} \frac{\alpha^{\ell}}{N}  \sum_{j, j^{\prime}=1}^N \left( \frac{s_{j^{\prime}}^{\rm in}- s_j^{\rm out} }{s_{j^{\prime}}^{\rm in} } \right) T_{jj^{\prime}} T^{\ell-1}_{j^{\prime} i}\,.
\end{align}
The leading contribution for small $\alpha$ makes the PageRank vector uniform across all nodes. Heterogeneity arises as $\alpha$ increases. Equation~\eqref{eq:formalK2} indicates that the contribution of each length-$\ell$ walk is proportional to $s_{j^{\prime}}^{\rm in}- s_j^{\rm out}$.
Each term on the right-hand side of Eq.~\eqref{eq:formalK2} vanishes when a network is regular in the weighted sense (i.e., when $s_i^{\rm in} = s_i^{\rm out} = s$, where $i \in \{1, \dots, N\}$). This yields $p_{i;\alpha}^*=1/N$ for any value of $\alpha$.

A strategy to minimize the dependence of the PageRank vector on $\alpha$ is to carefully choose the preference vector. One choice is $u_i = s_i^{\rm in}/\sum_{\ell=1}^N s_{\ell}^{\rm in}$ \cite{LambiotteRosvall2012PhysRevE}, inspired by the observation that the in-strength of a node is often correlated positively with $p_i^*$ for a DTRW on the original network (see Section~\ref{sub:stationary density DTRW}). With this choice of $u_i$, one uniformly randomly selects an edge rather than a node. One then teleports, uniformly at random, to one of the two end points of the selected edge. Substituting this preference vector into Eq.~\eqref{eq:formalB} yields
\begin{equation}\label{eq:formalK}
	p_{i;\alpha}^* = \frac{s_i^{\rm in}}{\sum_{\ell=1}^N s_{\ell}^{\rm in}} + \sum_{\ell=1}^{\infty} \frac{\alpha^{\ell}}{\sum_{\ell=1}^N s_{\ell}^{\rm in}} \sum_{j=1}^N \left( s_j^{\rm in}- s_j^{\rm out} \right)T_{ji}^{\ell}\,,
\end{equation}
which differs from Eq.~\eqref{eq:formalK2} in several respects. As $\alpha\to 0$, the components of the PageRank vector in Eq.~\eqref{eq:formalK} are given by the in-strength of the nodes. (The simplest --- and a rather popular --- measure of centrality in networks is simply to calculate node degrees and/or node strengths.) The $\ell$th-order contribution consists of a weighted mean of the walks of length $\ell$. One expresses their contribution to the PageRank vector in terms of the source node of a walk (i.e., $v_j$) in Eq.~\eqref{eq:formalK}. This contrasts with Eq.~\eqref{eq:formalK2}, where one instead expresses the contribution in terms of edges $(v_j, v_{j^{\prime}})$. A node $v_j$ that is the source of more probability flow than it receives as a destination
(i.e., $s_j^{\rm in} > s_j^{\rm out}$) makes a positive contribution to the PageRank vector, and a node $v_j$ with $s_j^{\rm in} < s_j^{\rm out}$ makes a negative contribution.
Equation~\eqref{eq:formalK} is independent of $\alpha$ when a network is balanced. (Recall from Section~\ref{sub:stationary density DTRW} that a directed network is balanced when $s_i^{\rm in}=s_i^{\rm out}$ for each $i$.)
In a balanced network, Eq.~\eqref{eq:formalK} reduces to $p_i^* = s_i^{\rm in}/\sum_{\ell=1}^N s_{\ell}^{\rm in}$.

Chung proposed a variant of PageRank called ``heat-kernel PageRank''
(which is defined for strongly connected networks) \cite{Chung2007PNAS,Chung2009InternetMath}. It is the probability density of a Poissonian node-centric CTRW at time $t$, where $t$ is the only parameter and it plays the role of $\alpha$ from the original PageRank. One uses a preference vector as an initial condition. Heat-kernel PageRank tends to the stationary density of a DTRW as $t\to\infty$. (For undirected networks, the components of the limiting stationarity density are thus proportional to the node strengths.)

We also note that various versions of PageRank and similar RW-based centralities for multilayer networks have been proposed \cite{ZhouOrshanskiy2007IeeIcdm,NgLiYe2011Kdd,Halu2013PlosOne,Dedomenico2015NatComm-ranking,Pedroche2016Chaos,Soleribalta2016PhysicaD}.


\subsubsection{Laplacian centrality}

PageRank is essentially the stationary density of a DTRW. The stationary density of the Poissonian edge-centric CTRW has also been employed as a centrality measure for directed networks (and, in fact, it has a longer history than PageRank \cite{Daniels1969Biom,Moon1970SiamRev,Berman1980SiamRev,Biggs1997BullLondMathSoc}). For strongly connected networks, such a ``Laplacian centrality'' is defined by the left eigenvector corresponding to the $0$ eigenvalue of the (combinatorial) Laplacian $L$. That is, it is given by $\bm p^*$ in Eq.~\eqref{eq:p^* CTRW unnormalized}.
This Laplacian centrality has been used, for example, to rank football teams \cite{Borm2002AnnOperatRes}, baseball players \cite{Saavedra2010PhysicaA}, and neurons \cite{MasudaKawamuraKori2009NewJPhys}. It has also been used in population ecology as a ``reproductive value'' \cite{Taylor1990AmNat,Taylor1996JMathBiol}.


\subsubsection{TempoRank}

One can extend the DTRW to temporal networks by using sequences $\{A(1), A(2), \ldots\}$ of adjacency matrices (see Section~\ref{sub:temporal networks}). Therefore, one can also extend PageRank to temporal networks. One such generalization is called ``TempoRank''  \cite{RochaMasuda2014NewJPhys}, and Katz centrality \cite{Grindrod2011PhysRevE,Grindrod_Higham_2013} and all eigenvector-based centralities \cite{Taylor2017MultiscaleModelSimul} have been generalized to such temporal networks.

In this section, we discuss TempoRank. We consider an undirected temporal network whose edge weights at each discrete time have (nonnegative) integer values. The latter assumption corresponds to a situation in which an event is an unweighted edge and each node pair can experience multiple events during the time window corresponding to a given matrix in the sequence. One can also image a sequence of networks, in which one has a time-independent view (or approximation) of a temporal network at a given instant in time.
This weighting assumes that a random walker at node $v_i$ that moves at discrete time $n$ selects each available edge (i.e., event) with the same probability and then traverse the chosen edge. Because we consider DTRWs, the walker moves at most once per time step. To avoid using a multilayer-network formalism, we also assume that there are no inter-layer edges between different matrices in the sequence.

To make the walk random even when just a single edge is available to a walker in a time period, we assume that, in each time period, a walker resists moving from node $v_i$ with probability $q$ per unit weight of an edge connected to $v_i$. For example, if $v_i$ is adjacent to a node with two events (i.e., edge weight equal to two) and to another node with three events at discrete time $n$, a walker visiting $v_i$ stays at the same node with probability $q^5$ at time $n$. A large $q$ entails slow diffusion, and the parameter $q$ allows one to explore situations in which diffusion is slower than the time scale of the dynamics of the network. We define the transition probability from node $v_i$ to node $v_j$ at discrete time $n$ as
\begin{equation}
	T_{ij}(n)=\begin{cases}
\delta_{ij} &  (s_i(n)=0\,, j \in \{1,\dots, N\})\,,\\
q^{s_i(n)} & (s_i(n)\ge 1\,, i=j)\,,\\
(A(n))_{ij}(1-q^{s_i(n)})/s_i(n) & (s_i(n)\ge 1\,, i\neq j)\,,
	\end{cases}
\label{eq:def T Rocha-Masuda}
\end{equation}
where $s_i(n) = \sum_{j=1}^N (A(n))_{ij}$ is the strength of $v_i$ at time $n$. Note that $\sum_{j=1}^N T_{ij}(n)=1$. From Eq.~\eqref{eq:def T Rocha-Masuda}, we see that a walker does not move with probability $q^{s_i(n)}$. Otherwise, it moves to a neighbor with a uniform probability of $1/s_i(n)$. By setting the probability of not moving to $q^{s_i(n)}$, one ensures that the probability of not moving from $v_i$ is unaffected by whether multiple edges are present simultaneously in a time period or if they are distributed over multiple times. For example, if $v_i$ is connected simultaneously to three other nodes by unweighted edges at time $n=1$ but isolated at times $n=2$ and $n=3$, the probability that a walker visiting $v_i$ does not move during $n=1$, $n = 2$, and $n = 3$ is equal to $q^3$. The probability is the same if $v_i$ is connected to one node at each of $n=1$, $n = 2$, and $n = 3$. Note that one can derive the former case (i.e., three edges simultaneously connected to $v_i$) from the latter case (i.e., one edge connected to $v_i$ at each time) by coarse-graining the temporal network (e.g., by regarding $A(3n-2)+A(3n-1)+A(3n)$ as a new adjacency matrix at a rescaled discrete time $n$). Our formulation mitigates the effect of temporal resolution (and time-window size) by equating the probability of not moving in the two cases.

The transition probability depends on time. When there are $n_{\max}$ time windows, the transition probability for one ``cycle'' (i.e., one time through the full time period in the temporal sequence of adjacency matrices) is defined as
\begin{equation}
	T^{\rm tp}\equiv T(1) T(2) \cdots T(n_{\max})\,.
\label{eq:T^{tp}}
\end{equation}
Using periodic boundary conditions (i.e., by having the last adjacency matrix $A(n_{\max})$ loop back to $A(1)$), the ``stationary density'' at node $v_i$ is given by the $i$th element of $\bm u(1)$, where
\begin{equation}
	\bm u(1) = \bm u(1) T^{\rm tp}\,.
\end{equation}
There is no stationary density in the present RW process in the conventional sense, because the network is changing in time. Due to the periodic boundary conditions, the stationary density of walkers at each node differs across time periods.
The vector $\bm u(1)$ represents the stationary density when the RW is observed right after time $n_{\max}$ (and before time $1$) in each cycle. One defines the TempoRank vector based on the running mean of the stationary density over all time periods. That is, it is given by ${\bm u}^{\text{avg}} \equiv \sum_{n=1}^{n_{\max}} \bm u(n)/n_{\max}$, where $\bm u(n)$ is the stationary density when the observation is made right after time $n-1$ (and before time $n$).


\subsubsection{Random-walk betweenness centrality}

In our discussions of ranking methods, we have discussed centrality measures (e.g., PageRank) that are derived from RWs. RWs are also useful for deriving variants of other familiar centrality measures, such as ``betweenness centrality''.

\begin{figure}[t]
\begin{center}
\includegraphics[height=3cm]{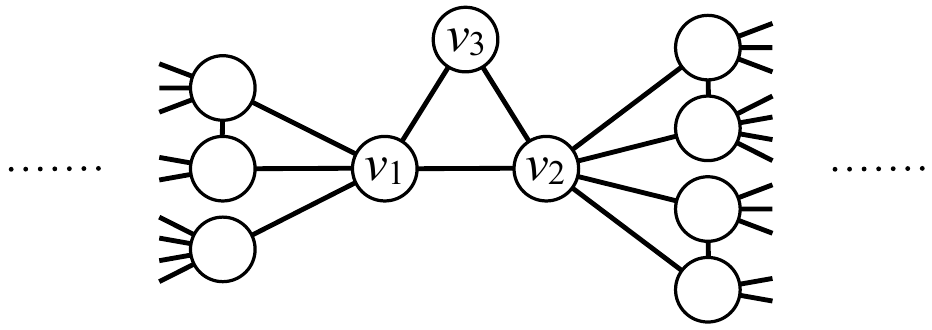}
\caption{A network with two clearly distinguished communities.}
\label{fig:failure of betweenness}
\end{center}
\end{figure}

Shortest-path betweenness centrality (i.e., geodesic betweenness centrality) of a node is defined from a normalized count of the shortest paths that pass through a focal node for all pairs of distinct source and target nodes in a network \cite{Freeman1977Sociom,Newman2010book}. Specifically, the shortest-path betweenness of node $v_i$ is
\begin{equation}
	b_i^{\rm geo} =
\sum_{i_{\rm s}\neq i_{\rm t}} \frac{(\text{number of shortest paths from } v_{i_{\rm s}} \text{ to } v_{i_{\rm t}} \text{ that pass through } v_i)}
{N(N-1)\times (\text{number of shortest paths from } v_{i_{\rm s}} \text{ to } v_{i_{\rm t}})}\,,
\end{equation}
where the nodes $v_i$, $v_{i_{\rm s}}$, and $v_{i_{\rm t}}$ are all distinct. However, restricting to strictly shortest paths can be problematic \cite{Newman2005SocNetw}. For example, consider the network in Fig.~\ref{fig:failure of betweenness} that includes two communities of densely-connected nodes. Nodes $v_1$ and $v_2$ have large betweenness-centrality values because any shortest path connecting one node in each community must pass through both $v_1$ and $v_2$. However, because such a shortest path does not pass through $v_3$, the shortest-path betweenness of node $v_3$ is $0$, yet $v_3$ may be more important than most other nodes in connecting different parts of the network (albeit to a lesser extent than $v_1$ and $v_2$). One can capture this intuition by allowing paths that are longer than the strictly shortest ones to contribute to the value of a betweenness centrality. One way to do this is to use RWs \cite{Newman2005SocNetw,Noh2004PhysRevLett}.

We now explain the ``RW betweenness centrality'' introduced in Ref.~\cite{Newman2005SocNetw}. Consider an undirected network. Similar to the definition of shortest-path betweenness centrality, we specify the starting node $v_{i_{\rm s}}$ and terminal node $v_{i_{\rm t}}$ of an RW. Intuitively, RW betweenness centrality of a node $v_i$ measures the number of times that a random walker starting from $v_{i_{\rm s}}$ passes through $v_i$ before reaching $v_{i_{\rm t}}$. If we do not specify $v_{i_{\rm t}}$, a walker wanders forever in the network, and the centrality of $v_i$ is proportional to $s_i$ [see Eq.~\eqref{eq:p_i^* propto k_i}]. In RW betweenness centrality, one still discounts long walks, because a walk terminates once a walker reaches $v_{i_{\rm t}}$.

The RW betweenness centrality of node $v_i$ as
\begin{equation}
	b_i^{\rm rw}\propto
\sum_{i_{\rm s} =1}^N \sum_{i_{\rm t}=1}^{i_{\rm s}-1} (\text{number of times that a walker starting at } v_{i_{\rm s}} \text{ and terminating at } v_{i_{\rm t}}
\text{ ``effectively'' visits } v_i)\,.
\label{eq:def RW betweenness centrality}
\end{equation}
Note that the ``effective'' number of transitions between nodes $v_i$ and $v_j \in \mathcal{N}_i$ is equal to the difference (in absolute value) between the number of times that a walker moves from node $v_i$ to node $v_j$ and the number of times that it moves from node $v_j$ to node $v_i$. An effective transition from $v_{\ell}$ to $v_i$ and then to a different node $v_j$ (with $j\neq \ell$) completes an effective visit to $v_i$.
Therefore, the number of effective visits to $v_i$ on the right-hand side of Eq.~\eqref{eq:def RW betweenness centrality} is given by $\sum_{j\in \mathcal{N}_i} \text{(number of effective transitions between } v_i \text{ and } v_j)/2$.

Because an RW on a network is related to a corresponding electric circuit on the same network \cite{Doyle1984book,Aldous2002book,LevinPeresWilmer2009book,Newman2010book,Blanchard2011book,Ibe2013book}, we also discuss a centrality based on electric circuits and then relate it to RW betweenness centrality $b_i^{\rm rw}$. Consider an electric circuit in which one injects a unit current at node $v_{i_{\rm s}}$ and drains it at $v_{i_{\rm t}}$. Suppose that each edge has a conductance of $A_{ij}$, and let $V_i$ denote the voltage at node $v_i$. Kirchhoff's current law at each $v_i$ implies that
\begin{equation}
	\sum_{j=1}^N A_{ij}(V_i-V_j)=\delta_{i,i_{\rm s}}-\delta_{i,i_{\rm t}}\,.
\label{eq:Kirchhoff}
\end{equation}
The left-hand side of Eq.~\eqref{eq:Kirchhoff} represents the current that flows from node $v_i$ to node $v_j$ for each $j \in \{1,\dots,N\}$.
Because
\begin{equation}
	\sum_{j=1}^N A_{ij}=s_i\,,
\end{equation}
we rewrite Eq.~\eqref{eq:Kirchhoff} as
\begin{equation}
	(D-A) {\bm V} = L \bm V = \bm I^{\rm curr}\,,
\label{eq:D-A}
\end{equation}
where ${\bm V}=(V_1,\; \ldots\; , V_N)^{\top}$, the quantity $\bm I^{\rm curr}$ is the column vector of size $N$ given by
\begin{equation}
	I^{\rm curr}_i=\begin{cases}
1\,, & (i=i_{\rm s})\,,\\
-1\,, & (i=i_{\rm t})\,,\\
0\,, & (i\not\in \{i_{\rm s}\,, i_{\rm t}\})\,,
	\end{cases}
\label{eq:s_i}
\end{equation}
and we recall that $L$ is the combinatorial Laplacian matrix.

Because $L$ does not have full rank, Eq.~\eqref{eq:D-A} does not have $N$ independent solutions, even though it consists of a set of $N$ linear equations with unknowns $V_i$ (with $i \in \{1, \dots, N\}$). Therefore, we delete an arbitrary $i_0$th row from $L$, corresponding to setting $V_{i_0}=0$, without loss of generality. As in Section~\ref{sub:MFPT}, we also delete the $i_0$th row and column from $D$ and $A$ to yield $(N-1)\times (N-1)$ matrices $\overline{D}^{(i_0)}$ and $\overline{A}^{(i_0)}$, respectively. Similarly, we remove the $i_0$th element from $\bm V$ and $\bm I^{\rm curr}$ to obtain $(N-1)$-dimensional vectors $\overline{\bm V}^{(i_0)}$ and $\overline{\bm I}^{\text{curr} (i_0)}$, respectively. Equation~\eqref{eq:D-A} is thus equivalent to
\begin{equation}
	(\overline{D}^{(i_0)} - \overline{A}^{(i_0)}) \overline{\bm V}^{(i_0)} = \overline{\bm I}^{\text{curr} (i_0)}\,.
\label{eq:D-A N-1}
\end{equation}
For a connected network, the matrix $\overline{D}^{(i_0)} - \overline{A}^{(i_0)}$ has full rank, and we obtain
\begin{equation}
	\overline{\bm V}^{(i_0)} = (\overline{D}^{(i_0)}-\overline{A}^{(i_0)})^{-1}\overline{\bm I}^{\text{curr} (i_0)}\,.
\label{eq:D-A N-1 inverted}
\end{equation}
We now reinsert the $i_0$th row and column of $(\overline{D}^{(i_0)}-\overline{A}^{(i_0)})^{-1}$ by filling them with $0$s, and we denote the resulting $N\times N$ matrix by $\overline{R} = (\overline{R}_{ij})$. Substituting Eq.~\eqref{eq:s_i} into Eq.~\eqref{eq:D-A N-1 inverted} then yields
\begin{equation}
	V_i=\overline{R}_{i, i_{\rm s}}-\overline{R}_{i, i_{\rm t}}\,.
\label{eq:Vi final}
\end{equation}
Note that Eq.~\eqref{eq:Vi final} satisfies the condition $V_{i_0}=0$. The total current that flows through node $v_i$ is
\begin{equation}
	\text{Current}_i^{i_{\rm s}, i_{\rm t}} = \begin{cases}
\displaystyle{\frac{1}{2}\sum_{j=1}^N A_{ij}\left|V_i-V_j
\right|
=\frac{1}{2}\sum_{j=1}^N A_{ij} \left| \overline{R}_{i, i_{\rm s}} - \overline{R}_{i, i_{\rm t}} - \overline{R}_{j, i_{\rm s}} + \overline{R}_{j, i_{\rm t}} \right|} &
(i\not\in \{ i_{\rm s}, i_{\rm t}\})\,,\\[3mm]
1 & (i \in \{ i_{\rm s}, i_{\rm t}\})\,.
	\end{cases}
\label{eq:I_i general}
\end{equation}
The division by $2$ in the first case of Eq.~\eqref{eq:I_i general} arises from the fact the same current is counted twice when it flows into and out of $v_i$.

One can show that RW betweenness centrality is equal to
\begin{equation}
	b_i^{\rm rw}= \frac{\displaystyle{\sum_{i_{\rm s}=1}^N \sum_{i_{\rm t}=1}^{i_{\rm s}-1}
\text{Current}_i^{i_{\rm s}, i_{\rm t}}}}{N(N-1)/2}\,.
\label{eq:current flow centrality}
\end{equation}
That is, it is the normalized frequency that a random walker visits node $v_i$ before it reaches $v_{i_t}$. To verify Eq.~\eqref{eq:current flow centrality}, let's consider a DTRW with an absorbing boundary at $v_{i_t}$. The transition-probability matrix consists of the elements
\begin{equation}
	T_{ij}^{\prime}=\begin{cases}
\frac{A_{ij}}{s_i} & (i\neq i_{\rm t})\,,\\
\delta_{i_{\rm t}j} & (i=i_{\rm t})\,.
	\end{cases}
\label{eq:Bij'}
\end{equation}
The matrix $T^{\prime}$ is equal to the transition-probability matrix of a DTRW with an absorbing boundary, so $T^{\prime}$ is equal to $D^{-1}A$ except in the $i_{\rm t}$th row. We remove the $i_{\rm t}$th row and column from $T^{\prime}$, $D^{-1}$, and $A$ to obtain
\begin{equation}
	\overline{T}^{\prime (i_{\rm t})}= \left(\overline{D}^{(i_{\rm t})}\right)^{-1} \overline{A}^{(i_{\rm t})}\,.
\end{equation}
Whenever the row sum of $\overline{T}^{\prime}$ is less than $1$, the walk is absorbed at
$v_{i_{\rm t}}$ with the residual probability.

Consider an RW that starts from node $v_{i_{\rm s}}$.
The probability that a random walker visits
$v_i$ (with $i\neq i_{\rm t}$) after $n$ steps is given by the $(i_{\rm s}, i)$th element
of $\left(\overline{T}^{\prime (i_{\rm t})}\right)^n$. (For clarity, we use the indices $1$, $\ldots$, $i_{\rm t}-1$, $i_{\rm t}+1$, $\ldots$, $N$ rather than $1$, $\ldots$, $N-1$ for the elements of $\overline{T}^{\prime}$.) Conditioned on this event, the probability that the walker moves to node $v_j$ in the next step is equal to $1/k_i$. The expected number of times that the walker steps from node $v_i$ to a neighboring node $v_j \in \mathcal{N}_i$ is
\begin{align}
	\sum_{n=0}^{\infty}\frac{\left(\left(\overline{T}^{\prime \left(i_{\rm t}\right)}\right)^n\right)_{i_{\rm s} i}}{k_i} &=
\frac{\left(\left[I - \overline{T}^{\prime (i_{\rm t})}\right]^{-1}\right)_{i_{\rm s} i}}{k_i}\notag\\
	&= i \text{th element of }\left(\overline{\bm I}^{\text{curr} (i_{\rm t})}\right)^{\top}
\left[I-\left(\overline{D}^{(i_{\rm t})}\right)^{-1} \left(\overline{A}^{(i_{\rm t})}\right)\right]^{-1} \left(\overline{D}^{(i_{\rm t})}\right)^{-1}\notag\\
	&= i \text{th element of }\left(\overline{\bm I}^{\text{curr} (i_{\rm t})}\right)^{\top}
\left(\overline{D}^{(i_{\rm t})} - \overline{A}^{(i_{\rm t})}\right)^{-1}\,.
\label{eq:RW v_i -> v_j}
\end{align}
Because $\overline{D}^{(i_{\rm t})}$ and $\overline{A}^{(i_{\rm t})}$ are symmetric matrices, the left-hand side of Eq.~\eqref{eq:RW v_i -> v_j} is also equal to the $i$th element of $\left[\left(\overline{\bm I}^{\text{curr} (i_{\rm t})}\right)^{\top}
\left(\overline{D}^{(i_{\rm t})} - \overline{A}^{(i_{\rm t})}\right)^{-1}\right]^{\top}
=\left(\overline{D}^{(i_{\rm t})} - \overline{A}^{(i_{\rm t})}\right)^{-1}\overline{\bm I}^{\text{curr}(i_{\rm t})}$. Therefore, Eq.~\eqref{eq:D-A N-1 inverted} guarantees that the quantity $\sum_{n=0}^{\infty}({[(\overline{T}^{\prime (i_{\rm t})})^n]_{i_{\rm s} i}}/{k_i})$ is equal to voltage $V_i$ when $v_{i_0} = v_{i_{\rm t}}$.
Finally, the ``effective'' number of transitions --- i.e., the difference between the number of times that a walker
moves from node $v_i$ to node $v_j$ and the number of times that it moves from node $v_j$ to node $v_i$ --- is equal to $|V_i-V_j|$.

We now consider ``RW centrality'' \cite{Noh2004PhysRevLett}, another a variant of RW betweenness centrality. This centrality quantifies the speed at which a walker starting from node $v_i$ reaches other nodes compared to the speed at which a walker starting from an arbitrary node reaches $v_i$. To formalize this idea, we use Eq.~\eqref{eq:T_ij general undirected},
which gives the MFPT $m_{ij}$ from node $v_i$ to node $v_j$, and we focus on undirected networks. One measures the importance of node $v_i$ relative to node $v_j$ by calculating
\begin{equation} \label{eq:T_ij-T_ji}
	m_{ij} - m_{ji} =
\left(\sum^N_{\ell=1}s_{\ell}\right)\times \left[
\left(\frac{R_{jj}^{(0)}}{s_j}-\frac{R_{ii}^{(0)}}{s_i}\right)
-
\left(\frac{R_{ij}^{(0)}}{s_j}-\frac{R_{ji}^{(0)}}{s_i}\right)\right]\,.
\end{equation}
For undirected networks, the following detailed balance, which extends Eq.~\eqref{eq:detailed balance basic}, holds \cite{Noh2004PhysRevLett}:
\begin{align}
	s_i p_{ij}(n) &= s_i \sum_{\ell_1, \ell_2, \ldots, \ell_{n-1} = 1}^N \frac{A_{i\ell_1}}{s_i} \frac{A_{\ell_1 \ell_2}}{s_{\ell_1}}
\times \frac{A_{\ell_{n-1}j}}{s_{\ell_{n-1}}}
\notag\\
	&= \sum_{\ell_1, \ell_2, \ldots, \ell_{n-1} = 1}^N \frac{A_{i\ell_1}}{s_{\ell_1}} \frac{A_{\ell_1 \ell_2}}{s_{\ell_2}}
\times \frac{A_{\ell_{n-1}j}}{s_j} s_j
	= s_j p_{ji}(n)\,.
\label{eq:detailed balance}
\end{align}
Substituting Eq.~\eqref{eq:detailed balance} into Eq.~\eqref{eq:R_ij^n def} yields
\begin{align}
	\frac{R_{ij}^{(0)}}{s_j} &=
\frac{\sum_{n=0}^{\infty}\left[p_{ij}(n)-p_j^* \right]}{s_j}
\notag\\
	&= \frac{\sum_{n=0}^{\infty}
\left[\frac{s_j p_{ji}(n)}{s_i}-\frac{s_j}{\sum^N_{\ell = 1}s_{\ell}}\right]}{s_j}\notag\\
	&= \frac{\sum_{n=0}^{\infty}
\left[p_{ji}(n)-\frac{s_i}{\sum^N_{\ell =1}s_{\ell}}\right]}{s_i}
	= \frac{R_{ji}^{(0)}}{s_i}\,.
\label{eq:vanish Noh}
\end{align}
We then apply Eq.~\eqref{eq:vanish Noh} to Eq.~\eqref{eq:T_ij-T_ji} to obtain
\begin{equation}
	m_{ij} - m_{ji} = C_{\rm rw}(j)^{-1} - C_{\rm rw}(i)^{-1}\,,
\label{eq:T_ij-T_ji C}
\end{equation}
where
\begin{align}
	C_{\rm rw}(i) &\equiv \frac{s_i}{R_{ii}^{(0)}\sum_{\ell =1}^N s_{\ell}} \notag \\
		&= \frac{s_i}{\sum^{\infty}_{n=0}\left[p_{ii}(n)-\frac{s_i}{\sum_{\ell =1}^N s_{\ell}}
\right]\sum_{\ell =1}^N s_{\ell}}
\end{align}
is defined to be the RW centrality.


\subsubsection{Discrete-choice models}\label{choice}


Discrete-choice models describe decisions between distinct alternatives \cite{Benakiva1985book,Train2003-2009book}.
Examples of discrete choices occur in everyday life; for example, one can choose to shop at a given store, use a specific mode of transportation, or root for the Los Angeles Dodgers instead of some other baseball team. In many applications, one faces the problem of ``rank aggregation'' \cite{Dwork2001WWW}, as it is necessary to aggregate preferences about an item over a set of alternatives, which one observes for different individuals, who have different subsets of alternatives.
For example, the Bradley--Terry--Luce (BTL) model defines the probability to select alternative $i$ (where $i\in \{ 1, \ldots, N\}$) over alternative $j$ in a pairwise comparison as
\begin{equation}
	p_{ij} = \frac{\gamma_i}{\gamma_i + \gamma_j}\,,
\end{equation}
where $\gamma_i >0$ is a latent parameter that encodes the attractiveness of alternative $i$ \cite{Bradley1952Biometrika,Luce1959book}.

The pairwise-choice Markov chain (PCMC) model is a discrete-choice model that uses the stationary density of a CTRW as the probability to select $i$ among several alternatives \cite{Ragain2016NIPS}.
In the PCMC model, one considers a Poissonian edge-centric CTRW on an $N$-node directed and weighted network.
An individual can choose an item from a subset $S$ of the $N$ alternatives (i.e., nodes). Instead of using the network's adjacency matrix $A$ to construct a transition-rate matrix for a CTRW on the entire network (see Eq.~\eqref{eq:master unnormalized CTRW}), the PCMC model uses $A$ to define a transition-rate matrix $Q_S=(q_{ij})$ on $S$.
The rows and columns of $Q_S$ are indexed by the elements in $S$, and they are defined by $q_{ij} = A_{ij}$ (for $j\neq i$) and $q_{ii} = -\sum_{j\in S\backslash i}q_{ij}$. For any set $S$, note that $Q_S$ does not require the diagonal elements of $A$, so we assume that they are $0$. The PCMC model uses the stationary density of the CTRW on $S$ as the probability that an individual chooses $i$ when $S$ is the set of alternatives. One can then estimate the matrix $A$ from, for example, empirical-choice data.

A generalization of the BTL model is the multinomial logit model (also called the Plackett--Luce moel) \cite{Luce1959book,Block1960ContProbStat,Plackett1975JRStatSocSerC}, which treats the case of a choice among more than two alternatives. The multinomial logit model defines the probability $p_{iS}$ to choose $i$ from $S$ as
\begin{equation}
	p_{iS} = \frac{\gamma_i}{\sum_{j\in S}\gamma_j}\,.
\end{equation}
This model is a PCMC model, where the adjacency matrix is determined by the BTL model, so $A_{ji} = \gamma_i / (\gamma_i + \gamma_j)$. A large $\gamma_i$ value makes $A_{ji}$ large, which in turn results in a large probability in-flow to the $i$th node and an increased probability that an individual chooses $i$. In fact, the vector $\bm p^* = (p_{iS})$, with $i\in S$, is the stationary density of the CTRW on $S$, because
\begin{align}
	\left(\bm p^* Q_S\right)_i =& \frac{1}{\sum_{\ell\in S}\gamma_\ell}
\left(\sum_{j\in S; j\neq i} \gamma_j A_{ji} - \gamma_i \sum_{j\in S; j\neq i}A_{ij}\right)\notag\\
=& \frac{\gamma_i}{\sum_{\ell\in S}\gamma_\ell}
\left(\sum_{j\in S; j\neq i} \frac{\gamma_j}{\gamma_i + \gamma_j} - \sum_{j\in S; j\neq i} \frac{\gamma_j}{\gamma_i + \gamma_j}\right) = 0
	\quad (i \in S)\,.
\end{align}
Consider a data set given in the form of $\mathcal D = \{ (i_\ell, S_\ell) | \ell=1, \ldots, \ell_{\max} \}$, where $S_{\ell}$ is the set of the items presented in the $\ell$th choice, $i_{\ell}\in S_{\ell}$ is the item chosen in the $\ell$th choice, and $\ell_{\max}$ is the number of choices. The PCMC in which the parameters (i.e., entries of $A$) are estimated by a maximum-likelihood method yields a better predictive performance than benchmark discrete-choice models on two empirical data sets \cite{Ragain2016NIPS}.




One can also derive the maximum-likelihood estimator of the multinomial logit model as the stationary density of
a Poissonian edge-centric CTRW \cite{Maystre2015NIPS}. The likelihood $\tilde{L}$ of the parameters $\bm \gamma \equiv \{\gamma_1$, $\ldots$, $\gamma_N\}$ given data $\mathcal D$ is
\begin{equation}
	\tilde{L}\left( \bm \gamma | \mathcal D \right) =
\prod_{\ell=1}^{\ell_{\max}}
	\frac{\gamma_{i_\ell}} {\sum_{i^{\prime}\in S_\ell} \gamma_{i^{\prime}}}\,.
\end{equation}
By maximizing the log likelihood, one obtains
\begin{align}
	\frac{\partial (\log \tilde{L})}{\partial \hat{\gamma}_i} =&
\frac{\partial}{\partial \hat{\gamma}_i} \sum_{\ell=1}^{\ell_{\max}} \left( \log \hat{\gamma}_{i_\ell} - \log\sum_{i^{\prime}\in S_\ell} \hat{\gamma}_{i^{\prime}} \right)\notag\\
=& \sum_{\ell=1; \ell\in \breve{W}_i}^{\ell_{\max}} \left( \frac{1}{\hat{\gamma}_i} - \frac{1}{\sum_{i^{\prime}\in S_\ell} \hat{\gamma}_{i^{\prime}}} \right) - \sum_{\ell=1; \ell\in \breve{L}_i}^{\ell_{\max}} \frac{1}{\sum_{i^{\prime}\in S_\ell} \hat{\gamma}_{i^{\prime}}}\notag\\
=& 0\quad (i \in \{1, \dots, N\})\,,
	\label{eq:Maystre 1}
\end{align}
where
$\breve{W}_i = \{ \ell | i\in S_{\ell} \text{ and } i \text{ is chosen}\}$, $\breve{L}_i = \{ \ell | i\in S_{\ell} \text{ and } i \text{ is not chosen}\}$, and $\hat{\gamma}_i$ (with $i\in \{1, \ldots, N\})$ is the maximum-likelihood estimator.
By multiplying $\hat{\gamma}_i$ by Eq.~\eqref{eq:Maystre 1}, one obtains
\begin{equation}
	\sum_{\ell=1; \ell\in \breve{W}_i}^{\ell_{\max}} \frac{\sum_{j\in S_{\ell}; j\neq i}\hat{\gamma}_j}{\sum_{i^{\prime}\in S_\ell} \hat{\gamma}_{i^{\prime}}}
- \sum_{\ell=1; \ell\in \breve{L}_i}^{\ell_{\max}} \frac{\hat{\gamma}_i}{\sum_{i^{\prime}\in S_\ell} \hat{\gamma}_{i^{\prime}}}
= 0\quad (i \in \{1, \dots, N\})\,.
	\label{eq:Maystre 2}
\end{equation}
Because $\breve{L}_i = \cup_{j=1; j\neq i}^N (\breve{W}_j \cap \breve{L}_i)$, one can rewrite Eq.~\eqref{eq:Maystre 2} as
\begin{equation}
	\sum_{j=1; j\neq i}^N \left[ \sum_{\ell=1; \ell\in \breve{W}_i\cap \breve{L}_j}^N \frac{\hat{\gamma}_j}{\sum_{i^{\prime}\in S_{\ell}} \hat{\gamma}_{i^{\prime}}}
- \sum_{\ell=1; \ell\in \breve{W}_j\cap \breve{L}_i}^N \frac{\hat{\gamma}_i}{\sum_{i^{\prime}\in S_{\ell}} \hat{\gamma}_{i^{\prime}}}
\right] = 0\quad (i \in \{1, \dots, N\})\,.
	\label{eq:Maystre 3}
\end{equation}
One rewrites Eq.~\eqref{eq:Maystre 3} as
	\begin{equation}
\sum_{j=1; j\neq i}^N \hat{\gamma}_i f({\mathcal D}_{j\succ i}, \bm{\hat{\gamma}})
= \sum_{j=1; j\neq i}^N \hat{\gamma}_j f({\mathcal D}_{i\succ j}, \bm{\hat{\gamma}})
\quad (i \in \{1, \dots, N\})\,,
	\label{eq:Maystre 4}
\end{equation}
where
\begin{equation}
	f(\mathcal D^{\prime}, \bm{\hat{\gamma}}) = \sum_{S\in \mathcal D^{\prime}} \frac{1}{\sum_{i^{\prime}\in S} \hat{\gamma}_{i^{\prime}}}\,,
	\label{eq:Maystre f}
\end{equation}
$\mathcal D^{\prime}\subset \mathcal D$ is a subset of the observation set $\mathcal D$,
and
${\mathcal D}_{i\succ j} = \{ (i_{\ell}, S_{\ell}) \in \mathcal D | \ell \in \breve{W}_i\cap \breve{L}_j\} \subset \mathcal D$ is the set of observations in which $i$ is preferred to $j$.
Equation~\eqref{eq:Maystre 4} implies that the maximum-likelihood estimator is the stationary density of the CTRW whose transition rate from the $j$th to the $i$th node is given by
$f({\mathcal D}_{i\succ j}, \bm{\hat{\gamma}})$. One interprets $f({\mathcal D}_{i\succ j}, \bm{\hat{\gamma}}) = \sum_{S\in \mathcal {\mathcal D}_{i\succ j}} \left(1/\sum_{i^{\prime}\in S} \hat{\gamma}_{i^{\prime}}\right)$ as the number of times $i$ is chosen over $j$ (taken into accounted by the sum $\sum_{S\in \mathcal {\mathcal D}_{i\succ j}}$), weighted by the strength of the alternatives in each observation
(which is taken into account with the term $1/\sum_{i^{\prime}\in S} \hat{\gamma}_{i^{\prime}}$).
Taking advantage of this relationship between the
CTRW and the maximum-likelihood estimator of the multinomial logit model has resulted in inference algorithms for the multinomial logit model that is faster and more accurate than previous methods for several data sets \cite{Maystre2015NIPS}.


For other methods of rank aggregation based on RWs, see Refs.~\cite{Dwork2001WWW,Negahban2012NIPS,Negahban2017OperRes}.



\subsection{Community detection} \label{sub:community}

A useful approach for studying networks is to examine mesoscale structures, of which the best-known type is ``community structure'' \cite{santo2016,porter2009,Fortunato2010PhysRep}. There are numerous methods to algorithmically detect communities (and many applications in which communities can be insightful), which are sets of densely connected nodes such that connections between different communities are relatively sparse. RWs provide a theoretical basis for understanding community structure and practical algorithms for detecting them. The main idea is that, if a given network has community structure, a random walker should be trapped within a community for a relatively long time before leaving it. This arises from the high density of edges within communities and the sparse connections across communities. Therefore, RWs that are observed on a short time scale should reveal intra-community structure in a network, and RWs that are observed on a long time scale should reveal global structure about the same network.

In this section, we introduce some algorithms for community detection that are based on RWs. For other RW-based algorithms and theoretical underpinnings, see papers such as Refs.~\cite{ShiMalik2000IeeeTransPatternAnal,Vandongen2001thesis,Eriksen2003PhysRevLett,ZhouH2003PhysRevE,Arenas2006PhysRevLett,Newman2006PhysRevE-collabo,Danon2008PhysRevE,ChengShen2010JStatMech,Lambiotte2011PhysRevE,Piccardi2011PlosOne,Jeub2015PhysRevE,Jeub2017NetwSci}.


\subsubsection{Markov-stability formulation of modularity\label{sub:modularity}}

It is common to use the ``modularity'' objective function $Q$ to quantify the quality of a partition of a network into nonoverlapping communities, and many community-detection methods are based on maximizing $Q$ \cite{santo2016}. Consider a partition of an undirected network into $N_{\rm CM}$ communities. Let ${\rm CM}_c$ denote the $c$th community (with $c \in \{1, 2, \ldots , N_{\rm CM}\}$). We use a variant (sometimes called the ``Newman--Girvan null model'') of an undirected configuration model \cite{Fosdick2016} that is defined as a random graph with a specified strength $s_i$ at each node. For this configuration model, the probability that nodes $v_i$ and $v_j$ are adjacent is approximately $P_{ij} \equiv s_is_j/(2M^{\prime})$, where $M^{\prime} = \sum_{i=1}^N s_i/2$ is the sum of the edge weight over all edges \cite{Newman2010book}. (Technically, $P_{ij}$ is a probability only for sufficiently small edge weights; otherwise, it is an expectation.) Note that $M^{\prime}=M$ for an unweighted network, where we recall that $M$ is the number of edges. Modularity is defined by
\begin{align}
	Q &= \frac{1}{2M^{\prime}}\sum_{c=1}^{N_{\rm CM}}
\left[\sum_{\begin{subarray}{c}
i,j=1;\\ v_i, v_j \in {\rm CM}_c\end{subarray}}^N
\left(A_{ij}-\frac{s_i s_j}{2M^{\prime}}\right)\right]\notag\\
	&= \frac{1}{2M^{\prime}} \sum_{i,j=1}^N \left(A_{ij}-\frac{s_i s_j}{2M^{\prime}}\right) \delta(g_i, g_j)\,,
\label{eq:Q def}
\end{align}
where $g_i$ is the community to which node $v_i$ has been assigned, and $\delta(g_i, g_j)=1$ if $g_i=g_j$ and $\delta(g_i, g_j)=0$ otherwise. The quantity $P_{ij}$ gives the elements of a null-model matrix, and a wide variety of different versions of the matrix $P = (P_{ij})$ have been examined \cite{sarzynska2016,Bazzi2016MultModelSimul}. More precisely, $P$ is not a ``null model'' but rather a ``null network'' (which is a network generated from a null model) \cite{Bazzi2016MultModelSimul}.
%

Methods based on modularity maximization suffer from the fact that $Q$ has a resolution limit, so using Eq.~\eqref{eq:Q def} does not allow one to detect dense communities of nodes that are smaller than a certain scale \cite{Fortunato2007PNAS,Good2010PhysRevE} (though some null models attempt to address this issue). Modularity maximization also implicitly favors communities of a particular size that depend on the size of the entire network (not only its internal structure), and methods based on maximizing $Q$ also have various other problematic features \cite{santo2016}.

One can use RWs to gain insights into modularity and its resolution issues. Modularity is closely related to ``Markov stability'', which quantifies the tendency for a random walker to stay inside a community for a long time. The Markov stability of a partition of a network is defined as the probability that a walker is in the same community at time $0$ and time $t$ in the equilibrium of the Poissonian node-centric CTRW \cite{Lambiotte2008arxiv,Schaub2012PlosOne,Beguerissediaz2014JRSocInterface,Lambiotte2015IEEETransNetwSciEng}.
See Refs.~\cite{Delvenne2010PNAS,Lambiotte2015IEEETransNetwSciEng} for a version of Markov stability derived from a DTRW.

The master equation is
\begin{equation}
	\frac{{\rm d}\bm p(t)}{{\rm d}t} = -\bm p(t) L^{\prime}\,,
\label{eq:master equation continuous-time RW}
\end{equation}
where we recall that $L^{\prime}$ is the random-walk normalized Laplacian matrix
[see Eq.~\eqref{eq:RW normalised Laplacian}]. The stationary density is given by Eq.~\eqref{eq:p_i^* propto k_i}.

Consider a pair of nodes, $v_i$ and $v_j$, that belong to the same community. Equation~\eqref{eq:master equation continuous-time RW} implies that, in the stationary state, the probability that a random walker visits $v_i$ and then $v_j$ after time $t$ is equal to $p_i^* (e^{-tL})_{ij}$. As with modularity maximization, one needs to compare this quantity with a null model. For Markov stability $R(t)$, the standard null model is given by the probability that a walker visits node $v_i$ at $t=0$ and node $v_j$ at $t=\infty$. This yields a null probability of $p_i^* p_j^*$. One thereby obtains a Markov stability of
\begin{equation} \label{eq:R(t) Lambiotte}
	R(t) = \sum_{i, j=1}^N
\left[\left( p_i^* e^{-tL^{\prime}} \right)_{ij} - p_i^* p_j^* \right]
\delta(g_i, g_j)\,.
\end{equation}

Because of the exponential factor $e^{-tL}$, Markov stability combines walks of various lengths between two nodes. The time $t$ acts as a resolution parameter, enabling one to zoom in and out to unravel multiscale structure in a network. A large value of $t$ gives large weightings to long walks and yields a small number of communities. In the limit $t\to\infty$, Markov stability is optimized by the bipartition given by the signs of the elements of the Fiedler vector (i.e., a type of spectral partitioning) if the corresponding eigenvalue is not degenerate \cite{Newman2006PhysRevE-collabo}. More generally, spectral partitioning is related to RWs on networks because it uses the eigenvectors of matrices such as the combinatorial Laplacian matrix or a modularity matrix \cite{Delvenne2013chapter,Yan2016PeerJComputSci}.

Because it is computationally expensive to calculate $e^{-tL^{\prime}}$ for large networks, we use a linear approximation $e^{-tL^{\prime}} \approx I - t L^{\prime}$. To simplify our exposition, we now assume the case of undirected networks for the rest of this section \cite{Lambiotte2015IEEETransNetwSciEng}. By substituting $p_i^* = s_i/(2M^{\prime})$ and $p_j^* = s_j/(2M^{\prime})$ into Eq.~\eqref{eq:R(t) Lambiotte}, we obtain
\begin{equation}
	R(t) = \frac{1}{2M^{\prime}} \sum_{i, j=1}^N \left[ t A_{ij} + (1-t)\delta_{ij}s_i + \frac{s_i s_j}{2M^{\prime}}\right]
\delta(g_i, g_j)\,.
\end{equation}
Because $\sum_{i,j=1}^N (1-t) \delta_{ij}s_i \delta(g_i, g_j) = \sum_{i=1}^N s_i$ does not depend on the partitioning of a network, maximizing $R(t)$ is equivalent to maximizing
\begin{equation}
	Q(\gamma) = \frac{1}{2M^{\prime}} \sum_{i,j=1}^N \left (A_{ij}-\gamma \frac{s_i s_j}{2M^{\prime}}\right) \delta(g_i, g_j)\,,
\label{eq:Q gamma}
\end{equation}
where $\gamma \equiv 1/t$. We ignore the constraint that $t$ is small (which is admittedly naughty mathematically) and thereby allow general values for $\gamma$ when maximizing $Q(\gamma)$. We also note that $Q(\gamma)$ was derived originally using the perspective of a Potts spin glass \cite{rb2006}, and recently it has been related to maximum-likelihood methods \cite{Newman2016PhysRevE-equivalent}.

When $\gamma=1$, Eq.~\eqref{eq:Q gamma} coincides with Eq.~\eqref{eq:Q def}. Therefore, modularity is an approximate variant of Markov stability. A large value of
$\gamma$ emphasizes the penalty for classifying nodes into the same community and results in many communities. The choice of the natural resolution parameter $\gamma$ is an important practical issue \cite{Delvenne2013chapter,bassett2013chaos}, and it can be examined from a maximum-likelihood approach \cite{Newman2016PhysRevE-equivalent}.


\subsubsection{Walktrap} \label{sub:walktrap}

In the Walktrap algorithm, one defines a measure of similarity between nodes based on DTRWs and uses it for community detection
\cite{Pons2006JGraphAlgoAppl}. (See Ref.~\cite{ZhouLipowsky2004LNCS} for a similar method that uses DTRWs.) Consider an undirected and unweighted network. Define the RW-based distance between two nodes, $v_i$ and $v_j$, by
\begin{equation}
	r_{ij} = \sqrt{\sum_{\ell=1}^N\frac{(T_{i\ell}^n - T_{j\ell}^n)^2}{k_{\ell}}}\,,
\label{eq:r_ij Walktrap}
\end{equation}
where $n$ is the number of steps in a DTRW. The distance $r_{ij}$ is small when a pair of random walkers --- one starting from $v_i$ and the other starting from $v_j$ --- visit each node with similar probabilities after $n$ steps. The denominator $k_{\ell}$ discounts the fact that a walker visits $v_{\ell}$ with a probability proportional to $k_{\ell}$ at equilibrium. Note that $n$ needs to be large enough for random walkers to be able to travel to any node. However, $n$ should not be too large, because $\lim_{n\to\infty}T_{i\ell}^n = \lim_{n\to\infty} T_{j\ell}^n = p_{\ell}^*$ implies that $r_{ij}$ is very close to $0$ for all $i, j \in \{1, \ldots, N\}$ when $n$ is large \cite{Fortunato2010PhysRep}.

We expect that a pair of nodes, $v_i$ and $v_j$, that are separated by a small distance $r_{ij}$ are likely to belong to the same community. One uses a standard agglomerative and hierarchical clustering algorithm on the distance matrix $r = (r_{ij})$. One starts from the partition composed of $N$ single-node communities and joins a pair of communities (so-called ``tentative communities'') with the smallest distance, one pair at time, to produce a series of partitions until the entire network is in a single community. In the merging process, one measures the distance between two communities $\text{CM}_c$ and $\text{CM}_{c^{\prime}}$ by the $r_{ij}$ value, normalized in some way, between $v_i, v_j\in \text{CM}_c\cup \text{CM}_{c^{\prime}}$. This agglomerative clustering algorithm is similar to a greedy algorithm to maximize modularity across partitionings with different numbers of communities \cite{Newman2004PhysRevE-fast}. In Walktrap, one merges a pair of communities under the restriction that they can be merged only when they are adjacent to each other by at least one edge.

Other community-detection methods also rely on defining a similarity measure between nodes.  An interesting approach is based on the concept of mean first-passage time $m_{ij}$ of a random walker (see Section \ref{sub:MFPT}) and its symmetrization $m_{ij}+m_{ji}$ (the so-called ``mean commute time'') \cite{saerens}.
The square root of the mean commute time has the desirable property of being a Euclidian distance between nodes. In this context, it is called the ``Euclidian commute-time distance''. It decreases when the number of paths between two nodes increases or when the length of any path between the two nodes decreases, and it can be derived from the pseudo-inverse of the combinatorial Laplacian matrix $L$ \cite{barnett}.



\subsubsection{InfoMap\label{sub:Infomap}}

InfoMap is another algorithm for community detection based on RWs \cite{Rosvall2008PNAS}. It is very popular and has been extended to the case of hierarchical algorithms \cite{RosvallBergstrom2011PlosOne}, memory networks \cite{Rosvall2014NatComm}, and multilayer networks \cite{Dedomenico2015PhysRevX}. In this section, we discuss the basic version of InfoMap.

Consider a DTRW on a network, which can be directed or weighted. If the network has meaningful community structure, a random walker tends to be trapped within a community for a long time before traveling to a different community. A trajectory of the RW is a sequence of the visited nodes (e.g., $v_3$, $v_6$, $v_3$, $v_1$, $v_8$, $\ldots$). Let's encode each node into a finite binary sequence (i.e., ``a code word'') and concatenate the code words to encode the
trajectory of a random walker. For example, if $v_1$, $v_2$, $v_3$, $v_4$, $v_5$, $\ldots$ are encoded into $000, 001, 010, 011, 100, \ldots$, then the trajectory $v_3$, $v_6$, $v_3$, $v_1$, $v_8$, $\ldots$ is encoded into $010101010000111 \cdots$. For unique decoding, one needs a ``prefix-free'' coding scheme. In other words, a code word cannot be a ``prefix'' (i.e., an initial segment) of another code word. For instance, if $v_1$ and $v_2$ are coded as $000$ and $0001$, respectively, then one's code is not prefix-free, because $000$ is an initial segment of $0001$.

The ``Huffman code'' is a popular prefix-free code that encodes individual symbols (i.e., nodes $v_i$) separately and tends to yield short binary sequences \cite{Huffman1952ProcIre}. It assigns a short code word to a frequently visited node. In a stationary state, the mean code word length per step of an RW is $\sum_{i=1}^N p_i^* \times \text{len}(i)$, where $\text{len}(i)$ denotes the length of the code word assigned to $v_i$.

If symbols (such as $v_i$ in our context) appear independently in each step of an RW, the Huffman code yields a mean code word length in each step that  is close to the theoretical lower bound set by the Shannon entropy
\begin{equation}
	H = -\sum_{i=1}^N p_i^* \log p_i^* \,.
\label{eq:Shannon entropy}
\end{equation}
However, the sequence of symbols is correlated in time, because it is produced by an RW. Consequently, a different coding scheme can yield a mean code length that is smaller than the Shannon entropy. InfoMap exploits community structure and uses a two-layer variant of the Huffman code to achieve this goal.
%
%
Because there are fewer nodes in a community than in an entire network, one can express a trajectory within each community using a shorter, different Huffman code that is local to individual communities. In practice, one constructs the two-layer Huffman code as follows:
\begin{enumerate}
\item When a random walker enters the $c$th community, one issues the (predetermined) code word that corresponds to entering community $\text{CM}_c$.
\item The walker moves around within community
$\text{CM}_c$ for some time. One records the trajectory during this period by the sequence of code words that corresponds to the sequence of visited nodes. One concatenates these code words, and they appear after the code word (obtained in the previous step) that corresponds to the entry to community $\text{CM}_c$.
\item The walker eventually exits $\text{CM}_c$. This event is represented by a special code word, which one places after the sequence of code words that one has obtained thus far.
\item The exit from $\text{CM}_c$ implies an immediate entry to a different community, which we denote by $\text{CM}_{c^{\prime}}$. Therefore, we concatenate the code word corresponding to the entry to $\text{CM}_{c^{\prime}}$ to the end of the sequence of code words that we have obtained thus far.
\item One uses the code words that are local to $\text{CM}_{c^{\prime}}$ to record the trajectory until the walker exits $\text{CM}_{c^{\prime}}$. Note that one can use the same code word to represent a node in $\text{CM}_c$ and a node in $\text{CM}_{c^{\prime}}$. This fact does not cause any problems, because one determines the current coding table from the entry and exit code words.
\item Repeat steps 3--5.
\end{enumerate}

\begin{figure}[tb]
\begin{center}
\includegraphics[scale=0.4]{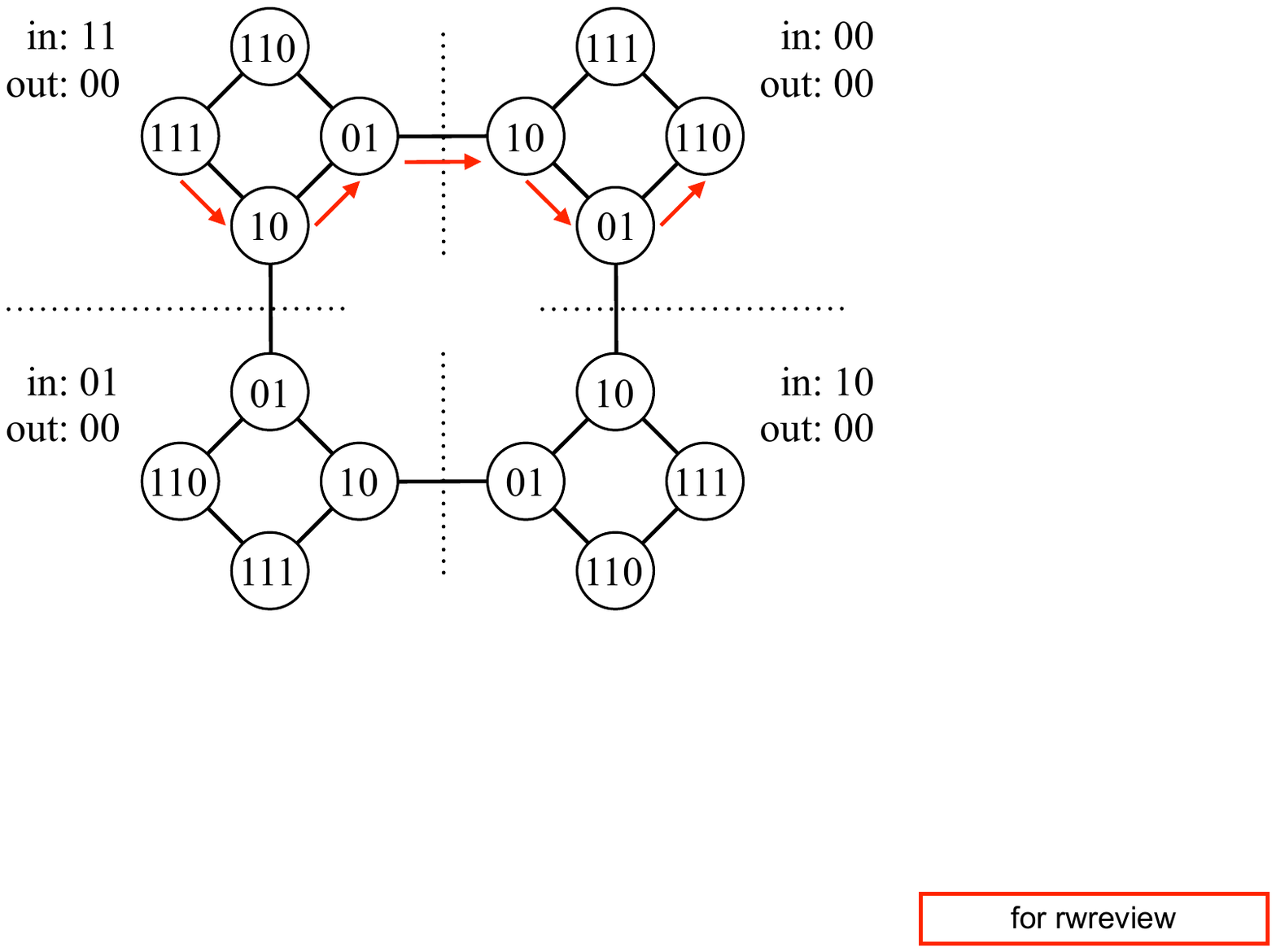}
\caption{Optimal partitioning from the InfoMap algorithm along with its resulting code words. We draw this example from a demonstration applet available at \cite{rosvall-website}.}
\label{fig:map equation demo flow}
\end{center}
\end{figure}

Let's consider the network in Fig.~\ref{fig:map equation demo flow}. The InfoMap algorithm partitions the network into four communities, whose boundaries we show with the dotted lines. The binary sequence at each node represents the local code word within the corresponding community. When a random walker enters or exits a community, one uses the corresponding ``in'' and ``out'' code word, respectively. For example, the trajectory indicated by the red arrows is encoded into 11 111 10 01 00 00 10 01 110. The first ``11'' indicates that the RW starts in the top left community, the subsequent ``111'' indicates that the walk starts at node ``111'' in this community, the ``00 00'' in the middle indicates that the walk exits this community (because of the first ``00'') and simultaneously enters the community to the right (because of the second ``00'').

In contrast to the original Huffman code, we need $2N_{\rm CM}$ additional code words to encode entry to and exit from communities. However, we can use a smaller code length when a random walker travels within a community because the code words local to a community are generally shorter than the code words of the original Huffman code. If a network has strong community structure, one expects that an RW within a community occupies a majority of steps if one optimally partitions the network into communities. Consequently, one expects the mean code length to be smaller using InfoMap than by using a straightforward Huffman code in networks with community structure. In practice, InfoMap optimizes a quality function, called the ``map equation''\index{map equation} (where the word ``equation'' is a misnomer), instead of constructing the optimized coding scheme. The map equation generalizes Eq.~\eqref{eq:Shannon entropy}. The resulting quality function provides a theoretical limit of how concisely one can encode an RW using a given partition. One can optimize this function using some computational heuristic.



\subsubsection{Local community detection}\label{sub:local}


Another approach to community detection is to use local algorithms. For example, given a node $v_i$ of interest, one can use a local algorithm to identify a relatively small community around $v_i$ by examining only the nodes that are adjacent to nodes that have been examined before. Local algorithms are particularly useful when a network is huge, and it is thus costly to apply a partitioning algorithm to the entire network. As discussed in Ref.~\cite{Jeub2015PhysRevE} (and in several references therein), they also provide a means to studying overlapping communities and to incorporate dynamical processes and seed sets into community detection.

Nibble is a local community-detection algorithm based on DTRWs \cite{Spielman2004STOC,Andersen2006FOCS,Spielman2013SiamJComput}. The idea is to examine nodes that are visited frequently by a random walker that starts from a node $v_i$. Specifically, Nibble uses the transition-probability matrix
\begin{equation}
	T_{\text{Nibble}} = \frac{D^{-1}A+I}{2}\,.
\label{eq:T Nibble}
\end{equation}
Equation~\eqref{eq:T Nibble} implies that a random walker obeys the usual DTRW with probability $1/2$ and does not move with probability $1/2$ in each time step. For each of the nodes, Nibble also reduces the probability of a visit to it to $0$ in each time step if it is smaller than some threshold.
Therefore, the probability that the random walker is still present in the network decreases in time.
The probability reduction ensures that the detected community does not become too large in a small number $n$ of steps. One terminates the DTRW after a certain number of steps according to a stopping criterion, which guarantees that the discovered set of nodes has a low conductance (see Eq.~\eqref{eq:Cheeger constant}) and is neither too small nor too large. Nibble can also be used as a building block for network-partitioning algorithms that run in $O(M)$ time \cite{Spielman2004STOC,Spielman2013SiamJComput}. (Recall that $M$ denotes the number of edges (see Table~\ref{tab:notation}).)





In the ``seed-set expansion problem'', one seeks to discover a local community that emanates from a small subset $S$ of a network's nodes.
 One expands the seed set to estimate the rest of a community by ranking the nodes outside $S$. Variants of personalized PageRank and heat-kernel PageRank are popular approaches for studying seed-set expansion \cite{Kloster2014SIGKDD,Kloumann2014SIGKDD,Kloumann2017PNAS}.
 Like Nibble, one starts a DTRW from a node $v_i \in S$, and one then examines $T^n_{ij}$, which gives the probability that a walker starting from $v_i$ visits node $v_j$ after $n$ steps.
 The score for $v_j$
  is given by a weighted sum of $T^n_{ij}$ over different lengths of walks. That is, the score is
 $\sum_{n=1}^{\infty} w_n T^n_{ij}$, where $w_n$ is the weight assigned to walks of length $n$ \cite{Kloumann2017PNAS}.


\subsubsection{Multilayer modularity} \label{multmod}

One can generalize Markov stability to multilayer networks to derive modularity functions for such networks, including temporal networks given in the form of a sequence of adjacency matrices (with interlayer edges that connect corresponding nodes in the sequence) \cite{Mucha2010Science,Bazzi2016MultModelSimul}.

As in Section~\ref{sec:multilayer}, consider a multilayer network in the (supra-adjacency) form of a weighted network on $N\ell_{\max}$ nodes, where $\ell_{\max}$ is the number of layers. One specifies a node by the pair $(v_i, \ell)$, where $i \in \{1, \dots,N\}$ indexes an entity and $\ell \in \{1, \dots, \ell_{\max}\}$ indicates a layer. The adjacency matrix in each layer $\ell$ (which can be, e.g., an aggregation over some time window of a temporal network) is $A(\ell)$, which we assume to be undirected for simplicity. The weight of the interlayer edge between nodes $(v_i, \ell)$ and $(v_i, {\ell}^{\prime})$ is $C_{i{\ell}^{\prime}\ell}$. We consider a multilayer network in which only nodes with the same index $i$ can be adjacent to each other, though multilayer networks also allow much more general structures \cite{Kivela2014JCompNetw}. (Note that an entity $v_i$ need not exist on all layers \cite{Mucha2010Science}.) For a multilayer network that represents a temporal network, the simplest choice is to connect the corresponding nodes (i.e., nodes with the same index $i$) across the adjacent layers symmetrically and uniformly, so $\omega = C_{i\ell {\ell}^{\prime}} = C_{i {\ell}^{\prime} \ell} >0$ when ${\ell}^{\prime} = \ell + 1$ for $\ell \in \{1, \dots, L-1\}$ and $C_{i\ell {\ell}^{\prime}} = 0$ for ${\ell}^{\prime}\neq \ell \pm 1$.

To derive an expression for multilayer modularity for these ``multislice'' networks, we generalize the RW interpretation of modularity for time-independent networks (see Section~\ref{sub:modularity}) to the case of multilayer networks \cite{Mucha2010Science}. Random walkers are allowed to move either between layers or within a layer. Consider a Poissonian node-centric CTRW on a multilayer network with $N \ell_{\max}$ nodes. The master equation is given by
\begin{equation}
	\frac{{\rm d}p_{i\ell}(t)}{{\rm d}t} = \sum_{{\ell}^{\prime}=1}^{\ell_{\max}} \sum_{j=1}^N
\frac{\left[A_{ij}({\ell}^{\prime}) \delta_{\ell {\ell}^{\prime}} + \delta_{ij}C_{j \ell {\ell}^{\prime}}\right] p_{j {\ell}^{\prime}}(t)}{\kappa_{j{\ell}^{\prime}}} - p_{i\ell}(t)\,,
\label{eq:master equation p_{it}}
\end{equation}
where $\kappa_{j{\ell}^{\prime}} = k_{j{\ell}^{\prime}} + c_{j{\ell}^{\prime}}$ is the strength of the $j$th node in the ${\ell}^{\prime}$th layer, $k_{j{\ell}^{\prime}} = \sum_{i=1}^N A_{ij}(\ell)$ is the intra-layer strength of the $j$th node in the ${\ell}^{\prime}$th layer, and $c_{j{\ell}^{\prime}} = \sum_{{\ell}^{\prime\prime}=1}^{\ell_{\max}} C_{j{\ell}^{\prime}{\ell}^{\prime\prime}}$ is the inter-layer strength of the same node.
The summand on the right-hand side of Eq.~\eqref{eq:master equation p_{it}} represents the rate at which a random walker moves from node $(v_j, {\ell}^{\prime})$ to node $(v_i, \ell)$. A move to $(v_i, \ell)$ is possible from the nodes $(v_j, \ell)$  in the same layer at a rate of $A_{ij}(\ell)/\kappa_{j{\ell}^{\prime}}$ and from the $i$th node in a different layer ${\ell}^{\prime}$ at a rate of $C_{j\ell {\ell}^{\prime}}/\kappa_{j{\ell}^{\prime}}$. If $C_{i\ell \ell^{\prime}} = C_{i\ell^{\prime} \ell}$ (with $i \in \{1, \dots,N\}$ and $\ell,\ell^{\prime} \in \{1, \dots, \ell_{\max}\}$),
the stationary density is given by
\begin{equation}
	p_{i\ell}^* = \frac{\kappa_{i\ell}}
{\sum_{{\ell}^{\prime}=1}^{\ell_{\max}} \sum_{i^{\prime}=1}^N \kappa_{i^{\prime}{\ell}^{\prime}}}
	\equiv \frac{\kappa_{i\ell}}{2\mu}\,.
\end{equation}

In the same manner as with monolayer networks, we examine the probability that a random walker visits node $(v_j, {\ell}^{\prime})$ at time $t=0$ and node $(v_i, \ell)$ at a small time $\Delta t$. Within the small time $\Delta t$, a walker initially at $(v_j, {\ell}^{\prime})$ can make at most a single step. Based on Eq.~\eqref{eq:master equation p_{it}}, the probability that the walker visits node $(v_j, {\ell}^{\prime})$ at time $0$ and node $(v_i, \ell)$ at small time $\Delta t$ is
\begin{equation}
	\left[\delta_{ij}\delta_{\ell {\ell}^{\prime}} +
\Delta t \left(\frac{A_{ij}(\ell) \delta_{\ell {\ell}^{\prime}} + \delta_{ij}C_{j\ell {\ell}^{\prime}}}{\kappa_{j{\ell}^{\prime}}} - \delta_{ij} \delta_{\ell {\ell}^{\prime}} \right) \right]
\frac{\kappa_{j{\ell}^{\prime}}}{2\mu}\,.
\label{eq:actual joint probability multilayer}
\end{equation}
Under the independence assumption, which sets the null model, the situation remains the same, but each intra-layer network is now replaced by a Newman--Girvan (NG) null network whose degree distribution is determined by the original set of adjacencies of the same layer \cite{Bazzi2016MultModelSimul}. The inter-layer transition probability, determined by $C_{j\ell {\ell}^{\prime}}$, remains the same. Under the independence assumption, the probability that a walker visits node $(v_j, {\ell}^{\prime})$ at time $t=0$ and node $(v_i, \ell)$ after a single move is
\begin{equation}
	\left( \frac{k_{i\ell}}{2M_{\ell}}\frac{k_{j{\ell}^{\prime}}}{\kappa_{j{\ell}^{\prime}}}\delta_{\ell {\ell}^{\prime}} + \delta_{ij}\frac{C_{j\ell {\ell}^{\prime}}}{c_{j{\ell}^{\prime}}}\frac{c_{j{\ell}^{\prime}}}{\kappa_{j{\ell}^{\prime}}} \right)
	\frac{\kappa_{j{\ell}^{\prime}}}{2\mu}\,,
\label{eq:rho_{it|jt'}p_{jt'}^*}
\end{equation}
where $M_{\ell} = \sum_{j=1}^N k_{j\ell}$.
In Eq.~\eqref{eq:rho_{it|jt'}p_{jt'}^*}, $\kappa_{j{\ell}^{\prime}}/(2\mu)$ is the probability that the random walker visits $(v_j,{\ell}^{\prime})$ at time $0$ at equilibrium. The quantity in parentheses represents the conditional probability that a walker visits node $(v_i, \ell)$ after a single move starting from node $(v_j, {\ell}^{\prime})$ at time $0$. A move occurs within the ${\ell}^{\prime}$th layer with probability $k_{j{\ell}^{\prime}}/\kappa_{j{\ell}^{\prime}}$. If an intra-layer move occurs, the walker moves to the $i$th node in the same layer with probability $k_{i{\ell}^{\prime}}/(2M_{{\ell}^{\prime}})$ according to the NG null model. Alternatively, the walker moves to a different layer with probability $c_{j{\ell}^{\prime}}/\kappa_{j{\ell}^{\prime}} = 1 - k_{j{\ell}^{\prime}}/\kappa_{j{\ell}^{\prime}}$. If an inter-layer move occurs, the walker moves to the $j$th node in the $\ell$th layer with probability $C_{j\ell {\ell}^{\prime}}/c_{j{\ell}^{\prime}}$.

By subtracting Eq.~\eqref{eq:rho_{it|jt'}p_{jt'}^*} from Eq.~\eqref{eq:actual joint probability multilayer} and then summing over nodes $(v_i, \ell)$ and $(v_j, {\ell}^{\prime})$ that belong to the same community, we obtain
\begin{align}
	Q = \frac{1}{2\mu} \sum_{i,j,\ell,{\ell}^{\prime}} \left[
(1-\Delta t)\delta_{ij}\delta_{\ell {\ell}^{\prime}} \kappa_{j \ell^{\prime}} + \Delta t A_{ij}(\ell)\delta_{\ell{\ell}^{\prime}} - \frac{k_{i\ell}k_{j{\ell}^{\prime}}}{2M_{\ell}}\delta_{\ell{\ell}^{\prime}}
	+ (\Delta t-1) \delta_{ij} C_{j\ell{\ell}^{\prime}}\right]
	  \times \delta(g_{i\ell}, g_{j{\ell}^{\prime}})\,,
\label{eq:Q multilayer Delta tilde(t)}
\end{align}
where $g_{i\ell}$ is the community to which node $(v_i, \ell)$ has been assigned. Because $\sum_{i,j,\ell,{\ell}^{\prime}} \delta_{ij}\delta_{\ell{\ell}^{\prime}} \kappa_{j \ell^{\prime}} \delta(g_{i\ell},g_{j{\ell}^{\prime}}) = \sum_{i, \ell} \kappa_{i \ell}$ is independent of the partitioning of the multilayer network and thus does not affect the maximization of $Q$, we ignore the first term on the right-hand side of Eq.~\eqref{eq:Q multilayer Delta tilde(t)}. By rescaling $C_{j\ell{\ell}^{\prime}}$ by a multiplicative factor of $(\Delta t-1)/\Delta t$, we can also ignore $(\Delta t - 1)$ in the fourth term. If we allow $\gamma \equiv 1/\Delta t$ to depend on the layer (see \cite{Mucha2010Science} for the justification), corresponding to different diffusion rates in different layers, we obtain the following formula for multilayer modularity:
\begin{equation}
	Q = \frac{1}{2\mu} \sum_{i,j,\ell,{\ell}^{\prime}}\left[ A_{ij}(\ell) \delta_{\ell {\ell}^{\prime}} - \gamma(\ell) \frac{k_{i\ell}k_{j{\ell}^{\prime}}}{2M_{\ell}}\delta_{\ell{\ell}^{\prime}}
+ \delta_{ij} C_{j\ell{\ell}^{\prime}}\right] \delta(g_{i\ell}, g_{j{\ell}^{\prime}})\,.
\label{eq:Q multilayer final}
\end{equation}

For simplicity, suppose that the inter-layer edge weight is uniform; that is, $\omega = C_{i\ell{\ell}^{\prime}}$ for any $i$, $\ell$, and $\ell^{\prime}$ whenever entity $v_i$ exists in both layers. If an entity $v_i$ does not exist in a layer, its associated interlayer edges have weight $0$ because they do not exist. If $\omega=0$, the different layers are independent networks. If $\omega$ is sufficiently large, all existing copies $(v_i,\ell)$ of each node $v_i$ (with $\ell \in \{1, \dots, \ell_{\max}\}$) are assigned to the same community because the third term on the right-hand side of Eq.~\eqref{eq:Q multilayer final} dominates the others. More generally, a large value of $\omega$ tends to yield a smaller number of communities. In contrast, a large $\gamma(\ell)$ value tends to yield a large number of communities. See Refs.~\cite{bassett2013chaos,Bazzi2016MultModelSimul,sarzynska2016,muchaporter2010} for illustrations and discussions.


\subsection{Core--periphery structure\label{sub:core-periphery}}

It is often insightful to decompose a network into one or more densely-connected cores along with sparsely-connected peripheral nodes. By definition, nodes in a core are heavily interconnected and also tend to be well-connected to peripheral nodes. By contrast, peripheral nodes are sparsely connected (or, ideally, not adjacent at all) to other peripheral nodes and tend to be adjacent predominantly to core nodes. This idea, whose intuition draws somewhat on the notion of pealing an onion (especially in the case of a single core), is also a mesoscale network structure, but it has a rather different character from community structure. See Ref.~\cite{csermely} for a review of core--periphery, and see the introduction of Ref.~\cite{Rombach2014SiamJApplMath} for a brief survey.

There is an RW-based algorithm to extract core--periphery structure from networks \cite{Dellarossa2013SciRep}. The idea is that if a random walker is located at a peripheral node, it is very unlikely to visit another peripheral node in the next time step in a DTRW. One defines a ``persistence probability'' $\alpha_S$ for a set of nodes $S$ by
\begin{equation}
	\alpha_S = \frac{\sum_{i, j\in S}p_i^*T_{ij}}{\sum_{i\in S}p_i^*}\,,
\label{eq:def persistence probability}
\end{equation}
where we recall that $p_i^*$ is the stationary density at node $v_i$, and $T_{ij}$ is the transition probability from $v_i$ to $v_j$ in a single move. Equation \eqref{eq:def persistence probability} is the steady-state probability that a DTRW starting from a node in $S$ remains in $S$ in the next time step. For an undirected network, we substitute $p_i^* = s_i/\sum_{\ell=1}^N s_{\ell}$ to reduce Eq.~\eqref{eq:def persistence probability} to
\begin{equation}
	\alpha_S = \frac{\sum_{i, j\in S}A_{ij}} {\sum_{i\in S}s_i}\,.
\end{equation}

Ideally, one obtains $\alpha_S=0$ for any set $S$ of nodes that includes only peripheral nodes. This condition is trivially satisfied when $S$ consists of a single node, and it becomes very difficult to satisfy as $S$ becomes large. Reference \cite{Dellarossa2013SciRep} used the following greedy algorithm. Start from a node with the smallest total node strength $s_i^{\rm in} + s_i^{\rm out}$. If there are multiple such nodes, we select one of them uniformly at random. For undirected networks, this reduces to selecting a node with the minimum node strength. The set $S$ is composed of a single node. One then adds one node to the set $S$ so that adding this node yields the smallest value of $\alpha_S$. Again, if there are multiple candidate nodes, we break the tie by selecting one of them uniformly at random. One continues this procedure and sequentially adds nodes to try to keep $\alpha_S$ small. One then assigns each node $v_i$ a coreness value of $\alpha_i$, which one sets as the value of $\alpha_S$ when $v_i$ is added. Nodes with larger values of $\alpha_i$ are deeper into a network core. One also defines a network's ``$\alpha$-periphery'' as the set of nodes that satisfy $\alpha_i \le \alpha$. Although the algorithm has randomness in it because of the tie-breakers, Ref.~\cite{Dellarossa2013SciRep} reported that the randomness had negligible effects on their results for empirical networks.




\subsection{Diffusion maps}

Dimension reduction is a type of compression that has numerous practical applications in data mining, image processing, visualization, and many other subjects \cite{LeeVerleysen2007book}.
Its aim is to find a transformation of a set of data points into a low-dimensional space in a way that preserves quantities of interest, such as distances between any pair of data points, preferably with a small number of free parameters.
``Diffusion maps'' are a framework of RW-based dimension reduction and encompass a wide variety of methods, such as kernel eigenmap methods, as special cases \cite{Coifman2005PNAS-1,Coifman2006ApplComputHarmonAnal}.
Diffusion maps are also useful for identifying synchronous clusters of nodes in synchronization dynamics \cite{Dedomenico2017PhysRevLett}.


Consider a DTRW on an undirected, weighted network constructed from a given set of data points, which one identifies with nodes.
The edge weight between nodes $v_i$ and $v_j$ is $A_{ij} =A_{ji}$, and it is given by a similarity value between the $i$th and $j$th data points.
In our terminology, the ``diffusion distance'' is defined by
\begin{align}
	\overline{d}_{ij}(n) &= \sqrt{\sum_{\ell=1}^N \frac{\left(T^n_{i\ell} - T^n_{j\ell}\right)^2} {p_\ell^*}}\notag\\
&= \sqrt{\sum_{\ell=1}^N \frac{\left(T^n_{i\ell} - T^n_{j\ell}\right)^2} {s_\ell} \times \sum_{\ell=1}^N s_{\ell}}\,,
\label{eq:diffusion distance}
\end{align}
which is the same as the distance measure used in the Walktrap algorithm, except for the normalization (see Eq.~\eqref{eq:r_ij Walktrap}).
Because $\overline{d}_{ij}(n)$ involves the summation of all walks of length $n$ starting from $v_i$ and the summation of such walks starting from $v_j$, Refs.~\cite{Coifman2005PNAS-1,Coifman2006ApplComputHarmonAnal} suggested that it is more robust to noise in data than when using $A_{ij}$ as a similarity or distance measure for dimension reduction.


Substituting Eq.~\eqref{eq:T^n eigen expansion} into Eq.~\eqref{eq:diffusion distance} yields
\begin{align}
	\overline{d}_{ij}(n) &=
\sqrt{\sum_{\ell=1}^N \frac{ \left[
\sum_{\ell^{\prime}=1}^N \lambda_{\ell^{\prime}}^n
\left(\frac{(u_{\ell^{\prime}})_i}{\sqrt{s_i}} - \frac{(u_{\ell^{\prime}})_j}{\sqrt{s_j}}\right)
(u_{\ell^{\prime}})_{\ell} \sqrt{s_{\ell}} \right]^2} {s_{\ell}}
\times \sum_{\ell=1}^N s_{\ell}}
\notag\\
&= \sqrt{\sum_{\ell=1}^N \left[
\sum_{\ell^{\prime}=1}^N \lambda_{\ell^{\prime}}^n
\left(\frac{(u_{\ell^{\prime}})_i}{\sqrt{s_i}} - \frac{(u_{\ell^{\prime}})_j}{\sqrt{s_j}}\right)
(u_{\ell^{\prime}})_{\ell} \right]^2
\times \sum_{\ell=1}^N s_{\ell}}\,,
\label{eq:diffusion distance 2}
\end{align}
where $\bm u_{\ell^{\prime}}$ is the eigenvector corresponding to the $\ell^{\prime}$th eigenvalue of $\tilde{A}$ (see Eq.~\eqref{eq:def tilde A}) and $\lambda_{\ell^{\prime}}$ is the $\ell^{\prime}$th largest eigenvalue of $\tilde{A}$ in terms of absolute value. Note that $\lambda_1 = 1$.
Using $\langle \bm u_{\ell^{\prime}}, \bm u_{\ell^{\prime\prime}}\rangle = \delta_{\ell^{\prime} \ell^{\prime\prime}}$, Eq.~\eqref{eq:diffusion distance 2} reduces to
\begin{align}
	\overline{d}_{ij}(n) &=
\sqrt{\sum_{\ell^{\prime}=1}^N \lambda_{\ell^{\prime}}^{2n}
\left(\frac{(u_{\ell^{\prime}})_i}{\sqrt{s_i}} - \frac{(u_{\ell^{\prime}})_j}{\sqrt{s_j}}\right)^2
\times \sum_{\ell=1}^N s_{\ell}}\,\notag\\
	&= \sqrt{\sum_{\ell^{\prime}=2}^N \lambda_{\ell^{\prime}}^{2n}
\left(\frac{(u_{\ell^{\prime}})_i}{\sqrt{s_i}} - \frac{(u_{\ell^{\prime}})_j}{\sqrt{s_j}}\right)^2
\times \sum_{\ell=1}^N s_{\ell}}\,.
\label{eq:diffusion distance 3}
\end{align}
To derive the last line in Eq.~\eqref{eq:diffusion distance 3}, we used
$\bm u_1 = (\sqrt{s_1}, \ldots, \sqrt{s_N})^{\top}$, corresponding to the stationary density (see Section~\ref{sub:relaxation time}).
By neglecting eigenmodes whose contributions are much smaller than the largest eigenmode in Eq.~\eqref{eq:diffusion distance 3} (i.e., $\bm u_2$), one defines a diffusion map by
\begin{equation}
	\Psi(i; n) = \frac{1}{\sqrt{s_i}}
\begin{pmatrix}
\lambda_2^n (u_2)_i\\ \vdots \\ \lambda_{\tilde{\ell}}^n (u_{\tilde{\ell}})_i
\end{pmatrix}\,,
	\label{eq:diffusion distance Psi}
\end{equation}
where $\tilde{\ell}$ is the largest index $\ell^{\prime}$ such that $\left|\lambda_{\ell^{\prime}}\right|^n > \overline{\delta} \left| \lambda_2\right|^n$, and $\overline{\delta}$ is a parameter.
Each component of $\Psi(i; n)$ is called a ``diffusion coordinate''. Equations~\eqref{eq:diffusion distance 3} and \eqref{eq:diffusion distance Psi} imply that, in ${\mathbb R}^{\tilde{\ell}-1}$, the Euclidean distance between two data points $i$ and $j$ is equal to the diffusion distance $\overline{d}_{ij}(n)$ with a tolerance of $\overline{\delta}$.



The properties of diffusion maps depend on the parameters $n$ and $\overline{\delta}$. A large value of $\overline{\delta}$ yields a small value of $\tilde{\ell}$ and hence results in a large dimension reduction. A diffusion map with a larger value of $n$ extracts geometry on a more global scale than one with a smaller value of $n$, so a collection of diffusion maps for different values of $n$ allows one to describe a data set with multliscale geometric properties.


\subsection{Respondent-driven sampling}

One often is interested in estimating a population mean of certain quantities, such as the fraction of infected individuals, the fraction of people who have a particular opinion, or demographics such as age. If a population is large, which is typical in the context of social surveys, it is impossible to record all individuals. In such situations, a common challenge is how to sample individuals in as unbiased manner as possible.

``Respondent-driven sampling'' (RDS) is a popular sampling method that uses edge-tracing in a social network \cite{Heckathorn1997SocProb,Salganik2004SocMet}. In RDS, one starts from a seed individual (i.e., a seed node). The seed individual recruits his/her neighbors to a survey by passing a coupon to each of them. The successfully recruited individuals then participate in the survey and in turn pass coupons to their neighbors who have not yet participated. To try to promote participation, individuals who participate are rewarded financially. One takes a weighted mean of the samples to derive an estimate of the quantity of interest (e.g., mean age of a population).

It is necessary to take a weighted mean because the probability of being recruited depends on the position of a person in a network. The so-called ``RDS II estimator'' is an efficient and realistic estimator \cite{Volz2008JOffStat}. Consider the case in which each respondent passes a single coupon to one of its uniformly randomly selected neighbors. One can then describe the recruitment process as a DTRW if one allows sampling with replacement for simplicity (i.e., if the same individual can be sampled more than once). Again for simplicity, let's also assume that the network is undirected and unweighted. The essential idea of the RDS II estimator is that one should discount the effect of a sampled node $v_i$ by a factor of its degree $k_i$, because $v_i$ is visited with probability $p_i^* \propto k_i$. Note that respondents have to report $k_i$ to be able to calculate this estimator, although empirically it is difficult to accurately collect the $k_i$ values of respondents \cite{Mccarty2001HumanOrg,Marsden2005chapter}.

We are interested in estimating the mean $\langle y\rangle$ of a quantity $y_i$ assigned to node $v_i$. We denote the set of sampled nodes by $S$ and the number of samples (i.e, the size of $S$) by $N_S$.
The estimator $\langle \hat{y}\rangle$ of $\langle y\rangle$ is
\begin{equation}
	\langle \hat{y} \rangle = \frac{1}{N_S} \sum_{v_i\in S} \frac{y_i}{N \hat{p}_i^*}\,,
\label{eq:<hat(y)>}
\end{equation}
where $\hat{p}_i^*$ is the estimate of the stationary density $p_i^*$. We set the discount factor on the right-hand side of Eq.~\eqref{eq:<hat(y)>} to be $N \hat{p}_i^*$, because it is normalized so that $\langle N \hat{p}_i^*\rangle = 1$. By assuming that we do not have access to the mean degree $\langle k\rangle$ of the entire network, we estimate it by calculating
\begin{equation}
	\hat{p}_i^* = \frac{k_i}{N\langle \hat{k}\rangle}\,,
\label{eq:hat(p_i^*)}
\end{equation}
where $\langle \hat{k}\rangle$ is an estimate of $\langle k\rangle$. We use
\begin{equation}
	\langle \hat{k}\rangle = \frac{\sum_{v_i\in S}\frac{k_i}{Np_i^*}}
{\sum_{v_i\in S}\frac{1}{Np_i^*}} = \frac{N_S}{\sum_{v_i\in S} \left(k_i\right)^{-1}}\,.
\label{eq:<hat(k)>}
\end{equation}
Combining Eqs.~\eqref{eq:<hat(y)>}, \eqref{eq:hat(p_i^*)}, and \eqref{eq:<hat(k)>} yields
\begin{equation}
	\langle \hat{y} \rangle = \frac{\sum_{v_i\in S}\left(k_i\right)^{-1}y_i}
{\sum_{v_i\in S}\left(k_i\right)^{-1}}\,.
\label{eq:<hat(y)> final}
\end{equation}

The estimated quantity $y$ can be either continuous-valued or discrete-valued.
Alternatively, one can
estimate the proportion of nodes $P_A$ that have a discrete type $A$ (e.g., an infected state) by setting $y_i$ to the indicator function (i.e., $y_i=1$ when $v_i$ is of type $A$ and $y_i=0$ otherwise). In this case, we obtain
\begin{equation}
	\hat{P}_A = \frac{\sum_{v_i\in A\cap S}(k_i)^{-1}} {\sum_{v_i\in S} (k_i)^{-1}}\,.
\end{equation}

Note that, even if one controls for the effect of $p_i^*$ in this manner, the estimator $\langle y\rangle$ is statistically biased in practice. For example, the estimator is inaccurate when networks have community structure \cite{Rocha2017ProcRStatSocA} or have multiple connected components \cite{Malmros2016arxiv}. Additionally, different techniques are required for directed networks, because Eq.~\eqref{eq:hat(p_i^*)} (or, more succinctly, $p_i^*\propto k_i$) does not hold for directed networks \cite{LuBengtsson2012JRStatSocA,Malmros2016JOffStat}.
Furthermore, actual sampling trajectories are non-backtracking, and one can incorporate this feature into RDS estimators \cite{Gile2011JAmStatAssoc}.

A strategy other than RDS II or other estimators of unbiased sampling of nodes is to use a ``Metropolis--Hasting RW'' \cite{Gilks1996book}. In such sampling, one modifies the edge weight of the original network to guarantee that the stationary density is the uniform density. This method has been used for sampling in peer-to-peer (P2P) and online social networks \cite{Lovasz1993Boyal,Stutzbach2009IeeeTransNetw,Gjoka2010IeeeInfocom}.


\subsection{Consensus probability and time of voter models\label{sub:voter model}}

\begin{figure}[tb]
\begin{center}
\includegraphics[scale=0.5]{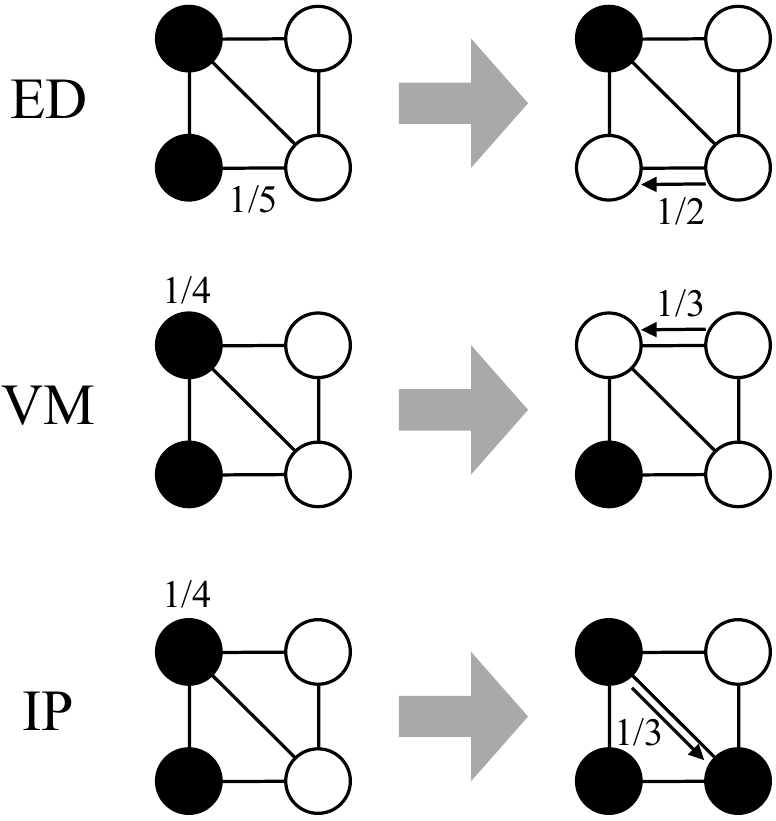}
\caption{Three updating rules for variants of the classical voter model on a network. For illustration, assume that we have an undirected and unweighted network. With edge-dynamics (ED), one first selects one of the $M=5$ edges with equal probability (i.e., with probability $1/5$ each). One then selects one of the two directions of the edge with equal probability $1/2$, and then one performs an opinion-updating step. In the most traditional voter model (VM), which has node dynamics, one selects one of the $N=4$ nodes with equal probability $1/4$. One then determines uniformly at random the neighbor from which the selected node imports its opinion. In the invasion process (IP), one first selects one of the $N=4$ nodes with equal probability $1/4$ (as in the VM). One then determines uniformly at random the neighbor to which the selected node exports its opinion.}
\label{fig:voter rules}
\end{center}
\end{figure}

Voter models are a prototypical family of models of opinion formation that are often defined in terms of a Markov process on a network \cite{Liggett1985book,Durrett1988book,Aldous2002book,Castellano2009RevModPhys,Redner2001book,Krapivsky2010book,Porter2016book}. In traditional voter models, each node assumes one of two opinions, which we call opinion $0$ and and opinion $1$, and the nodes' opinions evolve stochastically in time.
If two adjacent nodes have the opposite opinion, a local consensus of opinion 0 or opinion 1 between the two nodes occurs at some rate. We suppose that the local consensus dynamics on each edge obeys an independent Poisson process, so the nodes update their opinions asynchronously. For example, if a local consensus on the edge $(v_i, v_j)$ in an undirected network occurs according to a Poisson process at rate $\propto A_{ij}$, we say that voter dynamics obeys ``edge dynamics'' (ED) (see Fig.~\ref{fig:voter rules}) \cite{Antal2006PhysRevLett,Sood2008PhysRevE}.
(Note that people often use the term ``link dynamics'' (LD), because it is common in physics to use the term ``link'' for ``edge''.) On finite networks, the final state of a network is the perfect ``consensus'' of either opinion 0 or opinion 1 for every node. These two consensus configurations are the only absorbing states of the stochastic process. Note that consensus is sometimes also called ``fixation'' or ``coordination''.

The best-studied phenomena in voter models include the probability for a network to achieve consensus of a particular opinion and the mean time to achieve consensus. The consensus probability is the probability that a consensus of one opinion (e.g., opinion 0) is reached. With the complementary probability, a finite network achieves a consensus of the other opinion (e.g., opinion 1). When computing mean consensus time, one conditions on the consensus being reached. Both consensus probability and mean consensus time depend both on the initial configuration of opinions and on network structure.

The duality relationship between voter models and ``coalescing RWs'' (which are non-conservative) makes analysis of RWs a powerful approach for calculating consensus probability and mean consensus time \cite{Donnelly1983MPCPS,Liggett1985book,Durrett1988book,Aldous2002book}. By definition, a coalescing RW \cite{Griffeath1978ZWahr} starts by placing a random walker on each node in a network, and the walkers perform independent Poissonian edge-centric CTRWs. If different walkers meet at a node, they coalesce into one and continue as a single random walker. On a finite network, all walkers eventually coalesce into a single random walker.

When examining the dual process, we invert the time and direction of edges \cite{Donnelly1983MPCPS,Liggett1985book,Durrett1988book,Aldous2002book}. When proceeding backwards in time, two individuals sometimes ``collide'' in the dual process. Such a coalescence
event corresponds to two individuals sharing a common ancestor in the original opinion-formation process. After two individuals coalesce in the dual process, they behave as a single individual.

The duality relationship guarantees that the consensus probability $F_i$ for opinion 0 when node $v_i$ initially has opinion 0 and the other $N-1$ nodes initially have opinion 1 is given by the stationary density of the coalescing RW on the network that one obtains by reversing all edges in an original network. Because all walkers eventually coalesce into a single walker, $F_i$ is given by the stationary density of the usual RW on the edge-reversed network. If initially there are multiple nodes with opinion 0, then the consensus probability for opinion 0 is equal to the sum of $F_i$ over the nodes with initial opinion 0. The mean consensus time is equal to the mean time needed for all walkers to coalesce into one walker. This equality is useful for evaluating the mean consensus time for some networks, because the latter quantity is roughly approximated by the mean time for the first meeting of two independent walkers whose initial location is selected uniformly at random \cite{CooperElsasser2013SiamJDiscMath,Masuda2014PhysRevE,Iwamasa2014PhysRevE}. Similar to the MFPT, the mean time for two random walkers to meet is relatively easy to calculate.

Consider a directed network. As a convention, we assume that the directed edge from $v_i$ to $v_j$ indicates that $v_i$ can coax $v_j$ into $v_i$'s opinion. Even if the network is undirected, one has to distinguish three rules of opinion updating unless the network is regular
\cite{Antal2006PhysRevLett,Sood2008PhysRevE} (see Fig.~\ref{fig:voter rules}). We evaluate the consensus probability for these three types of voter dynamics using the duality relationship \cite{Donnelly1983MPCPS,MasudaOhtsuki2009NewJPhys}.

First, let's consider a variant of the voter model that focuses on the dynamics of edges \cite{Antal2006PhysRevLett,Sood2008PhysRevE}. Under these ``edge dynamics'' (ED), one selects a directed edge $v_i \to v_j$ (i.e., from node $v_i$ to node $v_j$) with probability $A_{ij}/\sum_{i^{\prime}, j^{\prime}=1}^N A_{i^{\prime}j^{\prime}}$ in each step, and then node $v_j$ copies $v_i$'s opinion with probability 1. One then advances time by $1/N$, so each node is updated once per unit time on average. The dynamics are equivalent to opinion dynamics in which each edge has a Poisson process with rate
$NA_{ij}/\sum_{i^{\prime}, j^{\prime}=1}^N A_{i^{\prime}j^{\prime}}$, and an event induces a local consensus event.
The dual process for ED is a coalescing RW on the edge-reversed network in continuous time. (In fact, it is a Poissonian edge-centric CTRW.) By modifying Eq.~\eqref{eq:master unnormalized CTRW}, a single random walker satisfies the following master equation:
\begin{equation}
	\frac{{\rm d}{\bm p}(t)}{{\rm d}t} = \bm p(t) (-D^{\rm rev}+A^{\top}) = - \bm p(t) L^{\rm rev}\,,
\label{eq:master LD}
\end{equation}
where $A^{\top}$ is the adjacency matrix of the edge-reversed network, $D^{\rm rev}$ is the diagonal matrix whose $(i, i)$th element is $s_i^{\rm in}$, and $L^{\rm rev}$ is the combinatorial Laplacian of the edge-reversed network. The consensus probability $F_i^{\rm ED}$ for each node is given by the equilibrium of Eq.~\eqref{eq:master LD}. That is,
\begin{equation}
	(F_1^{\rm ED},\; \ldots\; ,F_N^{\rm ED}) L^{\rm rev} = 0\,.
\label{eq:dual LD eigenvector}
\end{equation}

We can obtain an intuitive understanding of Eq.~\eqref{eq:dual LD eigenvector} by writing a recursive equation for the consensus probability when the process starts from a single node $v_i$ with opinion 0 (i.e., for $F_i^{\rm ED}$).
We obtain
\begin{equation}
	F_i^{\rm ED} =\sum_{j=1}^N \frac{A_{ij}}{\sum_{i^{\prime}, j^{\prime}=1}^N A_{i^{\prime} j^{\prime}}}
F_{\{i,j\}}^{\rm ED} +
\frac{\sum_{j=1}^N A_{ji}}{\sum_{i^{\prime}, j^{\prime}=1}^N A_{i^{\prime} j^{\prime}}}
\times 0 +
\frac{\sum_{i^{\prime}, j^{\prime} = 1; i^{\prime}\neq i, j^{\prime}\neq i}^N A_{i^{\prime} j^{\prime}}}{\sum_{i^{\prime}, j^{\prime}=1}^N A_{i^{\prime} j^{\prime}}}
	F_i^{\rm ED}\,,
\label{eq:Fi LD elementary}
\end{equation}
where $F_{\{i,j\}}^{\rm ED}$ is the probability that one reaches the consensus of opinion 0 starting from the configuration in which $v_i$ and $v_j$ but no other nodes have opinion 0. To prove that $F_{\{i,j\}}^{\rm ED} =F_i^{\rm ED}+F_j^{\rm ED}$, imagine that there are $N$ different opinions rather than two, and suppose that node $v_i$ (with $i \in \{1,\; \dots \; , N\}$) holds opinion $i$. One can express the probability that opinion $i$ or $j$ eventually occupies the entire network either as $F_{\{i,j\}}^{\rm ED}$ or as
$F_i^{\rm ED}+F_j^{\rm ED}$, so it follows that $F_{\{i,j\}}^{\rm ED} = F_i^{\rm ED}+F_j^{\rm ED}$. By substituting the latter relationship into Eq.~\eqref{eq:Fi LD elementary}, we obtain
\begin{equation}
	\sum_{j=1}^N A_{ij}F_j^{\rm ED} =F_i^{\rm ED} \sum_{j=1}^N A_{ji}\,,
\label{eq:dual LD}
\end{equation}
and we note that Eq.~\eqref{eq:dual LD} is equivalent to Eq.~\eqref{eq:dual LD eigenvector}.

The quantity $F_i^{\rm ED}$ is the stationary density of the Poissonian edge-centric CTRW on the edge-reversed network.
If the network is undirected, we obtain $L^{\rm rev} = L$ and $p_i^* = F_i^{\rm ED} = 1/N$ (with $i \in \{1,\dots, N\}$). Therefore, the likelihood of propagating an opinion does not depend on which node is the seed of the opinion. If the network is directed, we obtain a first-order approximation to the consensus probability of a node by applying Eq.~\eqref{eq:first-order kin/kout} for the edge-reversed network \cite{MasudaOhtsuki2009NewJPhys}:
\begin{equation} \label{eq:F_i^LD meanfield}
	F_i^{\rm ED} \approx \text{(const)} \times \frac{s_i^{\rm out}}{s_i^{\rm in}}\,.
\end{equation}
Equation~\eqref{eq:F_i^LD meanfield} is intuitive, because an out-edge indicates that $v_i$ can enforce its opinion on another node, and an in-edge indicates that $v_i$ listens to neighboring nodes.

In the traditional node-based ``voter model'' (VM) updating rule, one selects a node $v_i$ uniformly at random (i.e., with equal probability $1/N$) in each time step. One then selects an in-neighbor $v_j$ of $v_i$ with a probability that is proportional to the weight of the in-edge from that node (i.e., $=A_{ji}/s_i^{\rm in}$), and $v_i$ copies the opinion of $v_j$ with probability $1$. One then advances time by $1/N$ so that on average one node experiences one opinion update per unit time. One can map the dynamics of the VM updating rule to ED dynamics with a modified weighted adjacency matrix $A (D^{\rm rev})^{-1}$, whose $(i, j)$th element is equal to $A_{ij}/s_j^{\rm in}$. The master equation for a single random walker on the edge-reversed network is thus
\begin{equation}
	\frac{{\rm d}{\bm p}(t)}{{\rm d}t} = \bm p(t) (- I + (D^{\rm rev})^{-1}A^{\top})\,.
\label{eq:master VM}
\end{equation}
The equilibrium of the dynamics given by Eq.~\eqref{eq:master VM} gives the consensus probability $F_i^{\rm VM}$ for opinion $0$ when only node $v_i$ initially has opinion $0$. By setting the left-hand side of Eq.~\eqref{eq:master VM} to $0$, we obtain
\begin{equation}
	(F_1^{\rm VM},\; \ldots\; ,F_N^{\rm VM}) =
(F_1^{\rm VM},\; \ldots\; ,F_N^{\rm VM}) (D^{\rm rev})^{-1} A^{\top}\,,
\end{equation}
which is equal to the stationary density of a DTRW on the edge-reversed network. Because Eqs.~\eqref{eq:master LD} and \eqref{eq:master VM}, respectively, represent a Poissonian edge-centric CTRW and a DTRW on the same network, we obtain
\begin{equation}
	F_i^{\rm VM} = s_i^{\rm in} F_i^{\rm ED}
\end{equation}
for arbitrary networks (Section~\ref{sub:node-centric vs edge-centric}). When a network is undirected, the edge-reversed network is the same as the original network, and we thereby see that
\begin{equation}
	F_i^{\rm VM} = \frac{s_i}{\sum_{\ell}=1}^N s_{\ell}\,.
\end{equation}
When a network is directed, the first-order approximation is given by
\begin{equation}
	F_i^{\rm VM} \propto s_i^{\rm out}\,.
\end{equation}

In the so-called ``invasion process'' (IP) updating rule, one first selects a node $v_i$ uniformly at random (i.e., with probability $1/N$) at each time step to propagate its opinion to one of its out-neighbors. One then selects an out-neighbor $v_j$ of $v_i$ with probability $A_{ij}/s_i^{\rm out}$ (i.e., uniformly at random), and then node $v_j$ copies the opinion of $v_i$ with probability $1$. One then advances time by $1/N$. One can map IP dynamics to ED dynamics with the modified weighted adjacency matrix $D^{-1}A$, whose $(i, j)$th element is equal to $A_{ij}/s_i^{\rm out}$. The master equation for a single walker in the edge-reversed network is
\begin{equation}
	\frac{{\rm d}{\bm p}(t)}{{\rm d}t} = \bm p(t) (-  D^{\rm IP} + A^{\top}D^{-1})\,,
\label{eq:master IP}
\end{equation}
where $D^{\rm IP}$ is the diagonal matrix whose $(i, i)$th element is given by
$\sum_{j=1}^N \left(A_{ji}/s_j^{\rm out}\right)$. The consensus probability $F_i^{\rm IP}$ satisfies
\begin{equation}
	(F_1^{\rm IP},\; \ldots\;, F_N^{\rm IP}) = (F_1^{\rm IP},\; \ldots\; ,F_N^{\rm IP}) A^{\top} D^{-1} (D^{\rm IP})^{-1}\,.
\label{eq:p^* IP}
\end{equation}
For an undirected network, $p_i^* \propto 1/s_i$ solves Eq.~\eqref{eq:p^* IP}, so nodes with small strengths are good at disseminating their opinions. For a directed network, the first-order approximation to Eq.~\eqref{eq:p^* IP} is
\begin{align}
	F_i^{\rm IP} &= \sum_{j=1}^N \frac{F_j^{\rm IP} A_{ij}/s_i^{\rm out}}{\sum_{\ell=1}^N A_{\ell i}/s_{\ell}^{\rm out}} \notag \\
		&\approx \sum_{j=1}^N \frac{(\text{const})\times A_{ij}/s_i^{\rm out}}{ \sum_{\ell=1}^N A_{\ell i}/(\text{const})} \notag \\
	&\propto \frac{1}{s_i^{\rm in}}\,.
\end{align}


\subsection{DeGroot model} \label{sub:DeGroot model}

The ``DeGroot model'' is a linear deterministic model that describes opinion-formation dynamics towards consensus \cite{Abelson1964chapter,Degroot1974JAmStatAssoc,Jackson2008book}. Control theorists have studied it as an example of a decentralized consensus algorithm (or protocol) \cite{Olfatisaber2007IEEE}. Although the DeGroot model is not usually discussed as an application of RWs, there are relationships between the extent of a node's influence on the final collective opinion in the DeGroot model and the stationary density of RWs. Before proceeding with our discussion, note that a recent generalization of the DeGroot model combines the averaging rule of the former with an appraisal mechanism (See Ref.~\cite{friedkinbook} and references therein.) to describe the dynamics of individuals' self-appraisal and social power in a network \cite{friedkin2015}. For nonlinear opinion-formation dynamics that allow non-consensus steady states, see Refs.~\cite{Sznajdweron2000IntJModPhysC,Krapivsky2003PhysRevLett,Castellano2009RevModPhys,Schweitzer2009EurPhysJB,ShaoHavlin2009PhysRevLett}.

In the DeGroot model, the opinion of node $v_i$ at discrete time $n$ is given by a continuous variable $x_i(n)$. One assumes that node $v_j$ weighs the opinion $x_i(n)$ of node $v_i$ with weight $A_{ij}$ to determine its opinion in the next time step (i.e., $x_j(n+1)$). The normalization is $\sum^N_{i=1}A_{ij}=1$, and the dynamics are given by
\begin{equation}
	x_i(n)=\sum^N_{j=1}A_{ji} x_j(n-1)\qquad (i \in \{1, \dots, N\})\,.
\label{eq:DeGroot dynamics}
\end{equation}
In the DeGroot model, the column sum of $A$ is equal to $1$ for every column, and recall that the row sum of $T$ is equal to $1$ for every row in a DTRW. To see the correspondence between the two models, it is convenient to write Eq.~\eqref{eq:DeGroot dynamics} in vector form as follows:
\begin{equation}
	\bm x(n) = A^{\top} \bm x(n-1)\,,
\end{equation}
where $\bm x(n) = (x_1(n),\; \ldots \; ,x_N(n))^{\top}$. Because the row sum of $A^{\top}$ equals $1$, we can identify $A^{\top}$ with $T$. The DeGroot model and DTRWs are thus driven by the same matrix, so their dynamics are essentially the same. The only difference is that the state vector is multiplied on the left in the RW, but it is multiplied on the right in the DeGroot model. Up to rescaling, the models are characterized by the same eigenvalues and eigenvectors.

As long as the spectral gap of $T$ (i.e., $A^{\top}$) is positive, the stationary density of a DTRW is given uniquely by the left eigenvector of $T$ whose corresponding eigenvalue is $1$. Under the same condition, the asymptotic state of the DeGroot model is given by the corresponding right eigenvector of $A^{\top}$. This eigenvector is $\bm x^* = (x_1^*\;, \ldots \;, x_N^*)^{\top} \propto (1\;, \ldots \; , 1)^{\top}$, and it corresponds to a state with full consensus.

The initial opinion $x_i(0)$ of node $v_i$ affects the value of the final opinion $x_1^* = \cdots = x_N^*$ in consensus. If $x_1^* = \cdots = x_N^*$ is close to $x_i(0)$ (for a general set of initial conditions that we will specify below)
one interprets node $v_i$ as being influential.
To quantify this idea, we postulate that $\sum_{i=1}^N F_i^{\rm DG, disc} x_i(n)$ is conserved over time for positive constants
$F_i^{\rm DG, disc}$ (with $i \in \{1,\dots,N\}$), where the superscript ``disc'' stands for discrete time and $\sum_{i=1}^N F_i^{\rm DG, disc} = 1$ gives the normalization. If such a conserved quantity exists, one obtains
\begin{equation}
	\sum_{i=1}^N F_i^{\rm DG, disc} x_i(0) = \sum_{i=1}^N F_i^{\rm DG, disc} x_i^* = x_1^* = \cdots =x_N^*\,.
\label{eq:DeGroot discrete time final opinion}
\end{equation}
Equation~\eqref{eq:DeGroot discrete time final opinion} implies that $F_i^{\rm DG, disc}$ quantifies the influence of $v_i$ on the final opinion in consensus. By imposing this conservation law, one obtains
\begin{align}
	\sum^N_{i=1} F_i^{\rm DG, disc}x_i(n-1) &=
\sum^N_{i=1} F_i^{\rm DG, disc}x_i(n) \notag \\
	&=\sum^N_{i=1} F_i^{\rm DG, disc} \left(\sum_{j=1}^N A_{ji}x_j(n-1)\right)\,.
\label{eq:DeGroot discrete time conservation}
\end{align}
By requiring that Eq.~\eqref{eq:DeGroot discrete time conservation} holds for arbitrary $x_i(n-1)$ (with $i \in \{1,\dots ,N\}$), we obtain
\begin{equation}
	F_i^{\rm DG, disc}=\sum_{j=1}^N A_{ij}F_j^{\rm DG, disc}\,.
\label{eq:Fi^DG final}
\end{equation}
Equation~\eqref{eq:Fi^DG final} indicates that $F_i^{\rm DG, disc}$ is the stationary density of the DTRW whose transition-probability matrix is $A^{\top}$.

A continuous-time variant of the DeGroot model has similar relationships \cite{MasudaKawamuraKori2009NewJPhys}. Consider the continuous-time DeGroot model \cite{Olfatisaber2007IEEE}
\begin{equation}
	\frac{{\rm d}x_i(t)}{{\rm d}t}=\sum_{j=1}^N A_{ji}\left[x_j\left(t\right)
-x_i\left(t\right)\right]\,,
\label{eq:DeGroot dynamics continuous time}
\end{equation}
and note that we do not impose $\sum_{j=1}^N A_{ji}=1$.
The asymptotic state of Eq.~\eqref{eq:DeGroot dynamics continuous time} is given by $x_1^*= \cdots =x_N^*$. Similar to the discrete-time DeGroot model above, we rewrite Eq.~\eqref{eq:DeGroot dynamics continuous time} as
\begin{equation}
	\frac{{\rm d}\bm x(t)}{{\rm d}t} = \left(A^{\top} - D^{\rm rev}\right) \bm x(t) \equiv - L^{\rm rev}\bm x(t)\,.
\label{eq:DeGroot dynamics continuous time vector form}
\end{equation}
Recall that $D^{\rm rev}$ is the diagonal matrix whose ($i$, $i$)th element equals $s_i^{\rm in}$, and  $L^{\rm rev}$ is the combinatorial Laplacian matrix for the edge-reversed network. The left eigenvector of $L^{\rm rev}$ corresponding to eigenvalue $0$ gives the stationary density of the Poissonian edge-centric CTRW on the edge-reversed network. The corresponding right eigenvector gives the asymptotic state of the continuous-time DeGroot model. Moreover, this eigenvector is the consensus state $\bm x^* \propto (1,\; \ldots\;, 1)^{\top}$. Equation~\eqref{eq:DeGroot dynamics continuous time vector form} also has a fascinating interpretation as linear synchronization dynamics that results from linearizing nonlinear systems such as coupled Kuramoto oscillators \cite{Arenas2008PhysRep,rodrigues2016}. See, for example, the discussion in \cite{Dedomenico2017PhysRevLett}.

Equation~\eqref{eq:DeGroot dynamics continuous time vector form} yields
\begin{equation}
	\bm p^* \frac{{\rm d}\bm x(t)}{{\rm d}t} = - \left(\bm p^* L^{\rm rev}\right) \bm x(t) = 0\,,
\end{equation}
where $\bm p^* = (p_1^*, \; \ldots \; , p_N^*)$, and $p_i^*$ is the stationary density of the Poissonian edge-centric CTRW at node $v_i$ in the edge-reversed network. Therefore, $\bm p^* \bm x(t)$ is conserved, implying that $\sum_{i=1}^N p_i^* x_i(0) = \sum_{i=1}^N p_i^* x_i^* = x_1^* = \cdots = x_N^*$. We thereby see that $p_i^*$ quantifies the influence of node $v_i$ on the final opinion, similar to the case of the discrete-time DeGroot model.


\section{Conclusions and outlook\label{sec:outlook}}

Random walks
%
%
play a central role in network science. As we have seen in this review, RWs are at the core of numerous methods to extract information from networked systems, and they serve as a leading-order model for (conservative) diffusion processes on networks. Because conventional RWs are linear processes, they are amenable to analysis. For example, one can exploit methods from linear algebra to characterize dynamics in terms of modes relaxing on different time scales, and one can even derive analytical solutions (e.g., via recursive equations) for quantities such as mean first-passage time (MFPT).
The simplicity of RWs is crucial, because associated dynamical properties on networks can be analyzed exactly, allowing one to uncover mechanisms by which network structure affects dynamical processes, which is perhaps the primary goal of studying dynamical processes on networks \cite{Porter2016book}. Many nonlinear processes (e.g., reaction--diffusion systems) include terms related to linear diffusion, so studying RWs on networks also yields important insights into the linear stability (and weakly nonlinear regimes) of numerous nonlinear processes.

RWs have been studied thoroughly (especially on networks) for many decades, but there remains much exciting work to be done. In the following paragraphs, we discuss a few important directions in the study of RWs on networks. As with the rest of our paper, these suggestions are far from exhaustive, and we look forward to seeing new theory and applications of RWs. As we have discussed at length, RWs have connections both to many other processes and to a diverse variety of applications, and we look forward especially to new, unexpected connections that will come to light in the coming years.

One prominent research direction is ``non-backtracking RWs'', which have opened new perspectives in recent years in topics such as community detection \cite{Krzakala2013PNAS,Newman2013arxiv-backtracking,Bordevane2015Ieeeconf}, because of the convenient properties of their spectrum for sparse networks. Non-backtracking spreading processes have also been used in the examination of network centralities \cite{martin2014PRE}, percolation theory \cite{Karrer2014PhysRevLett,Hamilton2014PhysRevLett}, and the design of efficient immunization algorithms \cite{Morone2015Nature}. Non-backtracking RWs are a type of second-order Markov chain (see Section~\ref{sec:memory net}), and their further study may provide algorithms for clustering and other applications that are more efficient and/or realistic than current ones. As we have illustrated in this review, one can define different types of RWs on the same network, and different RWs lead to different processes, algorithms, and insights.

Intrinsically, community detection and other forms of clustering are a type of model reduction, as one seeks to represent a given network (or dynamical process on a network) using a smaller amount of information. InfoMap (see Section~\ref{sub:Infomap}) is a community-detection algorithm that is constructed explicitly on this principle. Related techniques include coarse-graining RWs in a way that preserves the spectral properties of relevant matrices \cite{Gfeller2007PhysRevLett,Gfeller2008PhysRevLett}, external equitable partitions \cite{Oclery2013PhysRevE}, and using computational group theory to find ``hidden'' symmetries in networks \cite{Pecora2014}.
More generally, RWs are at the heart of flow-based algorithms, and they have been exploited to examine node centralities (see Section~\ref{sub:ranking}), community structure (see Section~\ref{sub:community}), core--periphery structure (see Section~\ref{sub:core-periphery}), and the mapping of networks into a Euclidean feature space \cite{LiCampbell2017arxiv}. It may also be fruitful to exploit similar ideas to examine other types of network properties (e.g., ``role similarity'' \cite{Henderson2012Kdd,Beguerissediaz2014JRSocInterface}, ``rich clubs'' \cite{Zhou2004IEEECommLett,Colizza2006NatPhys}, and approximately multipartite structure \cite{Newman2007PNAS}). RWs have also been used for some studies of community structure in temporal and multilayer networks \cite{Mucha2010Science,Dedomenico2015PhysRevX,Jeub2015PhysRevE,Jeub2017NetwSci} as well as for examining diffusion processes and centralities in such networks \cite{naturephysicsspreading,Gomez2013PhysRevLett,Soleribalta2013PhysRevE,Taylor2017MultiscaleModelSimul,Halu2013PlosOne,Dedomenico2014PNAS,Soleribalta2016PhysicaD}, and much more remains to be discovered in such applications. In temporal networks, for example, it is important to consider the relative timescales of the network dynamics and the RW dynamics. Novel types of RWs also play an important role in examining higher-order network structure. Examples include the spacey RW \cite{Benson2017SiamRev,spacey-tensor2016}, RWs on hypergraphs \cite{lu2011}, and RWs on simplicial complexes \cite{Mukherjee2016RandStructAlgor}.

One can also combine RWs with other dynamical processes to model real-world phenomena in fascinating and insightful ways. For example, one can couple RWs to other processes in multilayer networks \cite{naturephysicsspreading,nicosia2014}, where it is important to study scenarios such as infection spreading coupled to human/animal mobility (and more generally to study diffusion dynamics coupled to other types of dynamics). One very successful family of models that combines multiple types of dynamics is metapopulation models of biological contagions, in which individuals move from one subpopulation to another in some way (e.g., according to an RW) and infection events occur within each subpopulation \cite{Colizza2007NatPhys,Vespignani2011NatPhys}. Metapopulation models, reaction--diffusion models \cite{Colizza2007NatPhys,ReactionDiffusion}, and many other dynamical processes on networks often feature diffusion in the form of a simple, memoryless Poisson process. The use of more complicated and realistic RW processes such as higher-order Markov chains (see Section~\ref{sec:memory net}) and CTRWs driven by non-Poissonian renewal processes (see Sections~\ref{sub:CTRW} and \ref{sub:CTRW nets}) may yield interesting results.

Various types of RWs continue to be employed actively for a diverse array of applications. We mentioned several examples in Section~\ref{sec:introduction}, and we now indicate a few more applications of different types of RWs. For example, a ``hungry RW'' (taking some inspiration from the arcade game Pac-Man) has yielded insights into anomalous diffusion in bacteria \cite{hungry2016}, a ``waddling RW'' allows one to devise an efficient sampler for estimating the frequency of small subgraphs in a network \cite{HanSethu2016IeeeConfDataMin}, L\'{e}vy flights can help capture features of animal foraging \cite{Viswanathan1999Nature,Humphries2010Nature}, multiplicative RWs are a useful approach for examining the dynamics of financial markets \cite{CampbellLoMackinlay1996book,Mantegna1999book}, self-avoiding RWs have helped improve understanding of polymer chains \cite{fisher1966,isic1992}, the stochastic dynamics of neuronal firing have been studied using Ornstein--Uhlenbeck processes (a type of CTRW with a leak term) \cite{Tuckwell1988book2,Gabbiani2010book}, and the dynamics of correlated novelties (and Kauffman's so-called ``adjacent possible'') have been modeled using an RW on a growing network (representing the growing space of possible innovations) \cite{strogatz2014}.

\begin{figure}[tb]
\begin{center}
\includegraphics[scale=0.074]{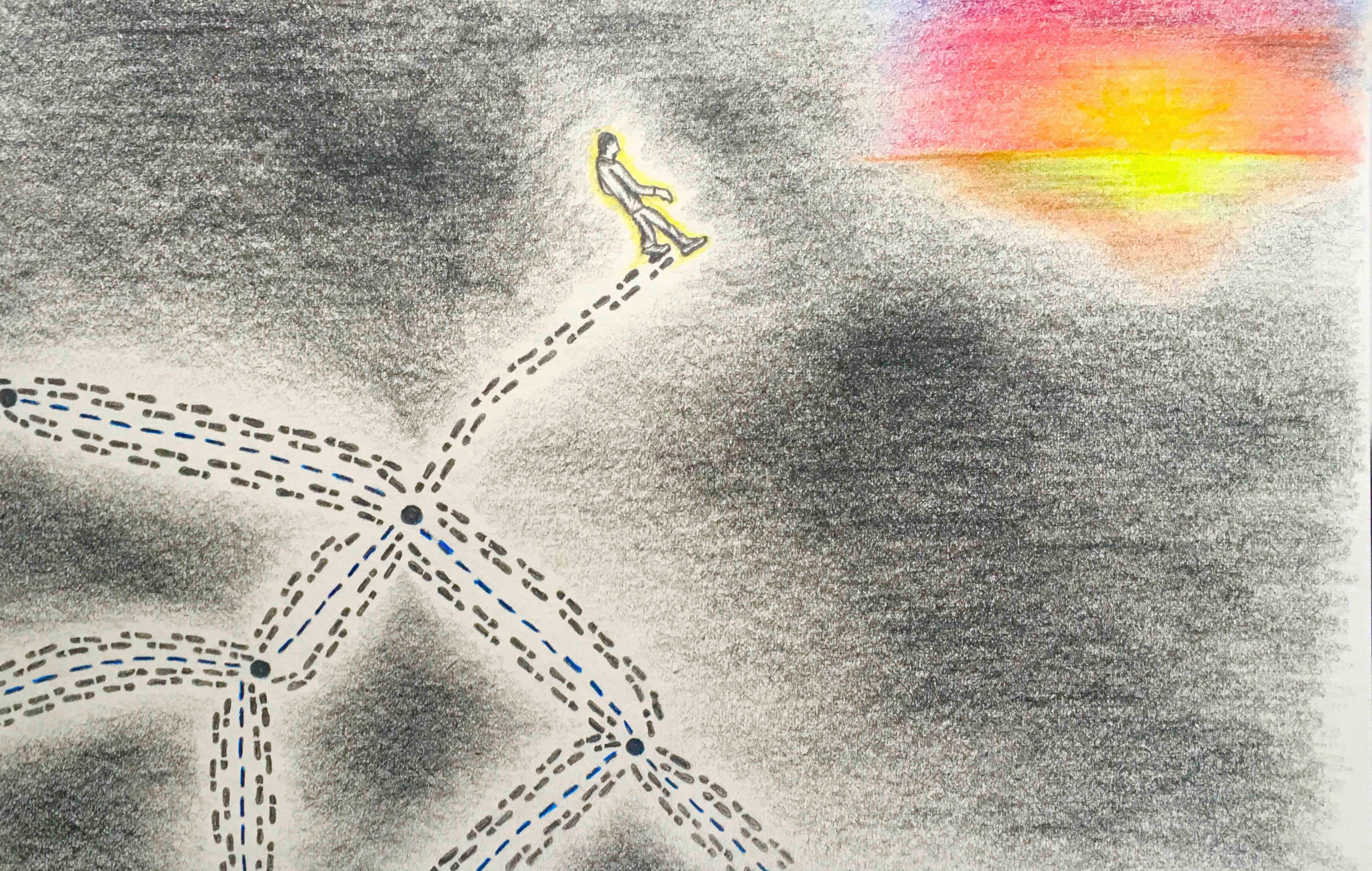}
\caption{The weary random walker retires from the network and heads off into the distant sunset. [This picture was drawn by Yulian Ng.]}
\label{art}
\end{center}
\end{figure}

In the coming years, we expect that RWs will continue to play a crucial role in physics, computer science, biology, sociology, and numerous other fields. The study of RWs continues to yield fascinating, important, and inspiring insights. Given how much random walkers have contributed to our scientific knowledge, they must be exhausted by now (see Fig.~\ref{art}).

\section*{Acknowledgments}

We thank Mariano Beguerisse-D\'{i}az, Michael T. Schaub, Taro Takaguchi, Johan Ugander, and Fabian Ying for valuable feedback on the manuscript. N.M. acknowledges the support provided through JST, CREST; and the JST, ERATO, Kawarabayashi Large Graph Project. R.L. acknowledges the support provided through Actions de Recherche Concert\'{e}e (ARC) Mining and Optimization of Big Data Models and the Belgian Network Dynamical Systems, Control, and Optimization (funded by the Interuniversity Attraction Poles Programme). We thank Yulian Ng for drawing Fig.~\ref{art}.


\section*{References}


\end{document}